\documentclass[aps,prd,superscriptaddress,nofootinbib,amsmath,amsfonts,preprintnumbers,groupedaddress,showpacs,10pt,english]{revtex4-1}
\usepackage{amsmath}
\usepackage{amssymb}
\usepackage{babel}
\usepackage{wrapfig}
\usepackage{cancel}
\newcommand{\nn}{\nonumber \\}
\newcommand{\e}{\mathrm{e}}
\usepackage{relsize,exscale}
\makeatletter

\usepackage{array,multirow,graphicx}
\usepackage{dcolumn}
\usepackage{newlfont}
\usepackage{bm}
\usepackage[colorlinks,citecolor=blue,urlcolor=blue,linkcolor=blue]{hyperref}
\usepackage[figtopcap]{subfigure}
\usepackage{color}

\allowdisplaybreaks[4]

\begin{document}

\tolerance=5000

\date{\today}

\title{ Rotating black hole in $f(R)$ theory}
\
\author{G.~G.~L.~Nashed}
\email{nashed@bue.edu.eg}
\affiliation {Centre for Theoretical Physics, The British University, P.O. Box
43, El Sherouk City, Cairo 11837, Egypt\\
\& \\
Int. Lab. Theor. Cosmology, Tomsk State University of Control
Systems and Radioelectronics (TUSUR), 634050 Tomsk, Russia}
\author{Shin'ichi~Nojiri}
\email{nojiri@gravity.phys.nagoya-u.ac.jp}
\affiliation{Department of Physics, Nagoya University, Nagoya 464-8602,
Japan \\
\& \\
Kobayashi-Maskawa Institute for the Origin of Particles and the Universe,
Nagoya University, Nagoya 464-8602, Japan }


\date{}

\begin{abstract}
In general, the field equation of $f(R)$ gravitational theory is very intricate, and therefore, it is not an easy task to derive analytical solutions.
We consider rotating black hole spacetime four-dimensional in the $f(R)$ gravitational theory and derive a novel black hole solution.
This black hole reduced to the one presented in \cite{Nashed:2020mnp} when the rotation parameter, $\Omega$, vanishes.
We study the physical properties of this black hole by writing its line element and show that it asymptotically behaves as the AdS/dS spacetime.
Moreover, we derive the values of various invariants finding that they do possess the central
singularity, and show that our black hole has a strong singularity compared with the black hole of the Einstein general relativity (GR).
We calculate several thermodynamical quantities and show that we have two horizons, the inner and outer Cauchy horizons in contrast to GR.
From the calculations of thermodynamics, we show that the outer Cauchy horizon gives satisfactory results for the Hawking temperature, entropy, and quasi-local energy.
Moreover, we show that our black hole has a positive value of the Gibbs free energy which means that it is a stable one.
Finally, we derive the stability condition analytically and graphically using the geodesic deviation method.
\end{abstract}

\maketitle
\section{Introduction}
The gravitational waves which were confirmed recently from the binary systems of black holes (BHs)
and the nulti-messenger signals from the first observation of the collision of two relativistic neutron stars have proved
some expectations of the strong-field regime of general relativity (GR) \cite{Abbott:2016blz,Abbott:2016nmj,Abbott:2017vtc,TheLIGOScientific:2017qsa}.
Furthermore, the event horizon telescope collaboration \cite{Mann:2018xkm} has published its first photo of a supermassive BH at the center of M87.
All the above observations ensure the existence of BHs which were ancient problems, like the central singularity and the information paradox, even more pressing.

It is known that astrophysical BHs are represented by the Kerr vacuum spacetime, which is a solution of the Einstein equation of GR.
This solution has a ring-shaped singularity at the center.
Because spacetime singularities are big issues in GR, many explanations have been directed to practical differences from the Kerr model.
Moreover, the Kerr spacetime has an instability that happens close to the inner Cauchy horizon constructing a new spacetime singularity
contrasting to the central singularity which is known as ``mass inflation'' \cite{Poisson:1989zz,Poisson:1990eh,Ori:1991zz}.

Shortly, the experimental data could help in investigating the essence of the origins of the gravitational waves which we are talented to discover.
The ring-down waveform of a BH is fully determined by using the Quasi-normal-modes (QNMs), which rely on the gravitational mass and angular momentum of the BH.
Therefore, every deviation from the standard result of GR may be associated with alternative theories of gravity or different nature of the source concerning
the case of the Kerr BH \cite{Simpson:2018tsi}.
In this respect, the possible presence of additional ``echoes'' in the ringdown waveform has been largely debated in the last years
(see the exhaustive review in Ref. \cite{Cardoso:2019rvt}) and alternatives to the BHs as gravastars \cite{Carballo-Rubio:2017tlh},
boson stars, or other exotic compact objects \cite{Mark:2017dnq,Cardoso:2016oxy,Hod:2017cga,Cardoso:2014sna,Sebastiani:2018ktb}
have been investigated.

Despite  the successes of Einstein's GR, which shows compatibility with observation, like the cosmic microwave background,
the solar system to the redshift of Type Ia Supernovae, the galaxy clustering, and the gravity waves \cite{Abbott:2016blz,Abbott:2016nmj,Abbott:2017vtc}.
However, GR still has defects with observational evidence of dark matter particle candidate \cite{Antoniadis:1985pj} which has intensive studies in last years.
Many modified gravitational theories have been constructed to solve the issue of dark matter \cite{Capozziello:2011et,Martin-Moruno:2015kaa,Bhattacharya:2016lup}.
These modified gravitational theories aimed to explain the accelerated cosmic expansion without involving a cosmological constant.
Moreover, the modified gravitational theories were able to investigate accelerated expansion by replacing the Einstein-Hilbert Lagrangian density that
depends on the Ricci scalar by a function of the Ricci scalar which is known as $f(R)$ gravitational theory.
Despite the failure of $f(R)$ in explaining the fine-tuning problem with the cosmological constant at the quantum level,
it is still considered as a viable theory that is capable to fit the observed data and can as well forecast something qualitatively novel and verifiable.
One merit of the $f(R)$ gravity in comparison with other modified gravitational theories is the absence of ghost instabilities \cite{Cardoso:2018tly,Nojiri:2017ncd,Nojiri:2010wj,DeFelice:2010aj}.

Amended theories should involve BH solutions similar to the Schwarzschild one to be compatible with GR prediction and must provide novel BH solutions
which could provide physical interest.
Because of this fact, the method to derive analytic or asymptote BH solutions is very important to explain if observations can coincide to
the amended gravity \cite{Capozziello:2007wc,Capozziello:2009jg}.
Within the frame of $f(R)$ there is a definite interest in spherically symmetric BH solutions.
There are many BH solutions derived using the constant Ricci scalar \cite{PhysRevD.74.064022}.
Additionally, spherically symmetric BH solutions, containing perfect fluid matter, have been examined \cite{PhysRevD.76.064021}.
Moreover, through the use of the Noether symmetry, different spherically symmetric BHs have been studied \cite{PhysRevD.76.064021}.
Furthermore, analytic solutions of static spherically symmetric spacetimes in $f(R)$ coupled to non-linear electrodynamics have been analyzed \cite{Hollenstein:2008hp}.
For those who are concerned with the static BHs we refer to \cite{PhysRevD.90.084011,Hendi:2011eg,Nashed:2005kn,PhysRevD.85.044012,Nojiri:2014jqa,Nojiri:2017kex,Shirafuji:1997wy,
PhysRevD.84.084006, Awad:2017tyz,Cembranos:2011sr,Hendi:2011eg,Nashed:2006yw,PhysRevD.80.104012,Azadi:2008qu,Nashed:2009hn,Nashed:2008ys,
Capozziello:2007id,PhysRevD.76.024020,Nashed:2007cu,Nashed:2015pda,PhysRevD.91.104004,2013CQGra..30l5003N,Nashed:2016tbj,2014CaJPh..92...76H,Awad:2017sau,
Hendi:2014mba,Nashed:uja,Nashed:2015qza,PhysRevD.81.124051,Nashed:2018piz,2018EPJP..133...18N,2018IJMPD..2750074N,PhysRevD.80.121501,Hendi:2012nj,
Myrzakulov:2015kda,PhysRevD.92.124019,Hanafy:2015yya,PhysRevLett.114.171601,PhysRevD.75.027502} and references there.
Through the Lagrangian multiplier, novel analytic solutions whose Ricci scalar are non-trivial are analyzed in \cite{Sebastiani:2010kv}.

Now let us focus on the BHs in $f(R)$ gravitational  theories \cite{Brevik:2004sd,Cognola:2005de,Saffari:2007zt,delaCruz-Dombriz:2009pzc,
Nashed:2014sea,Larranaga:2011fv,delaCruz-Dombriz:2012bni,Moon:2011hq}, where a special set of
solutions is derived under the constrains of the constant Ricci curvature scalar $R =\mbox{constant}=R_0$.
By taking the trace of the equation of motion of $f(R)$ then assigns this constant in terms of the function $f(R)$ and its derivative
at $R_0$. A comparison with GR and its BH solutions discovers, that the finite curvature scalar behaves essentially as a cosmological constant.
Thus, in a vacuum the Schwarzschild-AdS/dS and Kerr-AdS/dS BH solutions are regained, when specific rescalings are carried out.
Moreover, when adding a charge to the BH solutions by imposing an electromagnetic field, the
Reissner-Nordstr\"om-AdS/dS and the Kerr-Newman dS/dS BH solutions can be regained after
specific rescalings, since the trace of the energy-momentum tensor, vanishes identically.
It is the aim of this study, through the use of the field equation of $f(R)$, to derive new rotating BH solutions with the non-trivial Ricci scalar asymptotically
converging towards flat or AdS/dS spacetimes.

This study is organized as follows:
In Sec.~\ref{S2}, a summary of $f(R)$ gravity is presented.
In Sec.~\ref{S3}, a non-diagonal spherically symmetric ansatz that has three unknown functions is applied to the field equation of $f(R)$ theory,
and an exact solution is derived.
We solve the resulting non-linear differential equation in an exact way assuming a specific form of the first derivative of $f(R)$ in Sec.~\ref{S3}.
In Sec.~\ref{S55}, we discuss the significant physics of this BH and write its metric potential in an asymptotic form.
We show that this BH has an effective cosmological constant that comes from the form of $f(R)$ and shows that it behaves asymptotically
as AdS/dS when the effective cosmological constant does not vanish and when it is vanishing it behaves like a flat spacetime.
In Sec.~\ref{S66}, we study the thermodynamics behavior of this BH and show that it has two horizons, the inner and outer Cauchy horizons
in contrast with Einstein's GR.
Also in Sec.~\ref{S66}, we calculate some thermodynamical quantities like the Hawking temperature, entropy, quasi-local energy, and the Gibbs free energy.
We show that those thermodynamical quantities have acceptable behavior for the outer Cauchy horizon. In Sec.~\ref{S666},
We derive the stability condition of our BH analytically and graphically using the geodesic deviation.
In Sec.~\ref{S77}, we discuss the main results of our study and conclude.

\section{$f(R)$ gravitational theory}\label{S2}

In this paper we are interested in $f(R)$ gravity which was studied by \cite{1970MNRAS.150....1B},
see also \cite{Capozziello:2011et,2010deto.book.....A,Capozziello:2003gx,Capozziello:2002rd,Nojiri:2003ft,Carroll:2003wy}.
The action of the $f(R)$ gravitational theory in a vacuum has the form,
\begin{eqnarray}
\label{a1}
S=\frac{1}{2\kappa^2} \int d^4x \sqrt{-g}\left[ R+f(R) \right]\,,
\end{eqnarray}
where $\kappa^2$ is the gravitational constant, $R$ is the Ricci scalar, $f(R)$ is an analytic differentiable function of $R$, and $g$ is the determinant of the metric.
Equation~(\ref{a1}) shows that when $f(R)=0$, we return to the standard GR of Einstein.

The equation of motion of $f(R)$ theory is obtained by making the variations of the action given by Eq.~(\ref{a1}) w.r.t. the metric tensor $g_{\mu \nu}$
and the equation has the following form \cite{2005JCAP...02..010C,Koivisto:2005yc},
\begin{eqnarray}
\label{f1}
Q_{\mu \nu}\equiv R_{\mu \nu} [1+f_R]-\frac{1}{2}g_{\mu \nu} \left[ R+f(R) \right] +g_{\mu \nu} \Box f_R-\nabla_\mu \nabla_\nu f_R = 0\, ,
\end{eqnarray}
where $R_{\mu \nu}$ is the Ricci tensor that has the following form,
\begin{equation}
R_{\mu \nu}=R^{\rho}{}_{\mu \rho \nu}= 2\Gamma^\rho{}_{\mu [\nu,\rho]}+2\Gamma^\rho{}_{\beta [\rho}\Gamma^\beta{}_{\nu] \mu} \, ,
\end{equation}
where $\Gamma^\rho{}_{\mu \nu}$ is the second kind Christoffel symbol.
The d'Alembert operator $\Box$ is defined as $\Box= \nabla_\alpha\nabla^\alpha $ where $\nabla_\alpha W^\beta$
is the covariant derivatives of the vector $W^\beta$ and $f_R=\frac{df(R)}{dR}$.
The trace of equation (\ref{f1}) has the following form,
\begin{equation}
\label{f3}
Q=f_R-R-2f(R)+3\Box f_R = 0\,.
\end{equation}
Solving the above equation for $f(R)$ and substituting its value into Eq.~(\ref{f1}),
we get
\begin{eqnarray}
\label{f3ss}
R_{\mu \nu} \left[ 1+f_R \right] - \frac{1}{4}g_{\mu \nu}R \left[1+f_R \right] + \frac{1}{4}g_{\mu \nu}\Box f_R
 -\nabla_\mu \nabla_\nu f_R=0 \,.
\end{eqnarray}
If we assume that the Ricci tensor is covariantly constant, i.e.,  $\nabla^\nu R_{\mu\nu}=constant$ that yields $R_{\mu\nu}=\mathrm{constant}\times g_{\mu\nu}$,
Eq.~(\ref{f3})  reduces to an algebraic equation,
\begin{equation}
\label{f3B}
f_R-R-2f(R) = 0\,.
\end{equation}
If Eq.  (\ref{f3B}) has a real solution at $R=R_0$, then the  solution of the Einstein field equations
with cosmological constant $\Lambda$ that has the form:
\begin{equation}
\label{Ein}
R_{\mu\nu} - \frac{1}{2} g_{\mu\nu} R = - \frac{1}{2} g_{\mu\nu} \Lambda \, , \quad
\Lambda = \frac{R_0}{2} \,
\end{equation}
will include the (anti-de Sitter or de Sitter-)Schwarzschild space time and
the Kerr-(anti-)de Sitter space-time
\begin{align}
\label{KdSmetric1}
ds^2 = & \left( r^2+a^2 \cos^2{\theta} \right) \left[\frac{dr^2}{\Delta} + \frac{d\theta^2}{1+\frac{\Lambda}{3} a^2 \cos^2 \theta}
\right] + \frac{\sin^2\theta \left( 1+\frac{\Lambda}{3} a^2 \cos^2 \theta \right)}{r^2+a^2 \cos^2{\theta}} \left[\frac{a dt - \left( r^2+a^2 \right)
d\phi}{1 + \frac{\Lambda}{3} a^2}\right]^2 \nonumber \\
& - \frac{\Delta}{r^2+a^2 \cos^2{\theta}}\left[\frac{dt - a \sin^2\theta \:d\phi}{1 + \frac{\Lambda}{3} a^2}\right]^2\, , \nn
\Delta \equiv& \left( r^2+a^2 \right) \left(1-\frac{\Lambda}{3}r^2 \right) - 2Mr \, ,
\end{align}
and in case $R_0=0$, binary black holes with the cylindrical Weyl coordinates $(t,\rho,z,\phi)$
\cite{Astorino:2021dju},
\begin{align}
\label{bibhmetric}
{ds}^2 & = - V(\rho,z) {dt}^2 + \frac{\rho^2}{V(\rho,z)} {d\phi}^2 + f(\rho,z) \left({d\rho}^2 + {dz}^2 \right) \, , \nn
V =& \frac{\mu_1\mu_3}{\mu_2\mu_4}
\exp\left[2b_1z + 2b_2 \left( z^2 - \frac{\rho^2}{2} \right) \right] , \\
f =& \frac{16C_f \mu_1^3\mu_2^5\mu_3^3\mu_4^5}{W_{11}W_{22}W_{33}W_{44}W_{13}^2W_{24}^2Y_{12}Y_{14}Y_{23}Y_{34}}
\exp \Bigl[ -b_1^2\rho^2 + \frac{b_2^2}{2} \left( \rho^2 - 8z^2 \right) \rho^2 - 4b_1b_2 z \rho^2 \nn
& + 2 b_1 \left( -z + \mu_1 - \mu_2 + \mu_3 - \mu_4 \right)
+ b_2 \left( -2z^2 + \rho^2 + 4z \left( \mu_1 - \mu_2 \right) + \mu_1^2 - \mu_2^2
+ \left( \mu_3 - \mu_4 \right) \left( 4z + \mu_3 + \mu_4 \right) \right) \Bigr]\, , \nn
&W_{ij}=\rho^2+\mu_i\mu_j\, , \quad
Y_{ij}= \left( \mu_i-\mu_j \right)^2\, , \quad \mu_i=\sqrt{\rho^2+ \left( z-w_i \right)^2} - \left( z-w_i \right) \, .
\end{align}
Here $w_i$'s are constants, $w_1<w_2<w_3<w_4$, given by
\begin{equation}
\label{wis}
w_1 = z_1 - m_1\, , \quad
w_2 = z_1 + m_1\, , \quad
w_3 = z_2 - m_2\ , \quad
w_4 = z_2 + m_2\, ,
\end{equation}
where the parameters $m_i$ and $z_i$ are the mass and the position of the $i$-th black hole, respectively,
and $b_1$ and $b_2$ are the dipoles and quadrupole momenta of the external gravitational field polar expansion.
The constant $C_f$ is a gauge parameter.

In this paper, we consider the axially symmetric or rotating black hole space-time besides
the above solutions (\ref{KdSmetric1}) and (\ref{bibhmetric})


\section{An exact charged black hole solution }\label{S3}
Let us derive a black hole solution adopting the model $f(R)$.
Now we are interested in the axially symmetric solution as in Eqs. (\ref{KdSmetric1}) and (\ref{bibhmetric}).
Due to the technical reasons that we could be able to solve the resulting differential equations (\ref{f3ss}) exactly\footnote{
The other forms of the line-element (\ref{met}) which involve cross term $dtd\phi$ make the field equations (\ref{f3ss}) very complicated and not easy to solve. },
we use the following axially symmetric ansatz although it does not include the metric of the Kerr spacetime (\ref{KdSmetric1}),
\begin{equation}
\label{met}
ds^2=-\alpha(r)\,dt^2+\frac{dr^2}{\beta(r)}+r^2\,d\theta^2+\left(\frac{r^2\,\alpha(r)\,\sin^2\theta-\Omega^2\,\gamma(r)}{\alpha(r)}\right)\,d\phi^2
 -2\Omega\,\sqrt{\gamma(r)}\,dtd\phi\,,
\end{equation}
where $\alpha(r)$, $\beta(r)$, and $\gamma(r)$ are arbitrary functions and $\Omega$ is a constant parameter.
The above line element reduced to the diagonal spherically symmetric metric when the constant parameter $\Omega=0$ \cite{Nashed:2020mnp}.
The Ricci scalar of the line-element (\ref{met}) is given by:
\begin{align}
\label{r1}
R =& \frac {\cot^2 \theta}{8 \alpha^{3}\gamma{r}^{2}} \left\{4\, \alpha'\beta' {r}^{2} \alpha^{2}\gamma -4\, \alpha'^{2}\beta\gamma r^{2}\alpha-16\,\alpha^{3}\gamma
+16\, \alpha^{3}\beta\gamma+8\,\alpha''{r}^{2} \alpha^{2}\beta\gamma+16 \, \beta' \alpha^{3}r\gamma+ 16\,\beta\gamma\alpha' r \alpha^{2} \right\} \nonumber\\
& +\frac {1}{8\alpha^{3}\gamma{r}^{2} \sin^2 \theta}\left\{16\, \alpha^{3}\gamma
 -8\,\alpha'' {r}^{2} \alpha^{2}\beta\gamma-4\, \alpha'\beta'{r}^{2}\alpha^{2}\gamma+{ \Omega}^{2}\gamma'^{2}\alpha ^{2}\beta+4\, \alpha'^{2}\beta\gamma{r}^ {2}\alpha
 -16\, \alpha^{3}\beta\gamma+4\,\alpha'^{2}\beta \gamma ^{2}{ \Omega}^{2} \right. \nonumber\\
& \left. -16\,\beta'\alpha^{3}r\gamma-4\,{\Omega}^{2}\alpha'\gamma' \alpha\beta\gamma-16\,\beta\gamma \alpha' r\alpha^ {2}\right\} \,,
\end{align}
where $\alpha\equiv \alpha(r)$, $\beta\equiv \beta(r)$, $\gamma\equiv \gamma(r)$, $\alpha'\equiv \frac{d\alpha(r)}{dr}$,
$\beta'\equiv \frac{d\beta(r)}{dr}$, $\gamma'\equiv \frac{d\gamma(r)}{dr}$, and $\alpha''\equiv \frac{d^2\alpha(r)}{dr^2}$.
The expression of the Ricci scalar, $R$, in (\ref{r1}) coincides with that derived in \cite{Nashed:2020mnp} when $\Omega=0$.
Applying the ansatz (\ref{met}) to the field equations~(\ref{f3ss}), we get the differential equations, whose explicit forms are given in (\ref{df1}) of Appendix~\ref{AI}.
The system of differential equations reduce to that derived in \cite{Nashed:2020mnp} when $\gamma=0$ and coincides with GR differential equations
of diagonal line-element when $\Omega=0$ and $f(R)=0$. The analytic solutions of the above system take the form:
\begin{align}
\label{sol1}
\mbox{Case i)} \quad & \gamma(r)=0\,,\quad \alpha(r)=a_1\e^{\frac{-3c_0}{r^2}}\beta(r)\,,\quad F(r)=\frac{c_0}{r^2}\,,\nonumber\\
&\beta(r)=\frac{\e^{\frac{3c_0}{2r^2}}}{r} \left\{ \mathbb{H}a_2+\mathbb{H}_1r^3a_3+2\mathbb{H}_1r^3\int\frac{\e^{- \frac{3c_0}{2r^2}}\mathbb{H}}{r \left[
\left( 2c_0\mathbb{H}_2 - 3r^2\mathbb{H} \right) \mathbb{H}_1-2c_0\mathbb{H}\mathbb{H}_3 \right] }dr \right. \nonumber \\
& \qquad \qquad \qquad \left. -2\mathbb{H}\int\frac{\e^{- \frac{3c_0}{2r^2}}r^2 \mathbb{H}_1}{ \left( 2c_0\mathbb{H}_2 - 3r^2\mathbb{H} \right)\mathbb{H}_1
 -2c_0\mathbb{H}\mathbb{H}_3}dr \right\}\,,\nonumber\\
\mbox{Case ii)} \quad & \alpha(r)=c_1\,{\e^{^{-{\frac {3c_0}{{r}^{2}}}}}}\beta(r)\,,\quad F(r) =\frac{c_0}{r^2}\,,\nonumber\\
& \beta(r)=\frac{ 2\,\e^{\frac {3c_0}{2{r}^{2}}} \mathbb{H}_1{r}^{2}}{c_3c_2} \left[ \int \frac{\e^{\frac {3c_0}{2{r}^{2}}}}{\mathbb{H}_1{}^2
{r}^2 \left( {r}^{2}+c_0 \right)}{dr} \right. \nonumber \\
& \qquad \qquad \qquad \qquad \left. -c_3\,c_2\, \int \left\{\frac{\e^{\frac {3c_0}{2{r}^{2}}}}{\mathbb{H}_1{}^2{r}^2 \left( {r }^{2}+c_0 \right)}
\left( \int \frac{\mathbb{H}_1 \e^{-{\frac {3c_0}{{r}^{2}}}} \left( {r}^{2}+c_0 \right)}{ {r}^2}{dr}\right)\right\}dr+c_4 \right]\,,\nonumber\\
& \gamma(r)=-\frac{\mathbb{H}_1{}^2{r}^{4}}{4 \e^{\frac {3c_0}{{r}^{2}}} c_3{}^2} \left[ \int
\frac{\e^{\frac {3c_0}{2{r}^{2}}}}{ \mathbb{H}_1{}^2 {r}^2 \left( {r}^{2}+c_0 \right)} dr \right. \nonumber \\
& \qquad \qquad \qquad \left. -c_3\,c_2\,
\int \left\{\frac{\e^{\frac {3c_0}{2{r}^{2}}}}{\mathbb{H}_1{}^2{r}^2 \left( {r}^{2}+c_0 \right)}
\left( \int \left[ \mathbb{H}_1 {
\e^{-{\frac {3c_0 }{{r}^{2}}}}}+\frac{ \mathbb{H}_1\e^{-\frac {3c_0}{{r}^{2}}}c_0} {{r}^2}\right] dr\right)\right\} dr+c_4 \right]\,,
\end{align}
where $a_1$, $a_2$, $a_3$, $c_0$, $c_1$, $c_2$, $c_3$, and $c_4$ are constants and $\mathbb{H}=\mathrm{HeunC} \left( \frac{3}{2},\frac{3}{2},0,\frac{3}{8},\frac{9}{8},-\frac{c_0}{r^2} \right)$,
$\mathbb{H}_1=\mathrm{HeunC} \left( \frac{3}{2},-\frac{3}{2},0,\frac{3}{8},\frac{9}{8},-\frac{c_0}{r^2} \right)$,
$\mathbb{H}_2=\mathrm{HeunCPrime} \left(\frac{3}{2},\frac{3}{2},0,\frac{3}{8},\frac{9}{8},-\frac{c_0}{r^2} \right)$,
$\mathbb{H}_3=\mathrm{HeunCPrime} \left( \frac{3}{2},-\frac{3}{2},0,\frac{3}{8},\frac{9}{8},-\frac{c_0}{r^2} \right)$\footnote{
The $\mathrm{HeunC}$ function is the solution of the Heun Confluent equation which is defined as
\[
X''(r)-\frac{1+\beta-(\alpha-\beta-\gamma-2)r-r^2\alpha}{r(r-1)}X'(r)-\frac{\alpha(1+\beta)-\gamma-2\eta-(1+\gamma)\beta-r(2\delta+[2+\gamma+\beta])}
{2r(r-1)}X(r)=0\,.
\]
The solution of the above differential equation defines $\mathrm{HeunC} \left( \alpha,\beta,\gamma,\delta,\eta,r \right)$ for more details, interested readers
can check \cite{RONVEAUX2003177,MAIER2005171}.
The $\mathrm{HeunCPrime}$ is the derivative of the Heun Confluent function.}.
The first set of solution (\ref{sol1}) is obtained before and the applications to physics are explained in \cite{Nashed:2020mnp}.


Now we turn our attention to the second set which is a new solution.
We stress the fact that there is no coordinate transformation that can eliminate the cross term and the only way to get a diagonal metric is to set $\Omega=0$.
The second set of Eq.~(\ref{sol1}) shows that when $c_0=0$ we get:
\begin{equation}
\label{reda1}
\alpha(r)=\beta(r)\,,\quad \gamma=16\,\alpha^2\,,\quad \mathrm{and} \quad F(r)=0\, .
\end{equation}
Equation~(\ref{reda1}) shows that when $F(r)=0$ this gives $f(R)=\mathrm{constant}$ and in that case $\alpha(r)=\beta(r)=1+\frac{c_3}{r}$
provided that $c_1=1$ and $c_4=0$, which is related to the cosmological constant.
All the above data ensure that when $c_0=0$, we return to the GR BHs\footnote{
Note that when $c_0=0$, we get \cite{RONVEAUX2003177,MAIER2005171},
\begin{align}
& \mathbb{H}=\mathbb{H}_1=\mathrm{HeunC} \left( \frac{3}{2},\frac{3}{2},0,\frac{3}{8},\frac{9}{8},0 \right)
=\mathrm{HeunC} \left( \frac{3}{2},  -\frac{3}{2},0,\frac{3}{8},\frac{9}{8},0 \right)=\,1, \nonumber \\
& \mathbb{H}_2=\mathrm{HeunCPrime} \left( \frac{3}{2},\frac{3}{2},0,\frac{3}{8},\frac{9}{8},-\frac{c_1}{r^2} \right)=0\, ,\quad
\mbox{and} \quad \mathbb{H}_3=\mathrm{HeunCPrime} \left( \frac{3}{2},-\frac{3}{2},0,\frac{3}{8},\frac{9}{8},-\frac{c_1}{r^2} \right)=-\frac{3}{2}\, .
\nonumber
\end{align}
}.

The second set of the analytic solution (\ref{sol1}) satisfies the system of differential equations (\ref{df1}) including the trace of the field equations.
Using the second set of Eq.~(\ref{sol1}) in Eq.~(\ref{r1}), we get the Ricci scalar in the following form:
\begin{align}
\label{ri}
R=& \frac{4 \e^{\frac {21c_0}{2r^2}}}{c_3c_2 r^2 \mathbb{H}_1 \left( c_0+r^2 \right)^2}
\left\{c_2\,c_3\,\mathbb{H}_1 \e^{-\frac {21c_0 }{2r^2}}{c_0}^2 +10\, r^2 \mathbb{H}_1 \e^{-\frac {9c_0}{r^2}}{c_0}^{2}\mathbb{H}_3 c_4
+c_2\,c_3\,{r}^{4}\mathbb{H}_1 {\e^{-{\frac {21c_0}{2{r}^{2}}}}}-3c_0\,r{\e^{-{\frac {15c_0}{2{r}^{2 }}}}} \right. \nonumber \\
& +9\,{r}^{4} \mathbb{H}_1{}^{2}{\e^{-{\frac {9c_0}{{r}^{2}}}}}c_0c_4+ 6\,{r}^{6} \mathbb{H}_1{}^{2}{\e^{-{\frac {9c_0}{{r}^ {2}}}}}c_4
+c_2c_3r{\e^{-{\frac {15c_0}{2{r}^{2}}}}} \left[ 2{r}^{2}+3c_0 \right]\int \frac{\mathbb{H}_1{\e^{-{\frac {3c_0}{{ r}^{2}}}}} \left( {r}^{2}+c_0 \right)}{ {r}^{2}}{dr}
 -2\,{r}^{3 }{\e^{-{\frac {15c_0}{2{r}^{2}}}}} \nonumber\\
& +4\, {r}^{4}\mathbb{H}_1{\e^{-9\,{\frac {c_0}{{r}^{2}}}}}\mathbb{H}_3c_4+2\,c_2c_3\,{r}^{2}\mathbb{H}_1 {\e^{-21/2\, {\frac {c_0}{{r}^{2}}}}}c_0
+3\,{r}^{2}\mathbb{H}_1{}^{2}{\e^{-{\frac {9c_0}{{r}^{2}}}}}{c_0}^{2}c_4\nonumber\\
&+c_2c_3r^{2} {\e^{-{\frac {9c_0}{{r}^{2}}}}}\mathbb{H}_1{c_0} \left[3\mathbb{H}_1{c_0}+9{r}^{2} \mathbb{H}_1+4{r}^{2}\mathbb{H}_3+10{c_0}\mathbb{H}_3
+ \frac{6r^4\mathbb{H}_1}{c_0} + \frac{6c_0{}^2\mathbb{H}_3}{r^2} \right] \nonumber \\
& \qquad \qquad \times \left(\int {\frac{\e^{{\frac {3c_0}{2{r}^{2}}}}}{\mathbb{H}_1{}^{2}{r}^{2} \left( {r}^{2}+c_0 \right)}
\left({\int \frac{\mathbb{H}_1 {\e^{-{\frac {3c_0}{{r}^{2}}}}} \left( {r}^{2}+c_0 \right)}{ {r}^{2}}}{dr}\right)}\right){dr}\nonumber\\
& - \e^{- \frac {9c_0}{r^2}} \left[ 6\,\mathbb{H}_1\mathbb{H}_3 {c_0}^{3}+10{r}^{2}\mathbb{H}_1{c_0}^{2 }
+3\,{r}^{2}\mathbb{H}_1{}^{2} \left\{ {c_0}^{2}+ 3{r}^2c_0+2{r}^{4} - \frac{4}{3r^2} c_0\mathbb{H}_3 \right\} \right] \nonumber \\
& \qquad \qquad \left.  \times \int \frac{\e^{\frac{3c_0}{2 r^2}}}{\mathbb{H}_1{}^2 r^2 \left( r^2 +c_0 \right)} dr
+ 6\,\mathbb{H}_1 {\e^{-{\frac {9c_0}{{r}^{2}}}}} {c_0}^{3} \mathbb{H}_3 c_4 \right\} \,.
\end{align}
The above Ricci scalar vanishes when $c_0=0$ and $c_4=0$ and has a non-vanishing value when $c_4\neq 0$,
which is also a consistency check for the whole procedure.
The metric of the above solution takes the form
\begin{align}
\label{met5}
ds^2=&-\left\{\frac{2\,c_1\,{\e^{-{\frac {3c_0}{2{r}^{2}}}}}\mathbb{H}_1{r}^{2}}{ {c_2}{c_3}}
\left(c_2c_3 \int^r {\frac{\e^{{\frac {3c_0}{2{r_1}^{2}}}}}{\mathbb{H}_1{}^{2}{r_1}^{2} \left( {r_1}^{2}+c_0 \right)}
\left({\int^{r_1} \frac{\mathbb{H}_1 {\e^{-{\frac {3c_0}{{r_2}^{2}}}}} \left( {r_2}^{2}+c_0 \right)}{ {r_2}^{2}}}{dr_2}\right)}{dr_1} \right. \right. \nonumber \\
& \left. \left. \qquad \qquad - \int^r \frac{\e^{\frac {3c_0}{{r_1}^{2}}} dr_1}{\mathbb{H}_1{}^2 {r_1}^{2}\left( {r_1}^{2}+c_0 \right)}+c_4\right)\right\}
dt^2\nonumber\\
& +\frac{c_3c_2\e^{{-\frac {3c_0}{2{r}^{2}}}}\,dr^2}{2\mathbb{H}_1{r}^{2}\left(c_2c_3 \int^r {\frac{\e^{{\frac {3c_0}{2{r_1}^{2}}}}}{\mathbb{H}_1{}^{2}{r_1}^{2}
\left( {r_1}^{2}+c_0 \right)} \left({ \int^{r_1} \frac{\mathbb{H}_1 {\e^{-{\frac {3c_0}{{r_2}^{2}}}}} \left( {r_2}^{2}+c_0 \right)}{ {r_2}^{2}}}{dr_2}\right)}{dr_1}
 - \int^r \frac{ {\e^{{\frac {3c_0}{{r_1}^{2}}}}} }{\mathbb{H}_1{}^2 {r_1}^{2}\left( {r_1}^{2}+c_0 \right)}{dr_1}+c_4\right)}+r^2d\theta^2\nonumber\\
&+{r}^{2} \left\{ \sin^2 \theta-\frac{{\Omega}^{2}c_2\,{\e^{-{\frac {3c_0}{2{r}^{2}}}}}\mathbb{H}_1{r}^{2}}{8 {c_1}{c_3}}
\left(c_2c_3 \int^r {\frac{\e^{{\frac {3c_0}{2{r_1}^{2}}}}}{\mathbb{H}_1{}^{2}{r_1}^{2} \left( {r_1}^{2}+c_0 \right)}
\left({\int^{r_1} \frac{\mathbb{H}_1 {\e^{-{\frac {3c_0}{{r_2}^{2}}}}} \left( {r_2}^{2}+c_0 \right)}{ {r_2}^{2}}}{dr_2}\right)}{dr_1} \right. \right. \nonumber \\
& \left. \left. \qquad \qquad - \int^r \frac{ {\e^{{\frac {3c_0}{{r_1}^{2}}}}} {dr_1}}{\mathbb{H}_1{}^2 {r_1}^{2}\left( {r_1}^{2}+c_0 \right)}+c_4\right)\right\}d\phi^2\nonumber\\
&+\Omega\left\{\frac{{\e^{-{\frac {3c_0}{2{r}^{2}}}}}\mathbb{H}_1{r}^{2}}{c_3}
\left(c_2c_3 \int^r {\frac{\e^{{\frac {3c_0}{2{r_1}^{2}}}}}{\mathbb{H}_1{}^{2}{r_1}^{2} \left( {r_1}^{2}+c_0 \right)}
\left({\int^{r_1} \frac{\mathbb{H}_1 {\e^{-{\frac {3c_0}{{r_2}^{2}}}}} \left( {r_2}^{2}+c_0 \right)}{ {r_2}^{2}}}{dr_2}\right)}{dr_1} \right. \right. \nonumber \\
& \left. \left. \qquad \qquad - \int^r \frac{ {\e^{{\frac {3c_0}{{r_1}^{2}}}}} dr_1 }{\mathbb{H}_1{}^2 {r_1}^{2}\left( {r_1}^{2}+c_0 \right)}+c_4\right)\right\}
dtd\phi\,.
\end{align}
It is an easy task to show that the metric in Eq.~(\ref{met5}) coincides with the Schwarzschild-AdS/dS metric
when $c_0=0$, which gives $F=0$, and we recover the GR.
Equations~(\ref{sol1}),  (\ref{ri}), and (\ref{met5}) show that when $c_4=0$, we get:
\begin{equation}
\label{reda}
\alpha(r)=\beta(r)\, , \quad \gamma=\frac{{c_2}^2}{16}+\frac{1}{31r^2c_3{}^2}-\frac{c_2}{12rc_3} \, , \quad \mbox{and} \quad F(r)=0 \, .
\end{equation}
Equation~(\ref{reda}) shows that when $F(r)=0$, this gives $f(R)=\mathrm{constant}$ and in that case $\alpha(r)=\beta(r)=1-\frac{2}{3rc_2c_3}$
provided that $c_4=0$.
All the above data ensure that when $c_0=0$, we return to the GR BHs\footnote{
Note that when $c_0=0$  we get \cite{RONVEAUX2003177,MAIER2005171},
\begin{align}
& \mathbb{H}=\mathbb{H}_1=\mathrm{HeunC} \left( \frac{3}{2},\frac{3}{2},0,\frac{3}{8},\frac{9}{8},0 \right)
=\mathrm{HeunC} \left( \frac{3}{2}, -\frac{3}{2},0,\frac{3}{8},\frac{9}{8},0 \right)=1\, , \nonumber \\
& \mathbb{H}_2=\mathrm{HeunCPrime} \left( \frac{3}{2},\frac{3}{2},0,\frac{3}{8},\frac{9}{8},-\frac{c_1}{r^2} \right)=0\, , \quad \mbox{and}
\mathbb{H}_3=\mathrm{HeunCPrime} \left( \frac{3}{2},-\frac{3}{2},0,\frac{3}{8},\frac{9}{8},-\frac{c_1}{r^2} \right)=-\frac{3}{2} \, .
\nonumber
\end{align}
}.

\section{Physical significance of the black hole }\label{S55}

The form of the solution given by the second set of Eq.~(\ref{sol1}) is not easy to extract from it any physics, therefore,
we are going to study its asymptote behavior.
The functions $\alpha$, $\beta$, and $\gamma$ in the asymptotic form take the following asymptote:
\begin{align}
\label{metasy}
\alpha\approx & c_1\left[1-\frac{2c_4r^2}{c_2c_3}-\frac{2}{3c_2c_3r}-\frac{c_0}{r^2}+\frac{c_0}{c_2c_3r^3}+\frac{5c_0{}^2}{6r^4}-\frac{29c_0{}^2}{28c_2c_3r^5}
 - \frac{7c_0{}^3}{10r^6} + \cdots\right]\,,\nonumber\\
\beta\approx& -\frac{2c_4r^2}{c_2c_3}-\frac{6c_0c_4}{c_2c_3}+1-\frac{2}{3c_2c_3r}+\frac{c_0[2c_2c_3-9c_0c_4]}{c_2c_3r^2}-\frac{c_0}{c_2c_3r^3}
+ \frac{c_0{}^2[7c_2c_3-27c_0c_4]}{3c_2c_3r^4} \nonumber \\
& - \frac{29c_0{}^2}{28c_2c_3r^5}+\frac{9c_0{}^3[4c_2c_3-15c_0c_4]}{20c_2c_3r^6}+\cdots\,,
\nonumber\\
\gamma\approx & \frac{c_4{}^2r^4}{4c_3{}^2}-\frac{c_4c_2r^2}{4c_3}+\frac{c_4r}{6c_3{}^2}+\frac{c_2{}^2}{16}-\frac{c_2}{12c_3r}+\frac{c_4c_0c_2}{4c_3}-\frac{c_0c_4}{4rc_3{}^2}
 -\frac{5c_0{}^2c_2c_4}{24c_3r^2} - \frac{c_2{}^2c_0}{8r^2}+\frac{1}{36c_3{}^2r^2} \nonumber \\
& +\frac{29c_0{}^2c_4}{112c_3{}^2r^3}+\frac{5c_0c_2}{24c_3r^3}+\frac{7c_0{}^3c_4c_2}{40c_3r^4}+\cdots\,.
\end{align}
The asymptotic form of the metric (\ref{met5}) can be rewritten as:
\begin{align}
\label{me}
ds^2\approx &\left(\Lambda_\mathrm{eff}r^2+1-\frac{2m}{r}-\frac{c_0}{r^2}+\frac{3c_0m}{r^3}+\frac{5c_0{}^2}{6r^4}+\cdots\right)dt^2
 -\left(\frac{1}{\Lambda_\mathrm{eff}r^2}-\frac{3c_0}{\Lambda_\mathrm{eff}r^4}-\frac{1}{\Lambda_\mathrm{eff}{}^2r^4}-\frac{1}{3c_4\Lambda_\mathrm{eff}r^5}+\cdots\right)dr^2 \nn
& -r^2d\theta^2 + \left[r^2\sin^2\theta-\Omega^2c_2{}^2\left\{\frac{\Lambda_\mathrm{eff}r^2}{16}+\frac{1}{16}-\frac{m}{8r}-\frac{c_0}{16r^2}+\frac{3c_0m}{16r^3}
\right\}+\cdots\right] d\phi^2 \nonumber\\
& +\Omega\left(\frac{c_4r^2}{c_3}-\frac{c_2}{2}+\frac{1}{3rc_3}
+\frac{c_0c_2}{2r^2}-\frac{c_0}{2c_3r^3}-\frac{5c_2c_0{}^2}{12r^4}+\cdots\right)dtd\phi\,,\nonumber\\
\end{align}
where $\Lambda_\mathrm{eff}=-\frac{2c_4}{c_2c_3}$, $m=\frac{1}{3c_2c_3}$, and $c_1=1$.
The line-element (\ref{me}) shows clearly that the BH solution (\ref{sol1}) behave asymptotically as AdS/dS spacetime when the constant $c_4\neq0$
and when this constant vanishes, the line element behaves like a flat spacetime. Also, the line-element (\ref{me}) shows clearly that the parameter $c_0$ is the source of the difference from Einstein's GR.

Let us study now the regularity of the BH solution given by the second set of Eq.~(\ref{sol1}) by evaluating their scalar invariants that take the
the following form:
\begin{align}
\label{scal1}
R^{\mu \nu \lambda \rho}R_{\mu \nu \lambda \rho}=& 16\Lambda_\mathrm{eff}{}^2+\frac{120c_0\Lambda_\mathrm{eff}{}^2}{r^2}+\frac{96c_0{}^2\Lambda_\mathrm{eff}{}^2}{r^3}
 - \frac{24c_0\Lambda_\mathrm{eff} \left( 1-21c_0\Lambda_\mathrm{eff} \right)}{r^4}+\cdots\, , \nonumber\\
R^{\mu \nu}R_{\mu \nu}=& 36\Lambda_\mathrm{eff}{}^2+\frac{180c_0\Lambda_\mathrm{eff}{}^2}{r^2}+\frac{144c_0{}^2\Lambda_\mathrm{eff}{}^2}{r^3}
 -\frac{18c_0\Lambda_\mathrm{eff} \left( 1-16c_0\Lambda_\mathrm{eff} \right)}{r^4}+\cdots\, ,\nonumber\\
R =& 12\Lambda_\mathrm{eff}{}^2+\frac{30c_0\Lambda_\mathrm{eff}{}^2}{r^2}+\frac{12c_0{}^2\Lambda_\mathrm{eff}{}^2}{r^3}
 -\frac{3c_0 \left( 2-7c_0\Lambda_\mathrm{eff} \right){}^2}{r^4}+\cdots \, ,
\end{align}
where $R^{\mu \nu \lambda \rho}R_{\mu \nu \lambda \rho}$, $R^{\mu \nu}R_{\mu \nu }$, and $R$ are the Kretschmann scalars, the Ricci tensor square,
and the Ricci scalar, respectively.
Equations~(\ref{scal1}) show that the solutions, at $r=0$, have true singularities.
Also Eq.~(\ref{scal1}) shows clearly that when the parameter $c_0=0$, we get the invariants of Einstein's GR.
Now let us calculate the form of $f(r)$ associated with the second set of solution (\ref{sol1}) and we get:
\begin{align}
\label{fr}
f(r) =& \frac{8 \left( { r}^{2}+c_0 \right) c_0}{ \left(\mathbb{H}_1 \right){r}^{7} \left( {r}^{ 2}+c_0 \right) ^{3}{c_2}{c_3}}\int^r \bigg\{2\,r \left( 15c_0{r_1}^{2}+8{r_1}^{4} +9{c_0}^{2} \right)
{\e^{{\frac {3c_0}{{r_1}^{2}}}}} \nonumber \\
& -2c_3\,c_2\,r_1 \left( 15c_0{r_1}^{2}+8\,{r_1}^{4}+{c_0}^{2} \right) {\e^{{\frac {3c_0}{{r_1}^{2}}}}}
\int^{r_1} \frac{ \e^{{\frac {-3c_0}{{r_2}^{2}}}} \mathbb{H}_1{}^2\left( {r_2}^{2}+c_0 \right)}{{r_2}^2}{dr_2} \nonumber\\
& + 6 { \e^{{\frac {3c_0}{2{r_1}^{2}}}}} \left( {r_1}^{2}+c_0 \right) \mathbb{H}_1\left[ \left( 5{r_1}^{4}c_0+3{r_1}^{6} \right) \mathbb{H}_1
+ 4\mathbb{H}_3 \left( 15c_0{r_1}^{2}+{8}\,{r_1}^{4}+{c_0}^{2} \right) c_0 \right]
\int^{r_1} \frac{ {\e^{{\frac {3c_0}{{r_2}^{2}}}}} }{\mathbb{H}_1{}^2 {r_2}^{2}\left( {r_2}^{2}+c_0 \right)}{dr_2} \nonumber \\
& + 6 c_4{r_1}^{4} {\e^{{\frac {3c_0}{2{r_1}^{2}}}}} \left( {r_1}^{2}+c_0 \right)\left(3{r_1}^{2}+5c_0 \right) \mathbb{H}_1{}^{2}\nonumber\\
& -6c_2c_3 {\e^{{\frac {3c_0}{2{r_1}^{2}}}}}\left( {r_1}^{2}+c_0 \right)\mathbb{H}_1 \left[ \left( 5{r_1}^{4}c_0+3{r_1}^{6} \right) \mathbb{H}_1
+4\mathbb{H}_3 \left( 15c_0{r_1}^{2}+{ 8}{r_1}^{4}+{c_0}^{2} \right) c_0 \right] \nonumber \\
& \quad \times \int^{r_1} {\frac{\e^{{\frac {3c_0}{2{r_2}^{2}}}}}{\mathbb{H}_1{}^{2}{r_2}^{2} \left( {_2r}^{2}+c_0 \right)}
\left({\int^{r_2} \frac{\mathbb{H}_1 {\e^{-{\frac {3c_0}{{r_3}^{2}}}}} \left( {r_3}^{2}+c_0 \right)}{ {_3r}^{2}}}{dr_3}\right)}{dr_2} \nonumber\\
& -2\left( {r_1}^{2}+c_0 \right) \left[ {\e^{{\frac {3c_0}{2{r_1}^{2}}}}}c_4\, \left(15 c_0{r_1}^{2}+8\,{r_1}^{4}+{c_0}^{2} \right) c_0\mathbb{H}_3
 -c_2\,c_3\,{r_1}^{2} \left( {r_1}^{2}+c_0 \right) \right] \mathbb{H}_1 \bigg\} {dr_1}+c_5\,.
\end{align}
Equation (\ref{fr}) shows in a clear way that when $c_0=0$, we get $f(r)=c_5$, which means that we return to GR case.
In the next section, we are going to study in more detail the physics of the BH (\ref{sol1}).

\section{Thermodynamics of the second set of the black hole (\ref{sol1})}\label{S66}

\begin{figure}
\centering
\subfigure[~The plot of $r_\pm$ vs. the mass $m$ ]{\label{fig:1a}\includegraphics[scale=0.25]{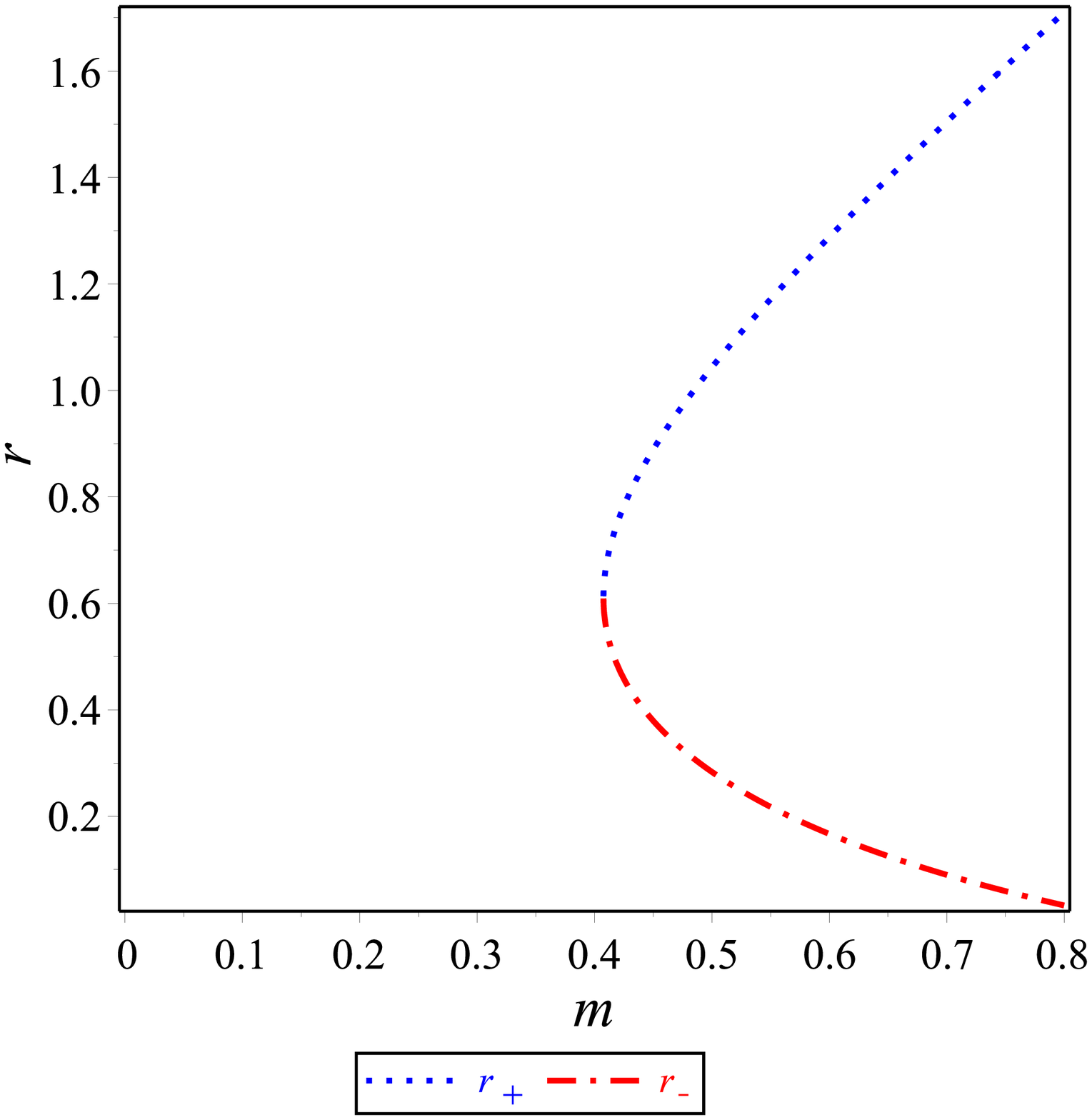}}\hspace{0.2cm}
\subfigure[~The plot of $r_\pm$ vs. the parameter $c_0$]{\label{fig:1b}\includegraphics[scale=0.25]{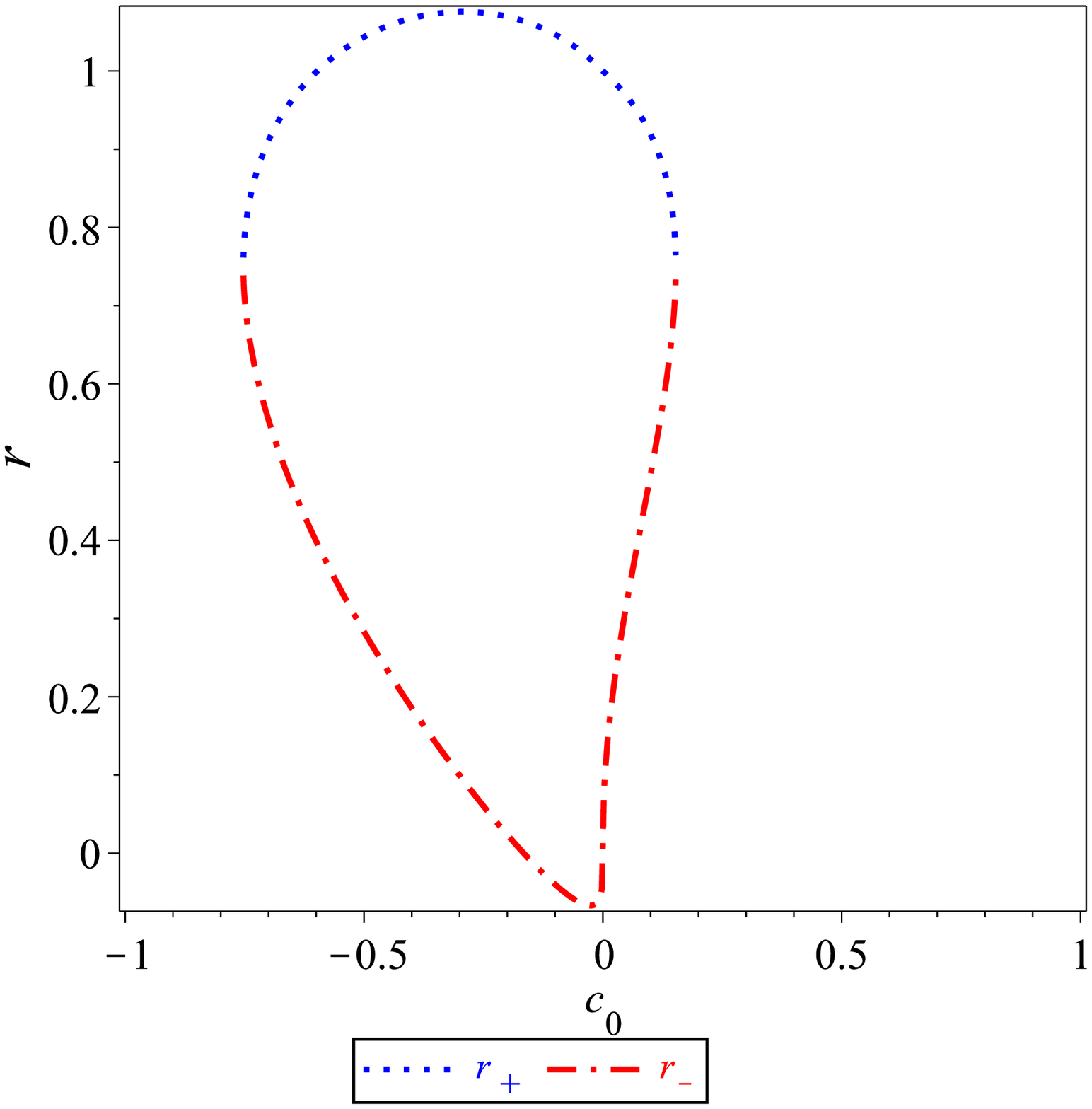}}\hspace{0.2cm}
\subfigure[~The metric potential $g_{00}$ vs. the radial coordinate $r$ ]{\label{fig:1c}\includegraphics[scale=0.25]{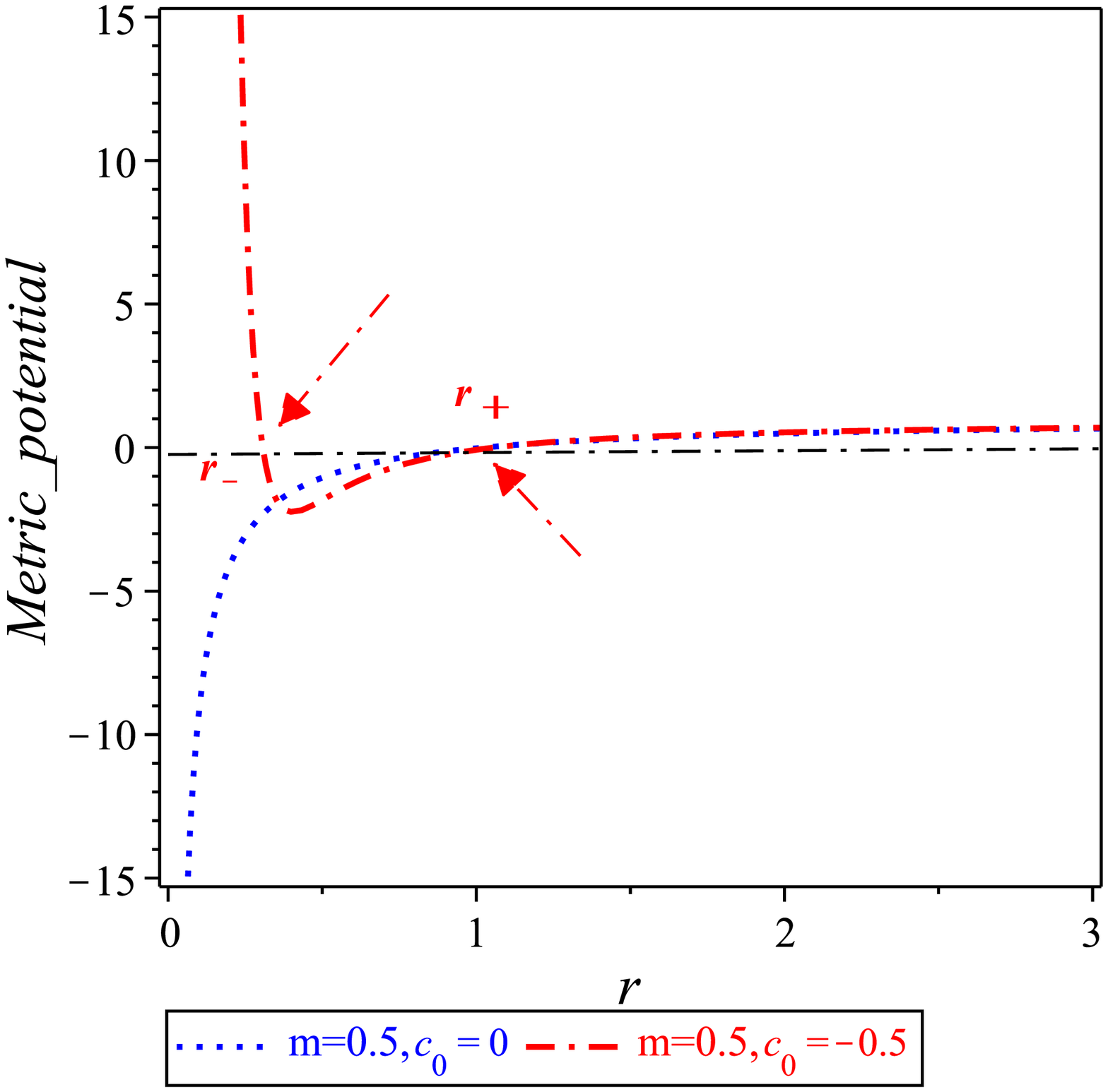}}\hspace{0.2cm}
\subfigure[~The naked singularity of the black hole (\ref{sol1}) ]{\label{fig:1d1}\includegraphics[scale=0.25]{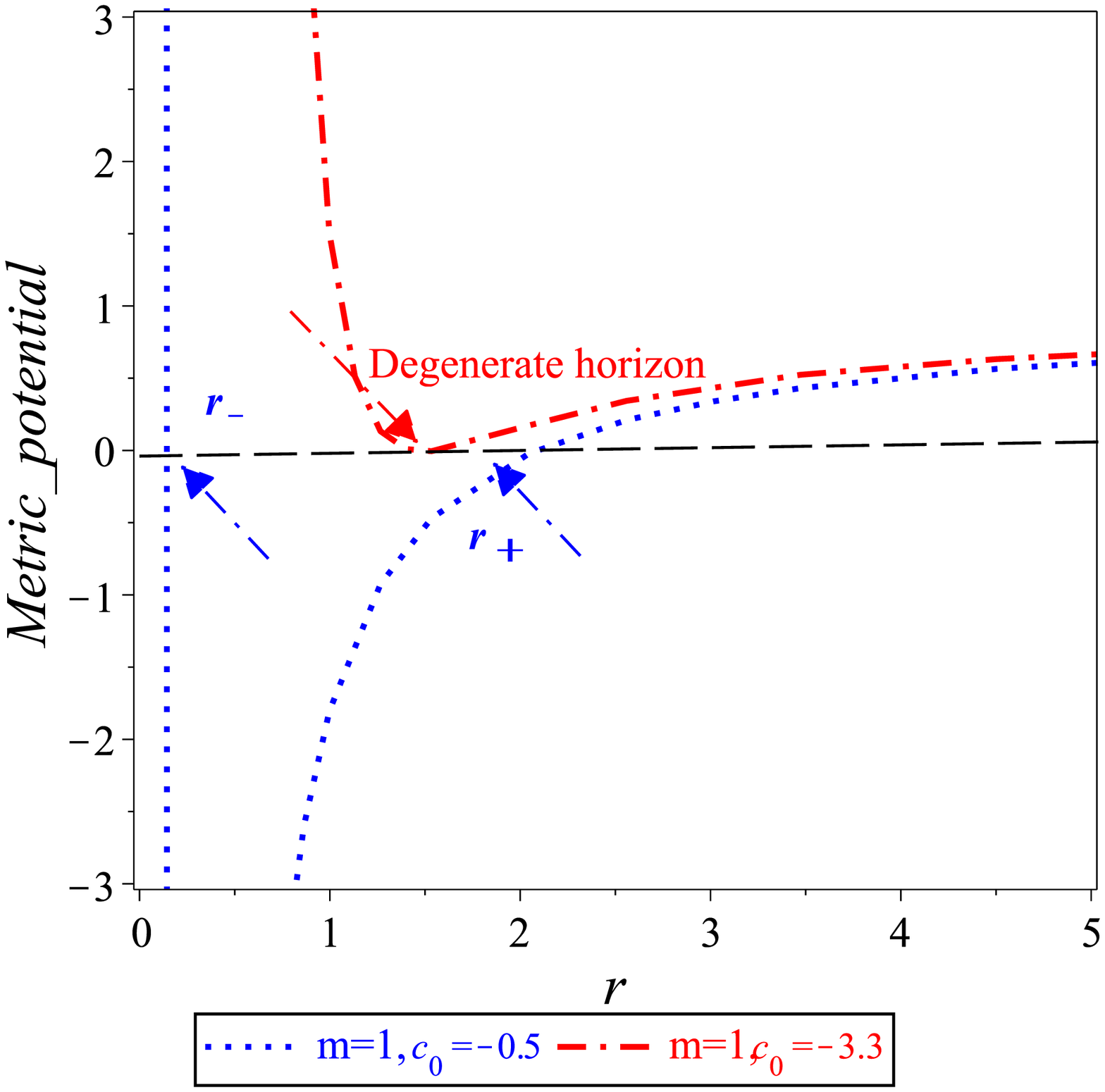}}
\caption{Schematic plots of the radial coordinate $r$ with respect to the $m$ Figure~\ref{Fig:1}~\subref{fig:1a}, with respect to $c_0$ Figure~\ref{Fig:1}~\subref{fig:1c},
the metric potential $g_{00}$ with respect to. $r$ of the black hole (\ref{sol1}) Figure~\ref{Fig:1}~\subref{fig:1c} and the naked singularity
Figure~\ref{Fig:1}~\subref{fig:1d1}.}
\label{Fig:1}
\end{figure}

Now we are going to explore the thermodynamics of the new black hole solutions derived in the previous section given by Eq.~(\ref{sol1}).
The Hawking temperature is defined as \cite{PhysRevD.86.024013,Sheykhi:2010zz,Hendi:2010gq,PhysRevD.81.084040},
\begin{equation}
\label{temp}
T_+ = \frac{\alpha'(r_+)}{4\pi} \, ,
\end{equation}
where the event horizon is located at $r = r_+$ which is the largest positive root of $\alpha(r_+) = 0$ that fulfils $\alpha'(r_+)\neq 0$.
The Bekenstein-Hawking entropy in the frame of $f(R)$ gravity is given as
\cite{PhysRevD.84.023515,PhysRevD.86.024013,Sheykhi:2010zz,Hendi:2010gq,PhysRevD.81.084040,Zheng:2018fyn},
\begin{equation}
\label{ent}
S(r_+)=\frac{1}{4}Af_{R}(r_+)\, ,
\end{equation}
where $A$ is the area of the event horizon.
The form of the quasi-local energy in the frame of $f(R)$ gravity is defined as
\cite{PhysRevD.84.023515,PhysRevD.86.024013,Sheykhi:2010zz,Hendi:2010gq,PhysRevD.81.084040,Zheng:2018fyn},
\begin{equation}
\label{en}
E(r_+)=\frac{1}{4} \int \left[ 2f_{R}(r_+)+r_+{}^2\left\{f \left( R \left(r_+ \right) \right) - R \left( r_+ \right)f_{R} \left( r_+ \right) \right\} \right]dr_+\, .
\end{equation}
At the horizon, one has the constraint $\alpha(r) = 0$, which gives,
\begin{equation}
\label{m33}
1-2\,{\frac {m}{r}}-{\frac {c_0}{{r}^{2}}} + \frac {3mc_0}{{r}^{3}} +\frac {5 {c_0}^{2}}{6{r}^{4}}
 - \frac {87 m{c_0}^{2}}{28 {r}^{5}} =0\,.
\end{equation}
Equation~(\ref{m33}) has two real roots, $r_\pm$ where $r_-$ is the inner Cauchy horizon and $r_+$ is the outer Cauchy horizon, and the other roots are imaginary.
The explicit form of the two real roots of Eq.~(\ref{m33}), $r_{\pm}$, are tedious, however, we plot their behavior in Figure~\ref{Fig:1}~\subref{fig:1a}
and \ref{Fig:1}~\subref{fig:1b} which shows the relation between the radial coordinate $r$ and the mass $m$,
and the relation between $r$ and the parameter $c_0$ that characterizes the higher order curvature terms.

To show how many horizons in the black hole of the second set of Eq.~(\ref{sol1}) has, we plot the metric potential $g_{00}$ concerning the radial coordinate
in Figure~\ref{Fig:1}~\subref{fig:1c}.
As Figure~\ref{Fig:1}~\subref{fig:1c} shows that in Einstein's GR, which corresponds to $c_0=0$, we have only one horizon and
when $c_0\neq 0$, which corresponds to higher-order curvature, we have two horizons.
In Figure~\ref{Fig:1}~\subref{fig:1d1}, we show the region where the black hole has a naked singularity.

\begin{figure}
\centering
\subfigure[~The entropy of the black hole solution (\ref{sol1}) with respect to the mass $m$ ]{\label{fig:2a}\includegraphics[scale=0.25]{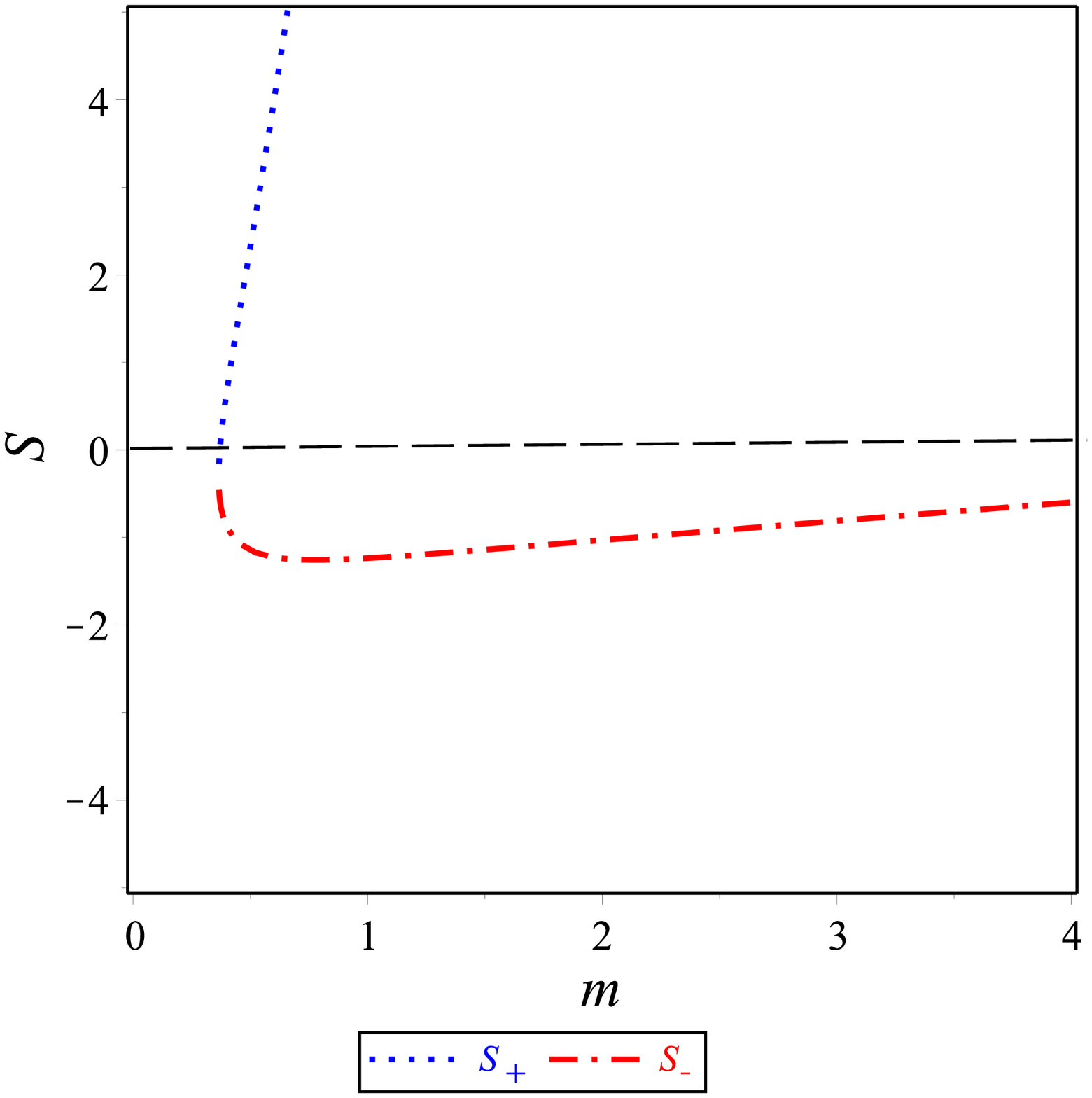}}\hspace{0.2cm}
\subfigure[~The entropy of the black hole solution (\ref{sol1}) with respect to the parameter $c_0$]{\label{fig:2b}\includegraphics[scale=0.25]{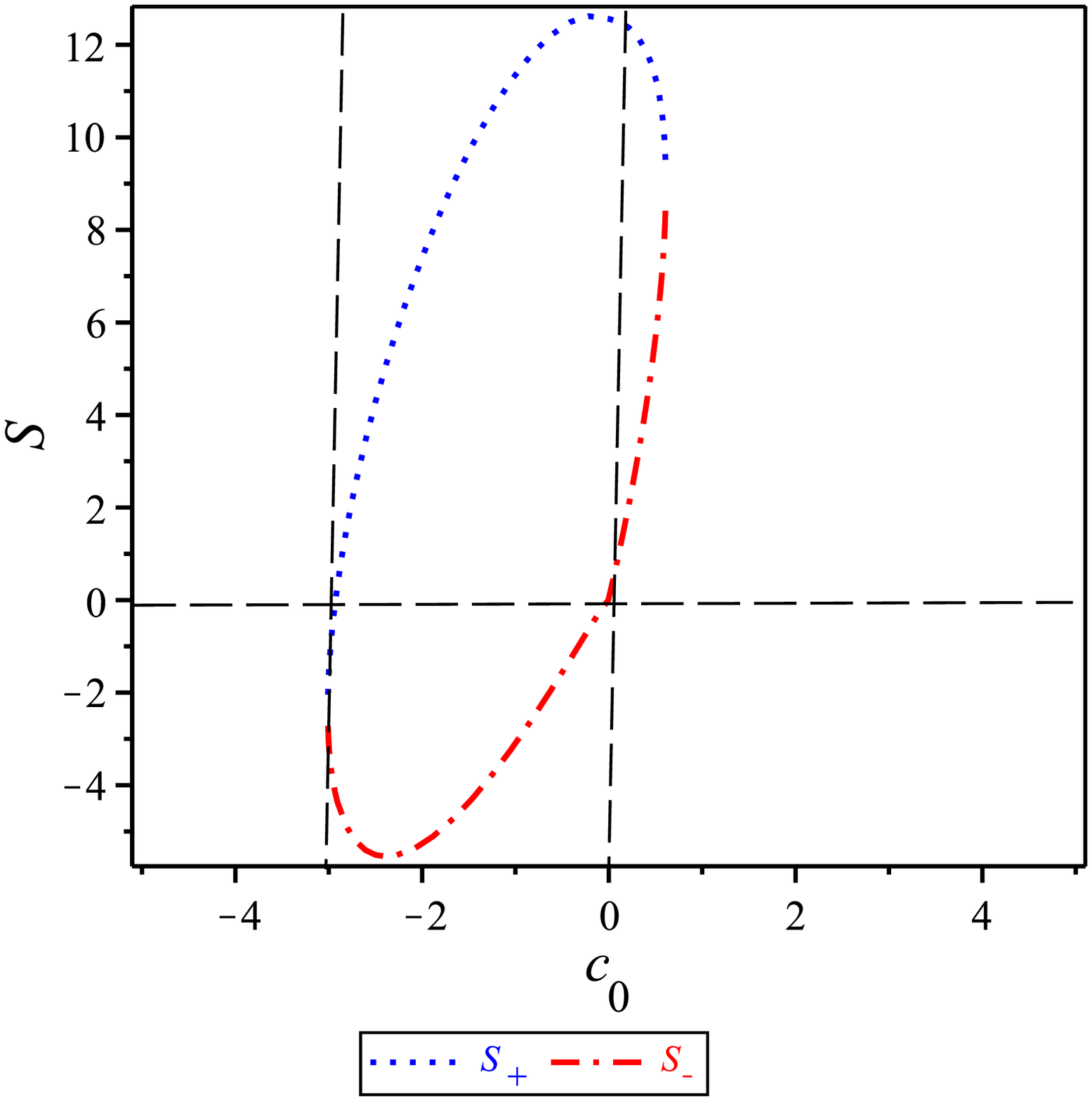}}\hspace{0.2cm}
\subfigure[~The temperature of the black hole solution (\ref{sol1}) with respect to the mass $m$ ]{\label{fig:2c}\includegraphics[scale=0.25]{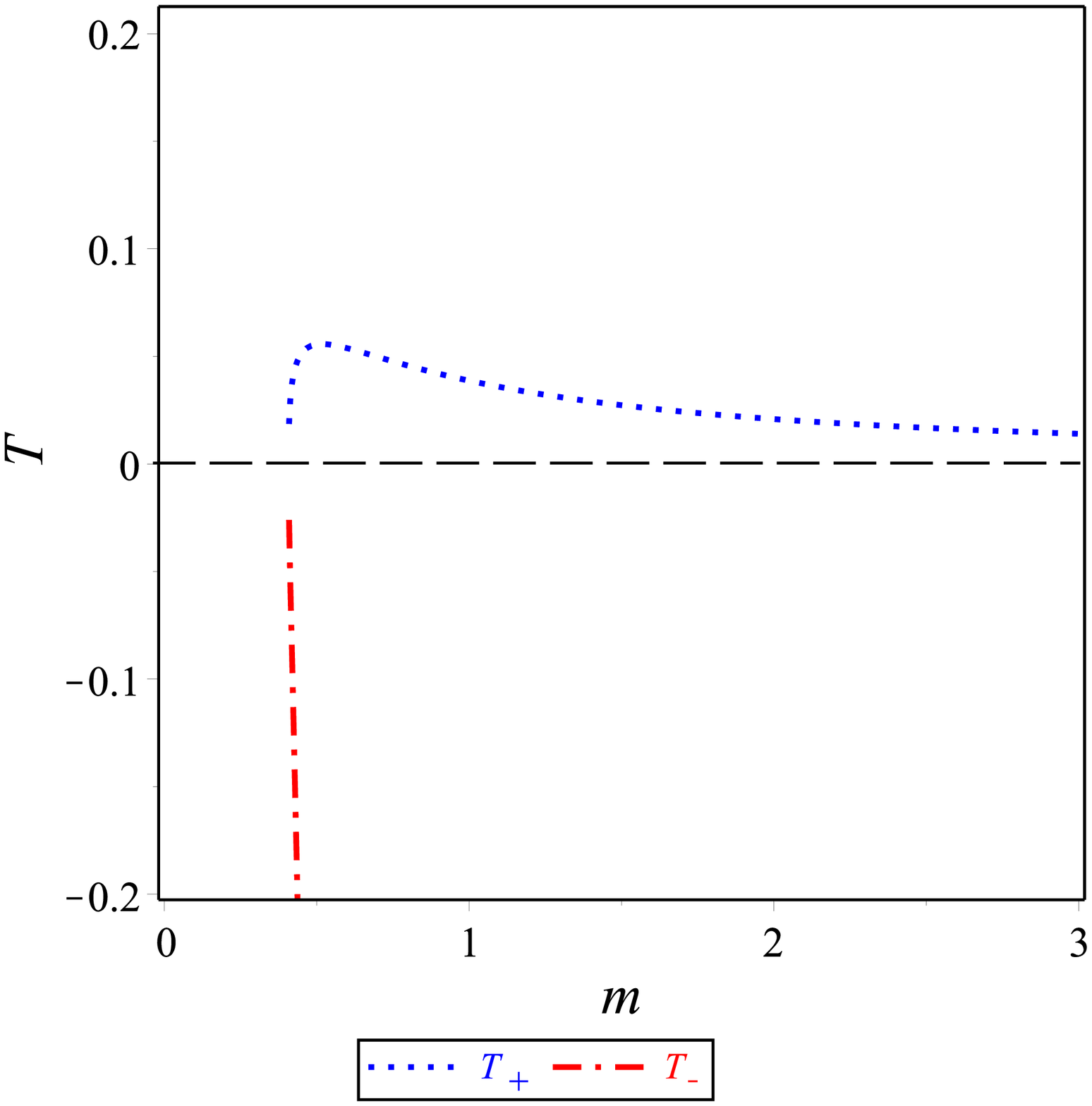}}\hspace{0.2cm}
\subfigure[~The temperature of the black hole solution (\ref{sol1}) with respect to the parameter $c_0$]{\label{fig:2d}\includegraphics[scale=0.25]{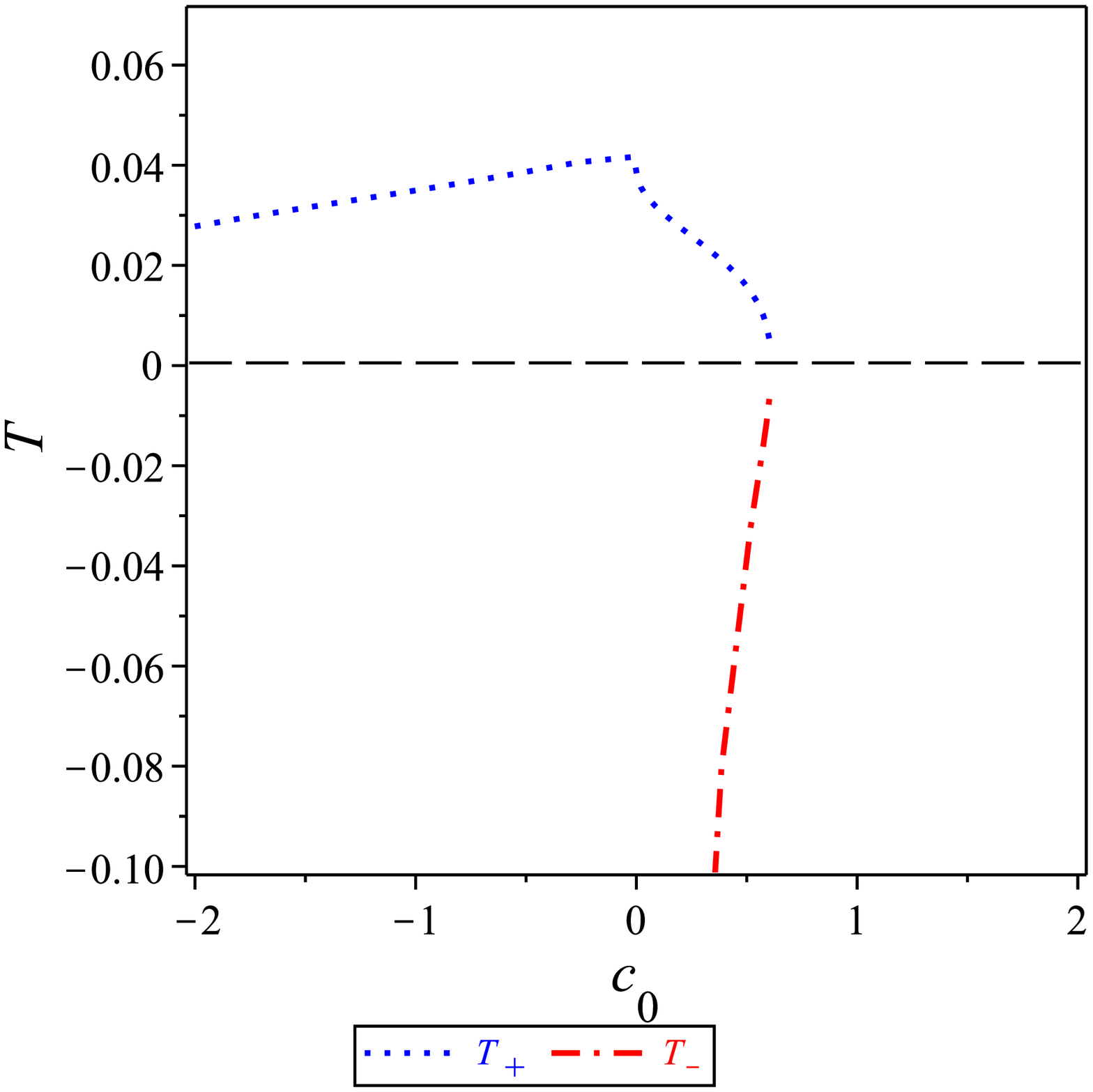}}
\subfigure[~The quasi-local energy of the black hole solution (\ref{sol1}) with respect to the mass $m$ ]{\label{fig:2e}\includegraphics[scale=0.25]{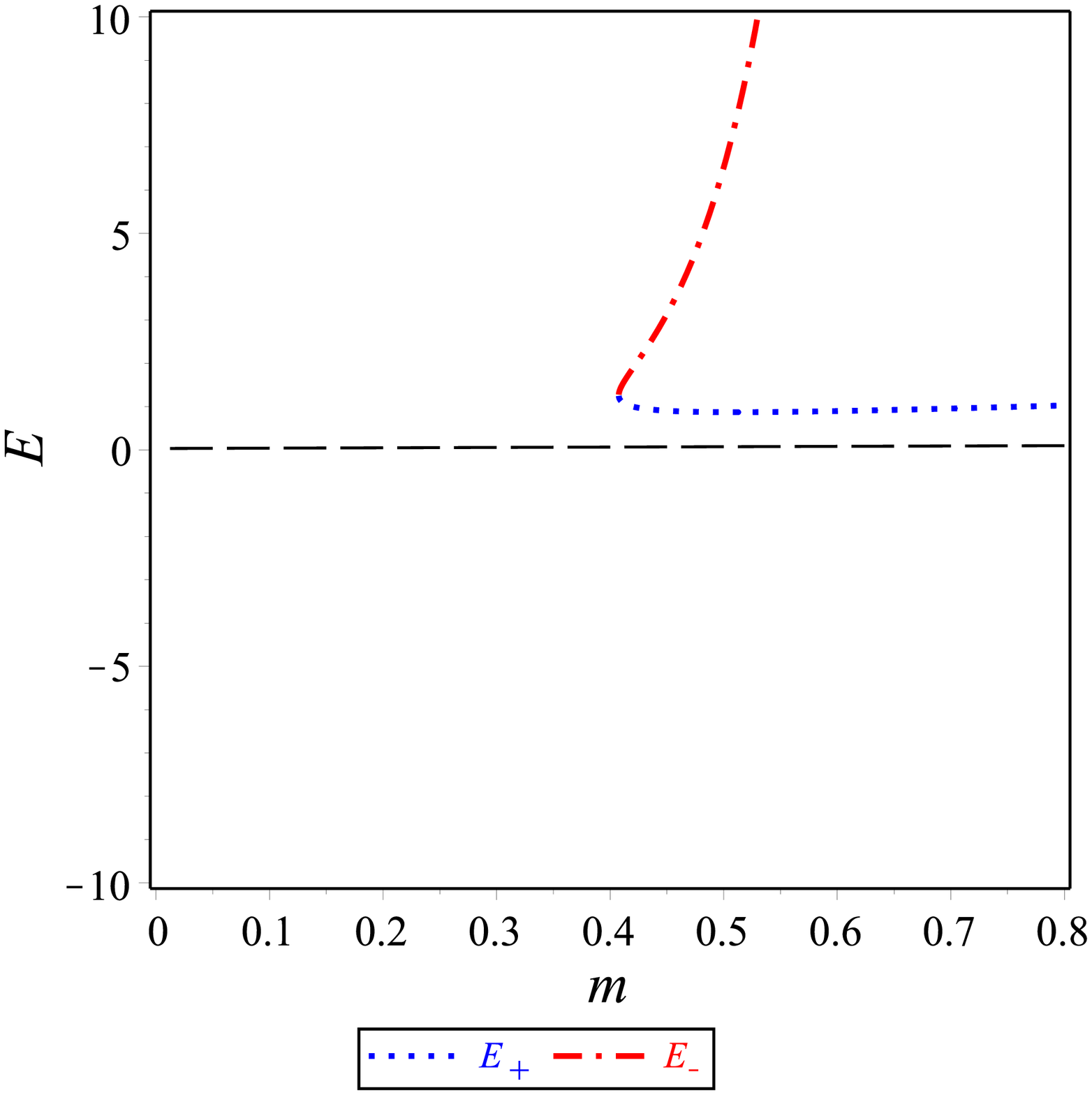}}\hspace{0.2cm}
\subfigure[~The quasi-local energy of the black hole solution (\ref{sol1}) with respect to the parameter $c_0$]{\label{fig:2f}\includegraphics[scale=0.25]{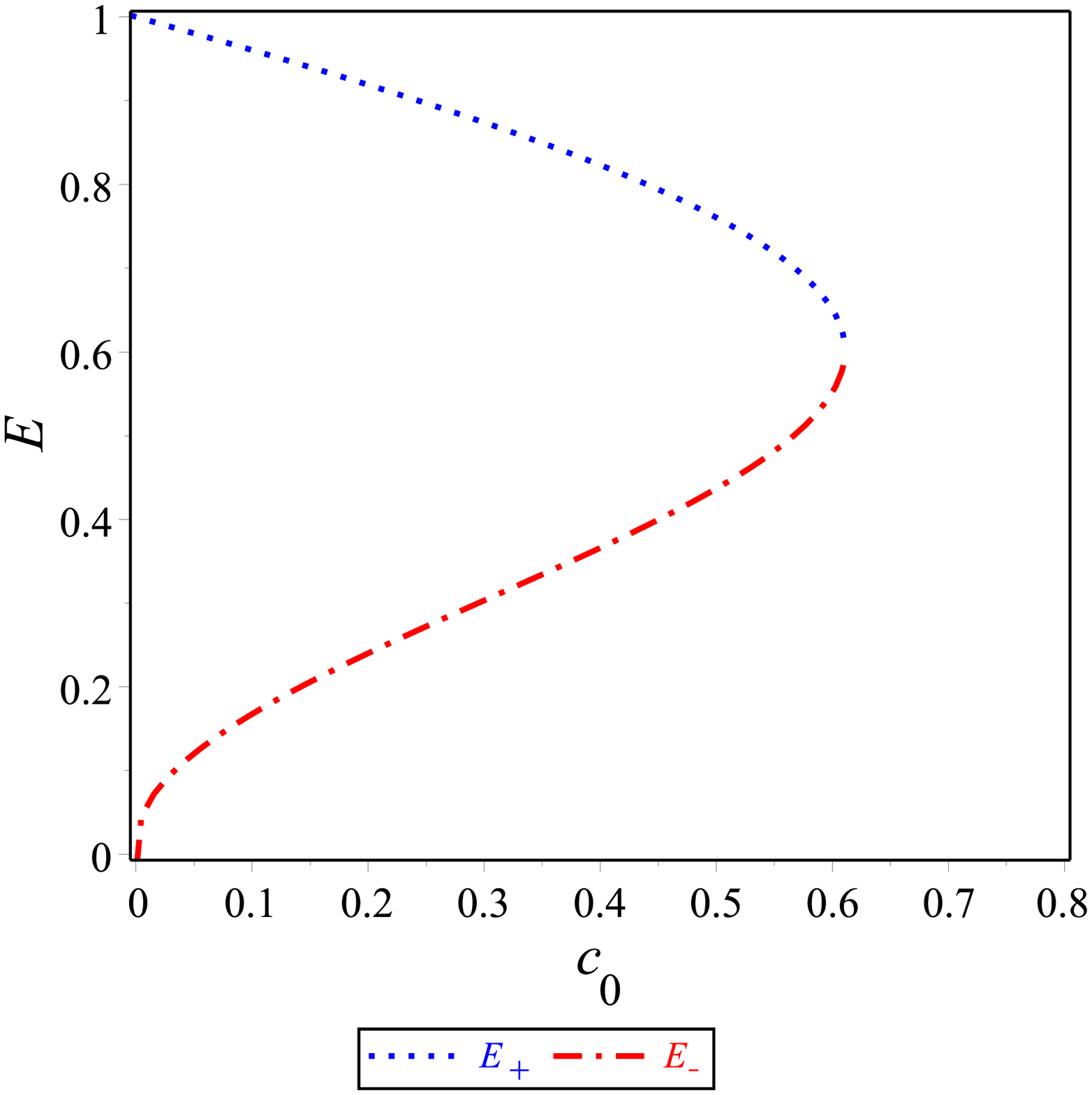}}
\caption{{ Schematic plot of the entropy Figure~\ref{Fig:2}~\subref{fig:1a}, Figure~\ref{Fig:2}~\subref{fig:2b},
the temperature Figure~\ref{Fig:2}~\subref{fig:2c}, Figure~\ref{Fig:2}~\subref{fig:2d} and the quasi-local energy of the black hole (\ref{sol1})
with respect to the mass $m$ and the parameter $c_0$, respectively, Figure~\ref{Fig:2}~\subref{fig:2e} and Figure~\ref{Fig:2}~\subref{fig:2f}.}}
\label{Fig:2}
\end{figure}

Using Eq.~(\ref{ent}), the entropy of the black holes (\ref{sol1}) is computed as,
\begin{equation}
\label{ent1}
S_+^{\tiny\mbox{Eq.~(\ref{sol1})}} =\pi\,{r_+}^2\left(1+\frac{c_0}{r_+{}^2}\right)\,, \quad S_-^{\tiny\mbox{Eq.~(\ref{sol1})}}=\pi\,{r_-}^2\left(1+\frac{c_0}{r_-{}^2}\right)\,.
\end{equation}
We plot Eq.~(\ref{ent1}) in Figures~\ref{Fig:2}~\subref{fig:2a} and \ref{Fig:2}~\subref{fig:2b}.
As these figures show that we have a positive entropy for the horizon outer Cauchy horizon, $r_+$, and under some constraints
the entropy of the inner Cauchy horizon, $r_-$, can take a positive value.
These constraints are given by $m>3.7$ and $c_0>0$.

\begin{figure}
\centering
\subfigure[~The free energy of the black hole solution (\ref{sol1}) with respect to the mass $m$ using
the horizon $r_+$ ]{\label{fig:3a}\includegraphics[scale=0.25]{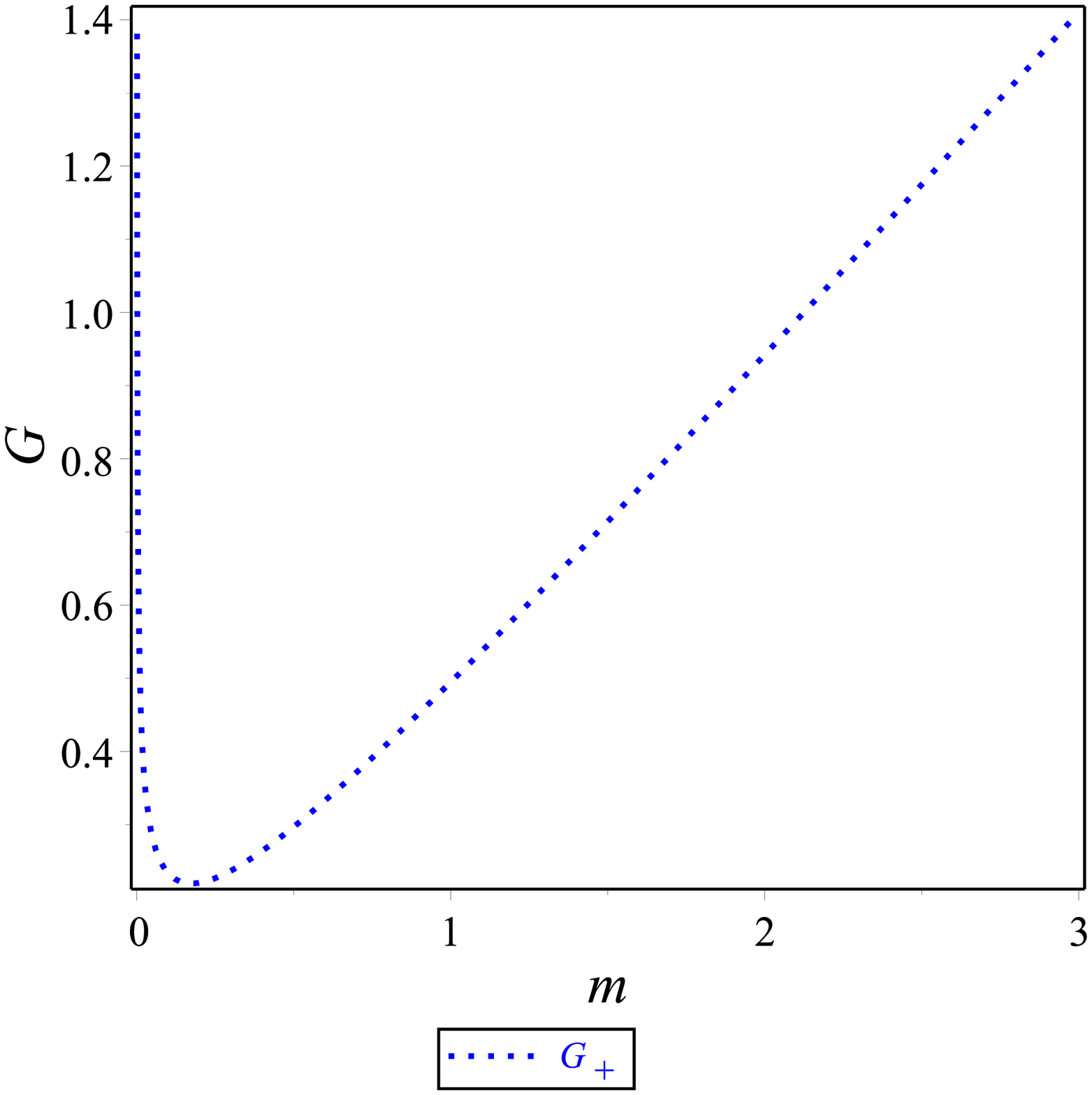}}\hspace{0.2cm}
\subfigure[~The free energy of the black hole solution (\ref{sol1}) with respect to the mass $m$ using
the horizon $r_-$]{\label{fig:3b}\includegraphics[scale=0.25]{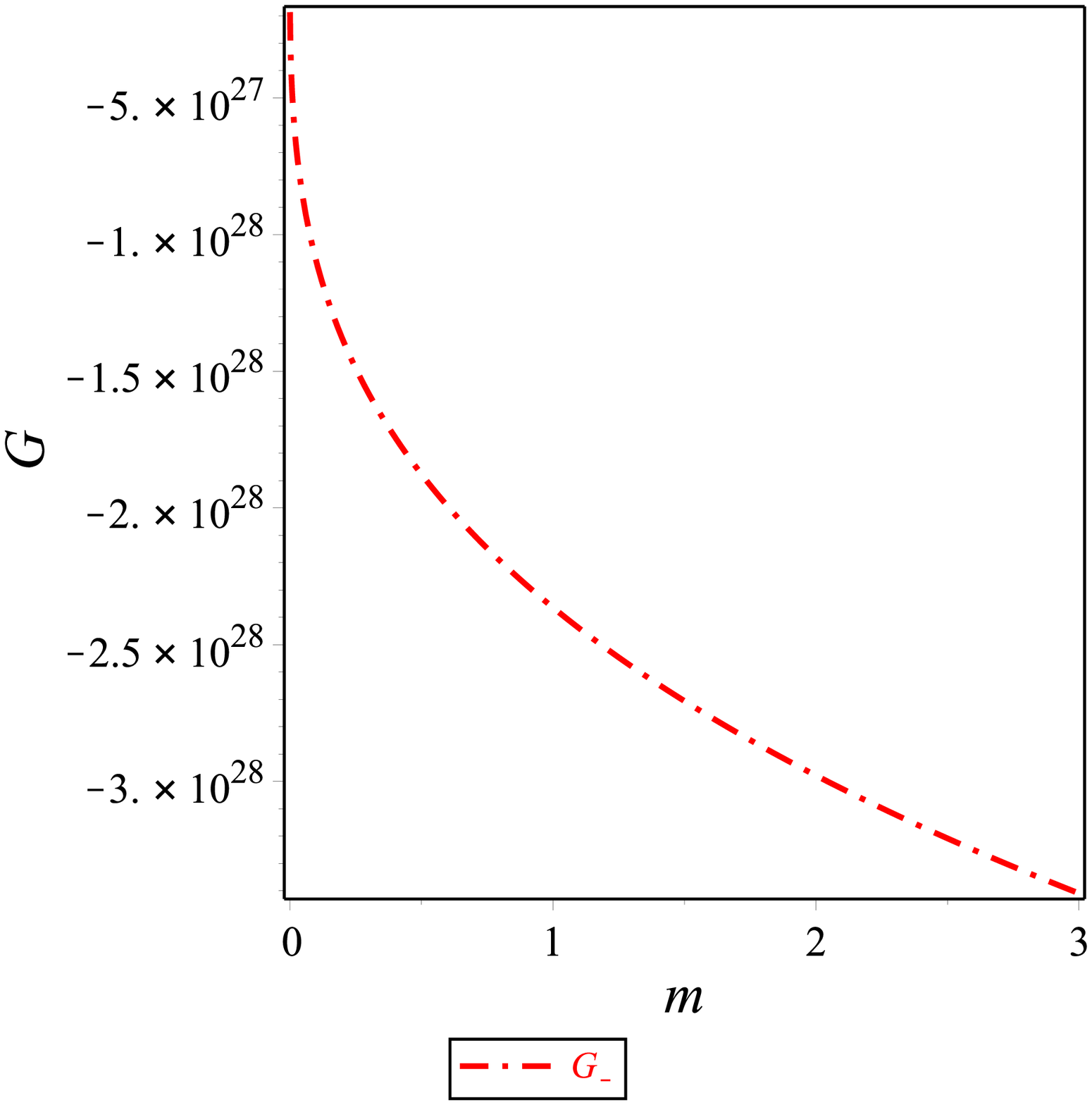}}\hspace{0.2cm}
\subfigure[~The free energy of the black hole solution (\ref{sol1}) with respect to the parameter $c_0$
using the horizon $r_+$ ]{\label{fig:3c}\includegraphics[scale=0.25]{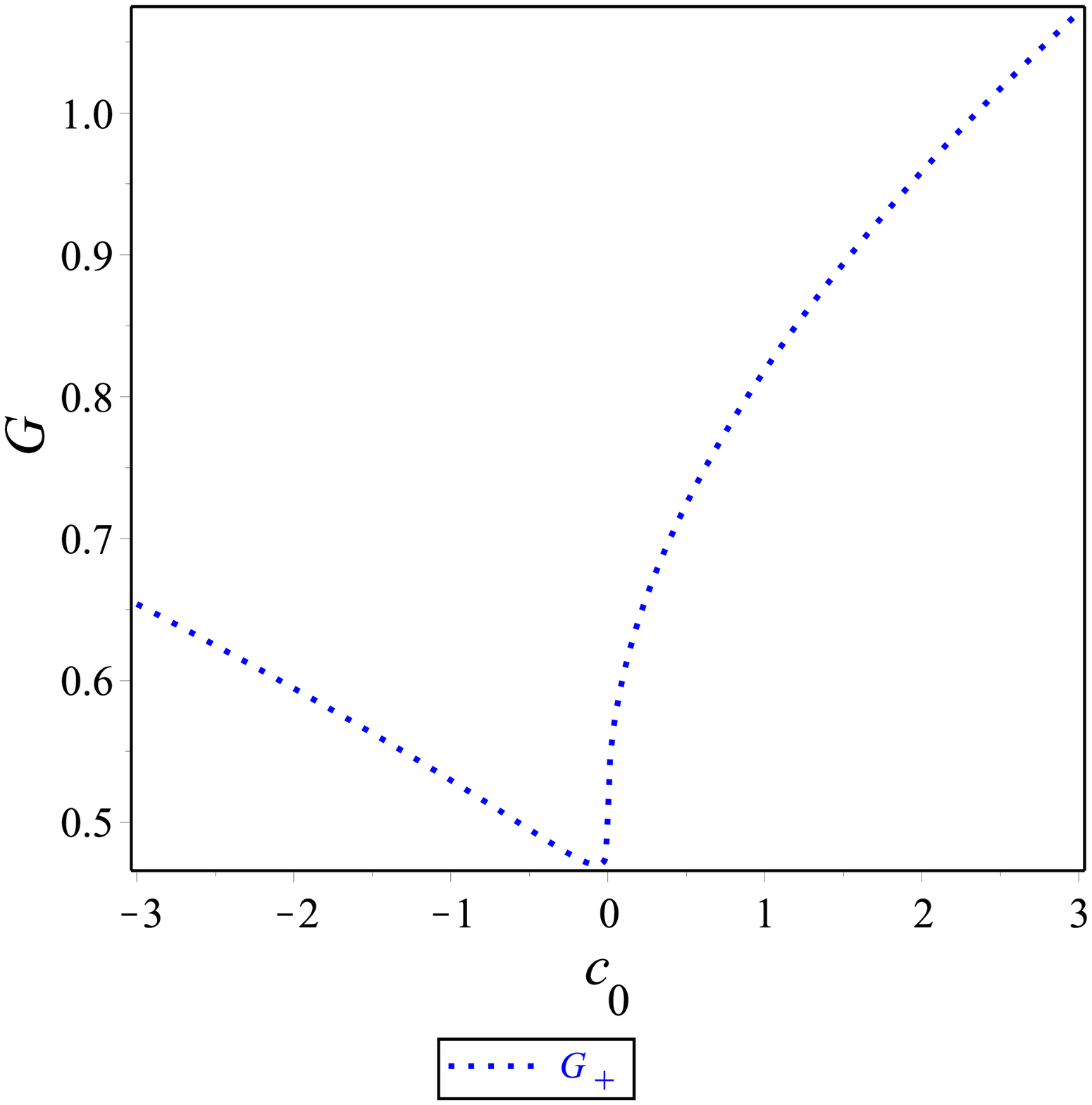}}\hspace{0.2cm}
\subfigure[~The free energy of the black hole solution (\ref{sol1}) with respect to the parameter $c_0$ using
the horizon $r_-$ ]{\label{fig:3d}\includegraphics[scale=0.25]{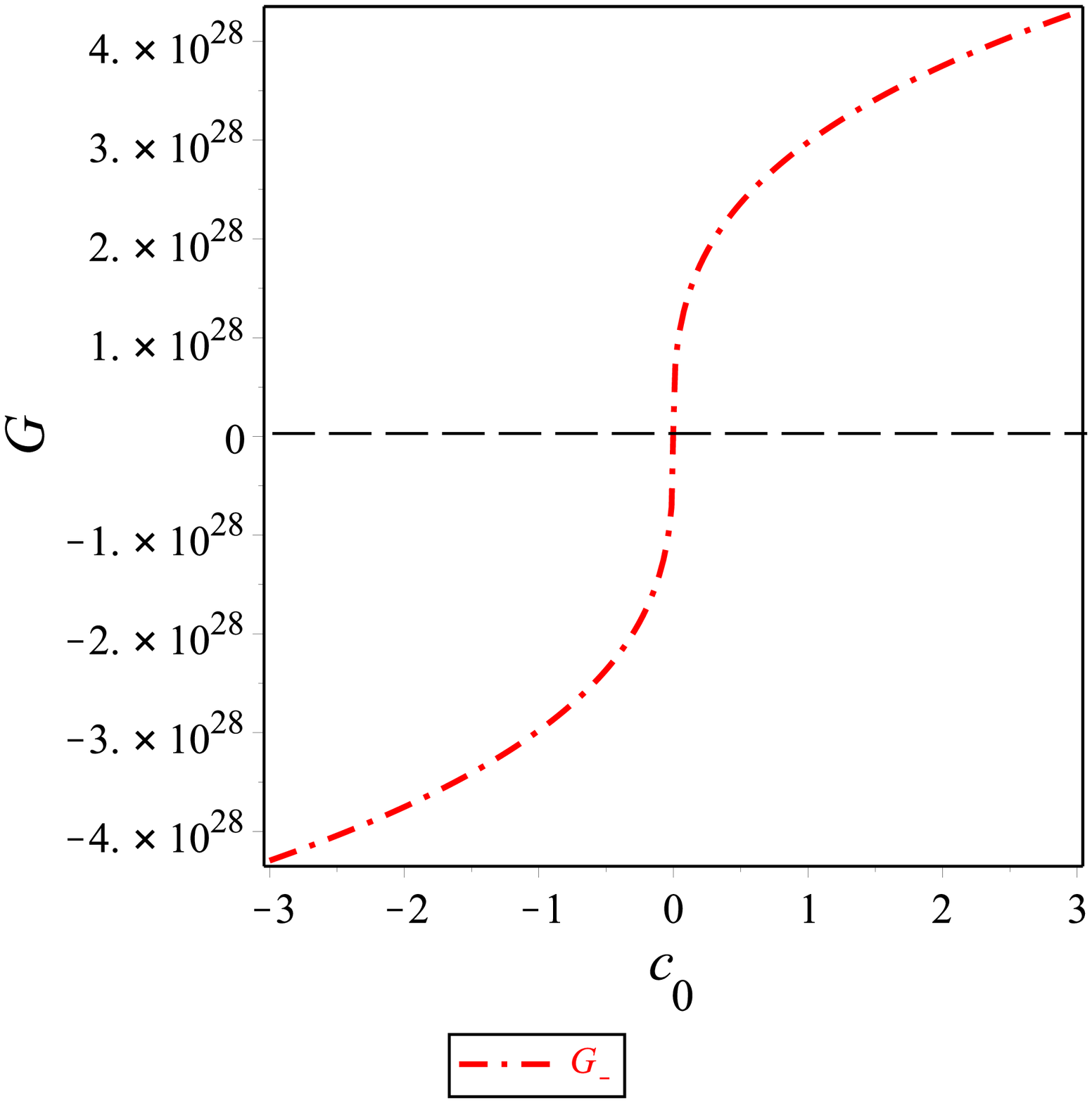}}
\caption{{Schematic plot of the free energy of the black holes (\ref{sol1}) with respect to the mass $m$
and the parameter $c_0$ respectively, Figure~\ref{Fig:3}~\subref{fig:3a}, Figure~\ref{Fig:3}~\subref{fig:3b},
Figure~\ref{Fig:3}~\subref{fig:3c} and Figure~\ref{Fig:3}~\subref{fig:3d}.}}
\label{Fig:3}
\end{figure}

The Hawking temperatures associated with the black hole solutions (\ref{sol1}) is plotted in Figures~\ref{Fig:2}~\subref{fig:2c} and
\ref{Fig:2}~\subref{fig:2d}. From these figures one can show that we have a positive temperature for $r_+$ and negative temperature for $r_-$.
 From Eq.~(\ref{en}), the quasi-local energy of the black hole (\ref{sol1}) is plotted in Figures~\ref{Fig:2}~\subref{fig:2e} and \ref{Fig:2}~\subref{fig:2f}.
 These figures show that we have always positive quasi-local energy for the black hole (\ref{sol1}).
The free energy in the grand canonical ensemble, also called the Gibbs free energy, can be defined as \cite{Zheng:2018fyn,Kim:2012cma}:
\begin{equation}
\label{enr}
G(r_+)=E(r_+)-T(r_+)S(r_+)\, . 
\end{equation}
where $E(r_+)$, $T(r_+)$, and $S(r_+)$ are the quasilocal energy, the temperature and entropy at the event horizons, respectively.
Using Eqs.~(\ref{temp}), (\ref{en}), and (\ref{ent1}) in (\ref{enr}), we plot the behavior of the Gibbs free energy in Figures~\ref{Fig:3}~\subref{fig:3a},
\ref{Fig:3}~\subref{fig:3b}, \ref{Fig:3}~\subref{fig:3c}, and \ref{Fig:3}~\subref{fig:3d}. From these figures we can conclude that
our black hole solution with the Cauchy outer horizon, $r_+$, is stable and the black hole with the inner Cauchy horizon, $r_-$, is not stable.

The behaviors of the Gibbs energy of our black hole that is presented in Figures~\ref{Fig:3}~\subref{fig:3a},
\ref{Fig:3}~\subref{fig:3b}, \ref{Fig:3}~\subref{fig:3c}, and \ref{Fig:3}~\subref{fig:3d} for particular values of the model parameters show,
for the black hole solution (\ref{sol1}), the Gibbs energy is positive when $r=r_+$ which means that we have a stable black hole.

\section{Stability of the BH of the second set of Eq.~(\ref{sol1}) using geodesic deviation}\label{S666}

The trajectories of a test particle in a gravitational field are described by the geodesic equations which has the form,
\begin{equation}
\label{ge}
\frac{d^2 x^\sigma}{ d\lambda^2} + \left\{ \begin{array}{c} \sigma \\ \mu \nu \end{array}  \right\}
\frac{d x^\mu}{d\lambda} \frac{d x^\nu}{d\lambda}=0 \, ,
\end{equation}
where $\lambda$ is an affine parameter along the geodesic.
The geodesic deviation takes the form \cite{1992ier..book.....D},
\begin{equation}
\label{ged}
\frac{d^2 \xi^\sigma}{d\lambda^2} + 2\left\{ \begin{array}{c} \sigma \\ \mu \nu \end{array}  \right\}
\frac{d x^\mu}{d\lambda} \frac{d \xi^\nu}{d\lambda} + \left\{ \begin{array}{c} \sigma \\ \mu \nu \end{array}  \right\}_{,\ \rho}
\frac{d x^\mu}{d\lambda} \frac{d x^\nu}{d\lambda}\xi^\rho=0 \, ,
 \end{equation}
where $\xi^\rho$ the deviation 4-vector.
Applying (\ref{ge}) and (\ref{ged}) into (\ref{met}),
we get the geodesic equations in the following from,
\begin{equation}
\frac{d^2 t}{d\lambda^2}=0\, , \quad \alpha'(r)\sqrt{\alpha\gamma}\left( \frac{dt}{d\lambda}\right)^2
+ \Omega \left( \gamma\alpha \right)' \frac{dt}{d\lambda} \frac{d \phi}{d\lambda}
 -\left[2r-\Omega^2\gamma'\right]\left( \frac{d \phi}{d\lambda}\right)^2=0\, , \quad
\frac{d^2\theta}{d\lambda^2}=0\, , \quad \frac{d^2 \phi}{d\lambda^2}=0\, .
\end{equation}
Using the circular orbit
\begin{equation}
\theta = \frac{\pi}{2}\, , \quad \frac{d\theta}{d\lambda}=0\, , \quad \frac{dr}{d\lambda}=0\, ,
\end{equation}
we get the geodesic deviation in the following form,
\begin{align}
\label{ged22}
0= &  \frac{d^2 \xi^1}{d\lambda^2} + \beta \alpha'(r) \frac{dt}{d\lambda} \frac{d\xi^0}{d\lambda}
 - \left( 2r-\Omega^2\gamma' \right) \beta \frac{d \phi}{d\lambda} \frac{d \xi^3}{d\lambda}
+ \frac{\beta \Omega \left( \alpha\gamma \right)'}{2\sqrt{\alpha \gamma}}
\left[ \frac{dt}{d\lambda} \frac{d \xi^3}{d\lambda} + \frac{d\phi}{d\lambda} \frac{d \xi^4}{d\lambda}\right] \nonumber \\
& +\left\{ \frac{\alpha'\beta'+\beta \alpha''}{2}\left( \frac{dt}{d\lambda}\right)^2
+\left[\frac{\beta\Omega^2\gamma''}{2} - \left( r-\frac{\Omega^2\gamma'}{2} \right)
\left(\beta+r\beta'\right) \right] \left( \frac{d\phi}{d\lambda}\right)^2 \right.  \nonumber\\
& \left. +\frac{ \Omega \left\{ 2\alpha\gamma \left[ \gamma\alpha'\beta'+\alpha\beta'\gamma'
+\beta\alpha' \gamma'+\beta\gamma\alpha''+\beta\alpha\gamma'' \right]
 -\beta\gamma^2\alpha'^2-\beta\alpha^2\gamma'^2 \right\}} {4\sqrt{\alpha^3\gamma^3}}
\left( \frac{d\phi}{d\lambda}\right) \left( \frac{dt}{d\lambda}\right) \right\}\xi^1 \, , \nonumber\\
0= & 2 \frac{d^2 \xi^0}{d\lambda^2} + \frac{ \left( 2r^2-\Omega^2\alpha \right) \alpha'(r)
+\Omega^2\alpha \gamma'}{r^2 \alpha} \frac{dt}{d\lambda} \frac{d \zeta^1}{d\lambda}
 - \frac{\Omega\alpha\left\{ \left( r^2\gamma\alpha'-\Omega^2\gamma^2 \right) \alpha'
+\left[ \left( \Omega^2\gamma+r^2 \right) \gamma'-4r\gamma \right] \alpha \right\}}{r^2 \alpha^{3/2}\sqrt{\gamma}}
\frac{d\phi}{d\lambda} \frac{d \zeta^1}{d\lambda}\,,\nonumber\\
& 0= \frac{d^2 \xi^2}{d\lambda^2} + \left( \frac{d\phi}{d\lambda} \right)^2 \xi^2\,, \quad
0=\frac{d^2 \xi^3}{d\lambda^2} + \frac{\Omega^2\left[\alpha'\gamma-\gamma'\alpha\right]+4r\alpha}{\alpha r^2} \frac{d\phi}{d\lambda}
\frac{d\xi^1}{d\lambda} + \frac{\Omega\left[\alpha'\gamma-\gamma'\alpha\right]}{2\sqrt{\alpha\gamma} r^2}
\frac{dt}{d\lambda} \frac{d \xi^1}{d\lambda}\, ,
\end{align}
where the functions $\alpha$, $\beta$, and $\gamma$ are defined in Eq.~(\ref{sol1}).

\begin{figure}
\centering
\subfigure[~The plot of $\sigma^2$ with respect to the radial coordinate $r$ when $\Lambda_\mathrm{eff}=0$]{\label{fig:1a}\includegraphics[scale=0.25]{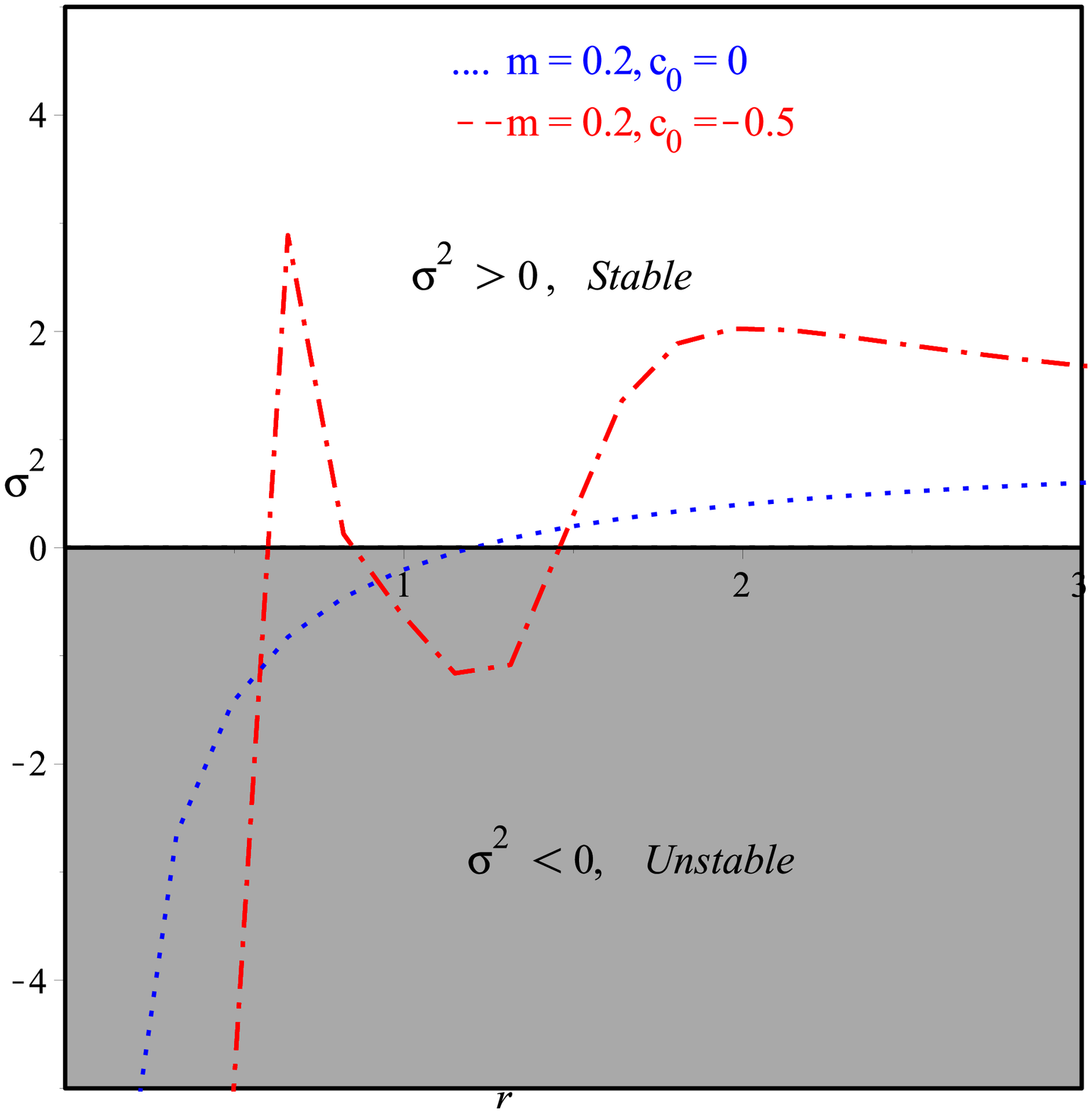}}\hspace{0.2cm}
\subfigure[~The plot of $\sigma^2$ with respect to the radial coordinate $r$ when $\Lambda_\mathrm{eff}\neq0$]{\label{fig:1d}\includegraphics[scale=0.25]{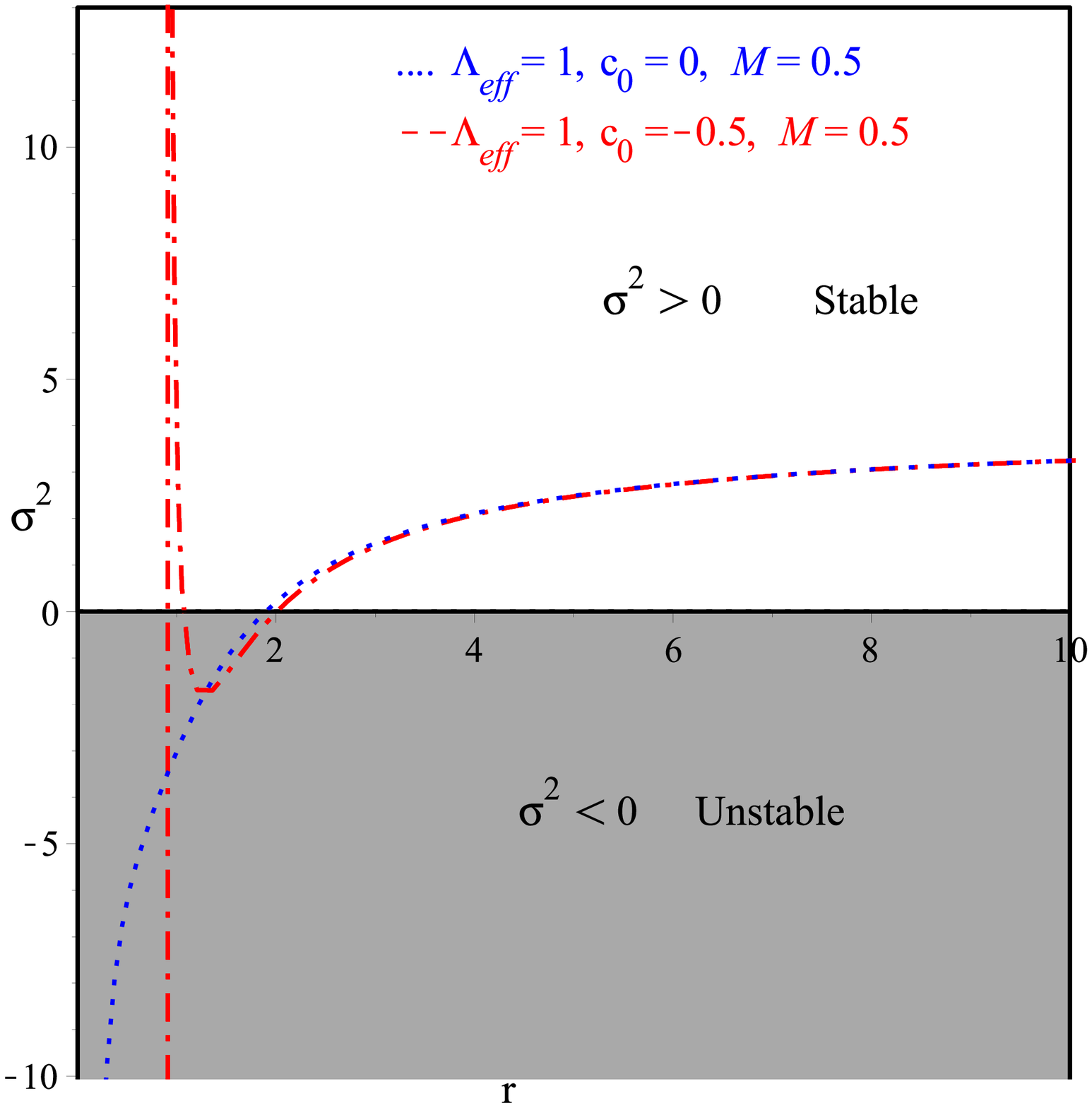}}
\caption{Schematic plots of the $sigma^2$ with respect to the radial coordinate $r$ of the black hole (\ref{sol1}) when $\Lambda_\mathrm{eff}=0$ and $\Lambda_\mathrm{eff}\neq0$.}
\label{Fig:4}
\end{figure}

The third equation of (\ref{ged22}) shows that it is a simple harmonic motion, which means that the motion in
the plan $\theta=\frac{\pi}{2}$ is stable.
Now we assume the solutions of the remaining equations of (\ref{ged22}) to have the form,
\begin{equation}
\label{ged33}
\xi^0 = \zeta_1 \e^{i \sigma \phi}\, , \quad \xi^1= \zeta_2\e^{i \sigma \phi}\, ,
\quad \mbox{and} \quad \xi^3 = \zeta_3 \e^{i \sigma \phi}\, ,
\end{equation}
where
$\zeta_1$, $\zeta_2$, and $\zeta_3$ are constants.
Substituting (\ref{ged33}) into (\ref{ged22}), we get $\sigma^2$ whose explicit form is given in (\ref{con1}) of Appendix~\ref{AII}.
The stability condition for the spacetime (\ref{sol1})  is $\sigma^2>0$ \cite{Misner:1974qy}.
Equation~(\ref{con1}) coincides with that derived in \cite{Nashed:2020mnp} when $\Omega=0$.
We draw the condition (\ref{con1}) in Figure~\ref{Fig:4} for the two cases,
$\Lambda_\mathrm{eff}=0$ and $\Lambda_\mathrm{eff}\neq0$.

\section{Discussion and conclusions }\label{S77}

The Kerr spacetime is the only spacetime that describes asymptotically flat and astrophysically stable BHs in GR \cite{Heusler:1997ui,Robinson:1975bv}.
In spite of this, various theoretical and observational problems propose new gravitational physics at the ultraviolet scale \cite{tHooft:1974toh},
hinting that astrophysical black holes may have non-Kerr hairs (e.g. \cite{Glampedakis:2005cf}).
These properties might theoretically investigate in a number of ways, one interesting possibility of which is that some fundamental symmetries are
broken, such as the Carter symmetry which implies the integrability of the geodesic equations \cite{Papadopoulos:2018nvd,Carson:2020dez}.

In this study, we focus on a class of rotating BH solutions in the frame of $f(R)$ gravitational theory.
After a brief review of $f(R)$ gravity, we apply its field equation to a line-element that has a cross term,
$dtd\phi$ that has a constant parameter $\Omega$ and contains three unknown functions.
When $\Omega=0$, we get a diagonal spherically symmetric line-element that studied in GR and $f(R)$ \cite{Nashed:2020mnp}.
Applying the field equation of $f(R)$ to the non-diagonal line-element, we derive their non-linear differential equations.
These field equations consist of six non-linear differential equations in four arbitrary functions.
It is very difficult to solve such a system without assuming a specific form of the first derivative of $f(R)$.

Assuming a specific form of the first derivative of $f(R)$, we solve the system of non-linear differential equations
and derive the exact form of the three functions of the line element.
This solution depends on the Heun Confluent function and its first derivative.
We show that the Ricci scalar associated with this solution has a dynamical form and becomes constant when the first derivative of $f(R)$ vanishing.
It is not easy to extract any physical properties from the metric potentials using the exact form of the solution.
Therefore, we calculate the asymptotic form of the metric potential and show that their line-element depends on an effective cosmological constant
and behaves asymptotically as AdS/dS and when the effective cosmological constant is vanishing the line element behaves like a flat spacetime.
We show that the metric potentials contain extra terms than those of Einstein's GR due to the contribution of the higher curvature terms.
Moreover, we calculated the invariants of this BH solution and show that these invariants have a true singularity at $r=0$.

To gain more physical properties of the BH solutions, we calculated the horizons of this solution and show that it has two horizons, inner and outer Cauchy horizons, in contrast to Einstein's GR which has only one horizon. Moreover, we calculated some thermodynamical quantities like the Hawking temperature, entropy, quasi-local energy, and Gibbs free energy. All these thermodynamical quantities show consistent physical behavior for the outer Cauchy horizon. Finally, by using the geodesic deviation equations, we discuss the stability condition analytically and graphically.

In conclusion, we derived for the first time a new rotating BH solution in the frame of $f(R)$.
This study shed light on the possibility to derive a solution similar to the Kerr solution of Einstein GR.
This will be tackle in our forthcoming study.

{\textbf{
It could be interesting to consider the merger of two black holes which are solution given in this paper or the merger of the
black hole and a neutron star as discussed in \cite{Astashenok:2021xpm,Astashenok:2021peo,Astashenok:2020qds} and investigate the gravitational wave generated by the merger.
We should note that there is a big controversy related to the gravitational event GW190814.
The reason is because one of the components could be too large to be a neutron star and too small to be a black hole.
Therefore it could be interesting if we can construct such an object in the framework in this paper.}}

\section*{Acknowledgments}

This work is supported by the JSPS
Grant-in-Aid for Scientific Research (C) No. 18K03615 (S.N.).

\appendix

\section{Differential Equations under Ansatz (\ref{met}) \label{AI}}

By applying the ansatz (\ref{met}) to the field equations~(\ref{f3ss}), we get the explicit forms of the differential equations,
\begin{align}
\label{df1}
Q_t{}^t=& \frac{\cot^2\theta}{32 \alpha^{3}\gamma {r}^{2}} \left\{ 16\,(1+F) \beta ' \alpha^{3}r\gamma -16\,(1+F) \alpha^{3}\gamma -8\,(1+F) \alpha'' {r}^{2}\alpha^{2}\beta \gamma
+4\,(1+F)\alpha'^{2}\beta \gamma {r}^{2}\alpha +16\,(1+F)\alpha^{3}\beta\gamma \right. \nonumber \\
& \left. -12\, \alpha ^{2}\gamma {r}^{2} F' \beta\alpha'  -4\,(1+F)\alpha'\beta'{r}^{2}\alpha^{2} \gamma -16\,(1+F) \beta \gamma\alpha' r\alpha^ {2}
+8\alpha^{3}\gamma r^2F'' \beta +4\alpha^{3}\gamma r^2 F'\beta' +16\alpha^3\gamma r F' \beta \right\} \nonumber \\
&+\frac {1 }{32\alpha^{3}\gamma {r}^{2} \sin^2 \theta}\left\{8F' \beta\alpha^2\gamma \Omega^2\gamma'-8\alpha^3\gamma r^2F''\beta
 -4\alpha^{3}\gamma r^2F'\beta'-16\,\alpha^3\gamma r F'\beta +28(1+F)\alpha'^2\beta\gamma^2\Omega^2  \right. \nonumber \\
& -16\,(1+F)\beta'\alpha^{3}r\gamma +(1+F) {\Omega}^{2} \gamma'^{2} \alpha^{2}\beta
 -16\,(1+F)\alpha^{3}\beta \gamma +4\,(1+F)\alpha'\beta' {r}^{2}\alpha^{2 }\gamma
 -4\,(1+F)\alpha'^{2}\beta\gamma {r}^{2}\alpha \nonumber\\
& +12\,\alpha^{2}\gamma {r}^{2} F'\beta \alpha' -24\,(1+F) {\Omega}^{2}\alpha'\gamma'\alpha \beta \gamma -16\,(1+F) \alpha \beta\gamma^{2} \alpha''{\Omega}^{2}
+4\,(1+F) \beta'\alpha^{2}\gamma {\Omega}^{2} \gamma' -8\,(1+F) \beta'\alpha\gamma^{2}\alpha'{\Omega}^{2}\nonumber\\
& \left. -16\,\alpha\gamma^{2}{\Omega}^{ 2}F'\beta\alpha' +8\,(1+F)\alpha'' {r}^{2}\alpha^{2}\beta \gamma +16\,(1+F) \alpha^{3}\gamma
+16\,(1+F) \beta \gamma\alpha' r\alpha^{2}+8\,(1+F) \alpha^{2}\beta \gamma {\Omega}^{2}\gamma''\right\} =0\,,\nonumber\\
Q_t{}^\phi=& \frac{\Omega}{8{r}^{2}\alpha^{2}{\gamma }^{3/2}\sin^{2}\theta} \left\{ 2\,F' \beta \gamma'\alpha^{2}\gamma +2\, \gamma^{2}(1+F)\alpha'^{2}\beta
+2\,(1+F)\gamma'' \alpha^{2}\gamma \beta +(1+F)\gamma'\beta' \alpha^{2}\gamma -4\,\gamma^{2}\alpha F' \beta\alpha' \right. \nonumber\\
& \left. -(1+F) \gamma'^{2} \alpha^{2}\beta
 -(1+F)\alpha'\gamma' \gamma \alpha \beta -4\,\gamma^{2}\alpha (1+F) \alpha'' \beta -2\, \gamma^{2}\alpha (1+F)\alpha'\beta'\right\} =0\,,\nonumber\\
Q_r{}^r=& \frac {\cot^2 \theta}{32\alpha^{3}\gamma {r}^{2} } \left\{8\,(1+F)\alpha''{r}^{2} \alpha^{2}\beta \gamma -4\,(1+F)\alpha'^{2}\beta \gamma {r}^{2}\alpha
-16\, \alpha ^{3}\gamma r F' \beta -16 \,(1+F) \beta \gamma\alpha' r \alpha^{2}+4\,(1+F)\alpha' \beta' {r}^{2} \alpha^{2}\gamma \right. \nonumber\\
& \left. +16\,(1+F) \alpha^{3}\gamma \gamma -4\, \alpha^{2}\gamma {r}^{2} F'\beta\alpha' -16\,(1+F) \alpha^{3}\beta+16\,(1+F)\beta'\alpha^{3}r\gamma
+24\,\alpha^{3}\gamma {r}^{2}F''\beta +12\, \alpha^{3}\gamma {r}^{2} F' \beta' \right\} \nonumber\\
& +\frac {1}{32\alpha^{3} \gamma {r}^{2} \sin^2 \theta} \left\{ 4\,(1+F)\alpha'^{2}\beta \gamma {r}^{2}\alpha \gamma
 -4\,(1+F)\alpha'\beta'{r}^{2} \alpha^{2}\gamma -12\,(1+F) {\Omega}^{2}\alpha'\gamma' \alpha \beta \gamma +3\,(1+F) {\Omega}^{2} \gamma'^{2} \alpha^{2}\beta \right. \nonumber\\
& -16\,(1+F) \alpha ^{3}\gamma +16\,(1+F) \alpha^{3}\beta-16\,(1+F)\beta' \alpha^{3}r\gamma
 +12\,(1+F) \alpha'^{2}\beta\gamma^{2}{\Omega}^{2}\beta+4\,\alpha^{2}\gamma {r}^{2}F' \beta\alpha' +16\, \alpha^{3}\gamma r F' \beta \nonumber \\
& \left. -8\,(1+F)\alpha''{r}^ {2} \alpha^{2}\beta \gamma +16\,(1+F) \beta \gamma\alpha' r\alpha^{2}-24\, \alpha^{3}\gamma {r}^{2}F''-12\, \alpha^{3}\gamma {r}^{2} F'\beta'\right\}
=0\,,\nonumber\\
Q_\theta{}^\theta=& \frac {\cot^2\theta}{32\alpha^{3}\gamma {r}^{2}} \left\{ 16\,(1+F) \alpha^{3}\gamma +8\,\alpha^{3}\gamma {r}^{2}F''\beta +4\,\alpha^{3}\gamma {r}^{2}F'\beta'
 -16\,\alpha^{3}\gamma r F' \beta +8\,(1+F)\alpha'' {r}^{2}\alpha^{2}\beta \gamma +4\,\alpha^{2}\gamma {r}^{2} F' \beta\alpha' \right. \nonumber \\
& \left. -4\,(1+F)\alpha'^{2}\beta \gamma {r}^{2}\alpha +4\,(1+F) \alpha'\beta'{r}^{2}\alpha^{2}\gamma-16(1+F) \alpha ^{3}\beta \gamma \right\} \nonumber\\
& +\frac {1}{32\alpha^{3}\gamma {r}^{2}\sin^2 \theta}\left\{4(1+F) \alpha'^{2}\beta \gamma {r}^{2}\alpha +16(1+F)\alpha^{3}\beta \gamma
 -8(1+F)\alpha''{r}^{2}\alpha^{2} \beta \gamma -4(1+F)\alpha' \beta'r^2\alpha^{2}\gamma -4\alpha^3\gamma r^2F' \beta' \right. \nonumber\\
& -4\,\alpha^2\gamma r^2F' \beta \alpha' +4(1+F) \alpha'^2\beta \gamma^2\Omega^2+(1+F) {\Omega}^{2}\gamma'^{2} \alpha^{2}\beta +16\,\alpha^{3}\gamma rF'\beta
 -16\,(1+F) \alpha^{3}\gamma \nonumber\\
& \left. -4\,(1+F) {\Omega}^{2} \alpha'\gamma' \alpha \beta \gamma -8\,\alpha^{3}\gamma {r}^{2}F' \beta\right\} =0\,,\nonumber\\
Q_\phi{}^t =& \frac{ \cot^2\theta}{\alpha^4{r}^2\gamma^{3/2}} \left\{ 8\, \gamma^{2}\alpha^{3}(1+F) +2\,(1+F) \gamma''\alpha^{3}\gamma {r}^{2}\beta
 -(1+F)\alpha'\gamma'\gamma \alpha^2\beta {r}^{2}+2\,F'\beta\gamma'\alpha^{3}{r}^{2}\gamma +4\,(1+F) \gamma^2\alpha^{2}\alpha' \beta r \right. \nonumber\\
& \left. -8\, \gamma^2\alpha^3(1+F) \beta - 8\, \gamma^{2}\alpha^{3}r F' \beta -(1+F) \gamma'^{2}\alpha^{3}{r}^{2}\beta +(1+F)\gamma'\beta'\alpha^3\gamma {r}^{2}
 -4\, \gamma^2\alpha^3(1+F) \beta' r \right\} \nonumber \\
& +\frac{1}{\alpha^4{r}^2\gamma^{3/2}\sin^2\theta}\left\{ (1+F)\gamma'^{2}\alpha^{3}{r }^{2}\beta -10\,(1+F) \gamma^{3}\alpha'^{2}\beta {\Omega}^{2}
+8\, \gamma^{2}\alpha^{3}r F'\beta -4\,(1+F) \gamma^{2}\alpha^{2}\alpha'\beta r+4\, \gamma^{2}\alpha^{3}(1+F)\beta' r \right. \nonumber\\
& +4\,{\Omega}^{2} \gamma^{3}\alpha (1+F) \alpha'' \beta +8\, \gamma^{2}\alpha^{3}(1+F) \beta -8\, \gamma^{2}\alpha^{3}(1+F) +4\,{\Omega}^{2} \gamma^{3}\alpha F'\beta\alpha'
 -2\,{\Omega}^{2} \gamma^{2}(1+F) \gamma'' \alpha^{2}\beta \nonumber\\
& - {\Omega}^{2}\gamma^{2}(1+F) \gamma'\beta'\alpha^{2} -(1+F)\gamma'\beta'\alpha ^{3}\gamma {r}^{2}+9\,(1+F) \gamma^{2}\alpha\alpha'\beta {\Omega}^{2}\gamma'
 -2\,F' \beta\gamma' \alpha^{3}{r}^{2}\gamma -2\,(1+F)\gamma'' \alpha^{3}\gamma {r}^{2} \beta \nonumber\\
& \left. +2\,{\Omega}^{2} \gamma^{3}\alpha (1+F)\alpha' \beta' -2\,{\Omega}^{2} \gamma^{2}F' \beta \gamma' \alpha^{ 2}-\gamma \alpha^{2}(1+F) {\Omega}^{2}\gamma'^{2}\beta
+(1+F) \alpha'\gamma' \gamma \alpha^{2}\beta {r}^{2} \right\} =0\,,\nonumber \\
Q_\phi{}^\phi=&\frac{\cot\theta}{ \alpha^{3}\gamma{r}^{2}} \left\{ 16\,(1+F)\alpha^{3}\gamma+8\,\alpha^{3}\gamma{r}^{2}F'' \beta+4\,\alpha^{3}\gamma {r}^{2}F'\beta'
 -16\,\alpha^{3}\gamma r F'\beta+8\,(1+F) \alpha'' {r}^{2}\alpha^{2}\beta\gamma+4\,\alpha^{2}\gamma{r}^{2}F'\beta\alpha' \right. \nonumber\\
& \left. -4\,(1+F)\alpha'^{2}\beta\gamma {r}^{2}\alpha+4\,(1+F)\alpha'\beta'{r}^{2}\alpha^{2}\gamma-16\,(1+F)\alpha^{3}\beta\gamma \right\} \nonumber \\
& +\frac{1}{\alpha^{3} \gamma{r}^{2} \sin^2\theta} \Bigg\{16\,(1+F)\alpha\beta\gamma^{2}\alpha'' {\Omega}^{2}-4\,(1+F) \beta'\alpha^{2}\gamma{\Omega}^{2}\gamma'+16\,(1+F){ \Omega}^{2}\alpha'\gamma'\alpha\beta \gamma +16\,\alpha \gamma^{2}{\Omega}^{2}F' \beta\alpha' \nonumber\\
& -8\,F' \beta\alpha^{2}\gamma {\Omega}^{2}\gamma'-4\,\alpha^{2}\gamma{r}^{2}F'\beta\alpha'+16\,(1+F)\alpha^{3}\beta\gamma-16 \,(1+F)\alpha^{3}\gamma
 -8\,(1+F)\alpha'' {r}^{2}\alpha^{2}\beta\gamma-8\,\alpha^{3}\gamma{r}^{2}F''\beta\nonumber\\
&+4\,(1+F)\alpha'^{2} \beta\gamma{r}^{2}\alpha-4\,(1+F)\alpha'\beta'{r}^{2}\alpha^{2} \gamma-8\,(1+F)\alpha^{2}\beta\gamma{\Omega}^{2}\gamma''
+(1+F){\Omega}^{2}\gamma'^{2}\alpha^{2}\beta \nonumber\\
& \left. -20\,(1+F)\alpha'^{2}\beta\gamma ^{2}{ \Omega}^{2}+8\,(1+F)\beta'\alpha \gamma^{2} \alpha' {\Omega}^{2}
+16\, \alpha^{3}\gamma r F' \beta-4\,\alpha^{3}\gamma{r}^{2} F' \beta'\right\}=0\,,
\end{align}
where $F\equiv F(r)=f_R \left( R \left(r\right) \right) = \left. \frac{df(R)}{dR} \right|_{R=R(r)}$.

\section{The   form of $\sigma^2$ \label{AII}}

The explicit form of $\sigma^2$ in (\ref{ged33}) is given by \begin{align}
\label{con1}
\sigma^2=& \frac{3}{4} \left\{ {\Lambda_\mathrm{eff}}^{2}[148176\,{r}^{14}m{c_0}^{2}-141120{r}^{16}mc_0]-1693440\,^{2}{r}^{9}{c_0}^{5}
-620928\,\Lambda_\mathrm{eff}\,{r}^{14}mc_0+472416\,\Lambda_\mathrm{eff}\,{r}^{12}m{c_0}^{2} \right. \nonumber \\
& +903168\,\Lambda_\mathrm{eff}\,{r}^{13}{m}^{2}c_0 -1161216\,\Lambda_\mathrm{eff}\,{r}^{11}{m}^{2}{c_0}^{2}+3640224\,\Lambda_\mathrm{eff}\,{r}^
{10}{c_0}^{3}m-3308256\,\Lambda_\mathrm{eff}\,{r}^{9}{m}^{2}{c_0}^{3} \nonumber \\
& +6398280\,\Lambda_\mathrm{eff} \,{r}^{8}{c_0}^{4}m-8839152\,\Lambda_\mathrm{eff}\,{r}^{7}{m}^{2}{c_0}^{4}
 -451584\,{\Lambda_\mathrm{eff}}^{2}{r}^{13}{c_0}^{3}-677376\,{\Lambda_\mathrm{eff}}^{2}{r}^{11}{c_0}^{4}+
282240\,{\Lambda_\mathrm{eff}}^{2}{r}^{18}m \nonumber \\
& -37632\,{\Lambda_\mathrm{eff}}^{2}{r}^{15}{c_0}^{2} + 413952\,\Lambda_\mathrm{eff}\,{r}^{16}m+75264\,\Lambda_\mathrm{eff}\,{r}^{15}c_0
 -25088\,\Lambda_\mathrm{eff}\,{r}^{13}{c_0}^{2}-451584\,\Lambda_\mathrm{eff}\,{r}^{15}{m}^{2} \nonumber \\
& -564480\,\Lambda_\mathrm{eff}\,{r}^{11}{c_0}^{3}-915712\,\Lambda_\mathrm{eff}\,{r}^{9}{c_0}^{4}+47040\,{r}^{12}mc_0
 -179312\,{r}^{10}m{c_0}^{2} -451584\,{r}^{11}{m}^{2}c_0+1056384\,{r}^{9}{m}^{2}{c_0}^{2} \nonumber \\
& +1476048\,{r}^{8}{c_0}^{3}m-3867360\,{r}^{7}{m}^{2}{c_0}^{3}+1582448\,{r}^{6}{c_0}^{4}m-2864592\,{r}^{5}{m}^{2}{c_0}^{4}\nonumber\\
& +790272\,{m}^{3}{r}^{10}c_0 - 1899072\,{m}^{3}{r}^{8}{c_0}^{2}+4257792\,{m}^{3}{r}^{6}{c_0}^{3}-344520\,{m}^{3}{r}^{4}{c_0}^{4}
 -75264\,\Lambda_\mathrm{eff}\,{r}^{17}-75264\,{\Lambda_\mathrm{eff}}^{2}{r}^{19} \nonumber \\
& -18816\,{r}^{14}m +12544\,{r}^{11}{c_0}^{2}+150528\,{r}^{13}{m}^{2}- 163072\,{r}^{9}{c_0}^{3}-200704\,{r}^{7}{c_0}^{4}-225792\,{m}^{3}{r}^{12}
+ 31360\,{c_0}^{5}{r}^{5} \nonumber \\
& +235200\,{c_0}^{6}{r}^{3}-9877545\,{m}^{3}{c_0}^{6} + 24012\,{c_0}^{5}{r}^{3}{m}^{2}+1907388\,{m}^{3}{c_0}^{5}{r}^{2}
 -2569980\,{c_0}^{6}{r}^{2}m+9082800\,{c_0}^{6}r{m}^{2} \nonumber \\
& -261240\,{c_0}^{5}{r}^{4}m+956340 \,{c_0}^{6}{r}^{4}m\Lambda_\mathrm{eff}
 -2269134\,{m}^{2}{c_0}^{6}{r}^{3}\Lambda_\mathrm{eff}-6248340 \,{c_0}^{7}{r}^{2}m\Lambda_\mathrm{eff} \nonumber \\
& +14305410\,{m}^{2}{c_0}^{7}r\Lambda_\mathrm{eff}+8287272\,{\Lambda_\mathrm{eff}}^{2}{r}^{8}m{c_0}^{5}+1710072\,{r}^{6}m{c_0}^{5}\Lambda_\mathrm{eff}
+\Lambda_\mathrm{eff}\left[256284\,{m}^{2}{r}^{5}{c_0}^{5} \right. \nonumber \\
& \left. \left. +2349648\,{\Lambda_\mathrm{eff}}{r}^{12}m{c_0}^{3}+3841992\,{\Lambda_\mathrm{eff}}{r}^{10}m{c_0}^{4}-319872\,{r}^{7}{c_0}^{5}
 - 141120\,{c_0}^{6}{r}^{5}+705600\,{c_0}^{7}{r}^{3}\right]\right\} \nonumber\\
& \times \left\{{r}^{5} \left( -84\,\Lambda_\mathrm{eff}\,{r}^{7}-84\,{r}^{5}+168\,m{r}^{4}+84\,c_0{r}^{3}
 - 252\,mc_0{r}^{2}-70\,{c_0}^{2}r+261\,m{c_0}^{2} \right) \right. \nonumber \\
& \left. \times \left( 168\,\Lambda_\mathrm{eff} \,{r}^{7}+168\,m{r}^{4}+168\,c_0{r}^{3}-756\,mc_0{r}^{2}
 -280\,{c_0}^{2}r + 1305\,m{c_0}^{2} \right) \right\}^{-1}\, .
\end{align}
The stability condition for the spacetime (\ref{sol1}) \cite{Misner:1974qy} is $\sigma^2>0$.


\begin{thebibliography}{106}%
\makeatletter
\providecommand \@ifxundefined [1]{%
 \@ifx{#1\undefined}
}%
\providecommand \@ifnum [1]{%
 \ifnum #1\expandafter \@firstoftwo
 \else \expandafter \@secondoftwo
 \fi
}%
\providecommand \@ifx [1]{%
 \ifx #1\expandafter \@firstoftwo
 \else \expandafter \@secondoftwo
 \fi
}%
\providecommand \natexlab [1]{#1}%
\providecommand \enquote  [1]{``#1''}%
\providecommand \bibnamefont  [1]{#1}%
\providecommand \bibfnamefont [1]{#1}%
\providecommand \citenamefont [1]{#1}%
\providecommand \href@noop [0]{\@secondoftwo}%
\providecommand \href [0]{\begingroup \@sanitize@url \@href}%
\providecommand \@href[1]{\@@startlink{#1}\@@href}%
\providecommand \@@href[1]{\endgroup#1\@@endlink}%
\providecommand \@sanitize@url [0]{\catcode `\\12\catcode `\$12\catcode
  `\&12\catcode `\#12\catcode `\^12\catcode `\_12\catcode `\%12\relax}%
\providecommand \@@startlink[1]{}%
\providecommand \@@endlink[0]{}%
\providecommand \url  [0]{\begingroup\@sanitize@url \@url }%
\providecommand \@url [1]{\endgroup\@href {#1}{\urlprefix }}%
\providecommand \urlprefix  [0]{URL }%
\providecommand \Eprint [0]{\href }%
\providecommand \doibase [0]{http://dx.doi.org/}%
\providecommand \selectlanguage [0]{\@gobble}%
\providecommand \bibinfo  [0]{\@secondoftwo}%
\providecommand \bibfield  [0]{\@secondoftwo}%
\providecommand \translation [1]{[#1]}%
\providecommand \BibitemOpen [0]{}%
\providecommand \bibitemStop [0]{}%
\providecommand \bibitemNoStop [0]{.\EOS\space}%
\providecommand \EOS [0]{\spacefactor3000\relax}%
\providecommand \BibitemShut  [1]{\csname bibitem#1\endcsname}%
\let\auto@bib@innerbib\@empty
\bibitem [{\citenamefont {Nashed}\ and\ \citenamefont
  {Nojiri}(2020)}]{Nashed:2020mnp}%
  \BibitemOpen
  \bibfield  {author} {\bibinfo {author} {\bibfnamefont {G.~G.~L.}\
  \bibnamefont {Nashed}}\ and\ \bibinfo {author} {\bibfnamefont
  {S.}~\bibnamefont {Nojiri}},\ }\href {\doibase 10.1103/PhysRevD.102.124022}
  {\bibfield  {journal} {\bibinfo  {journal} {Phys. Rev. D}\ }\textbf {\bibinfo
  {volume} {102}},\ \bibinfo {pages} {124022} (\bibinfo {year} {2020})},\
  \Eprint {http://arxiv.org/abs/2012.05711} {arXiv:2012.05711 [gr-qc]}
  \BibitemShut {NoStop}%
\bibitem [{\citenamefont {Abbott}\ \emph
  {et~al.}(2016{\natexlab{a}})\citenamefont {Abbott} \emph
  {et~al.}}]{Abbott:2016blz}%
  \BibitemOpen
  \bibfield  {author} {\bibinfo {author} {\bibfnamefont {B.~P.}\ \bibnamefont
  {Abbott}} \emph {et~al.} (\bibinfo {collaboration} {LIGO Scientific,
  Virgo}),\ }\href {\doibase 10.1103/PhysRevLett.116.061102} {\bibfield
  {journal} {\bibinfo  {journal} {Phys. Rev. Lett.}\ }\textbf {\bibinfo
  {volume} {116}},\ \bibinfo {pages} {061102} (\bibinfo {year}
  {2016}{\natexlab{a}})},\ \Eprint {http://arxiv.org/abs/1602.03837}
  {arXiv:1602.03837 [gr-qc]} \BibitemShut {NoStop}%
\bibitem [{\citenamefont {Abbott}\ \emph
  {et~al.}(2016{\natexlab{b}})\citenamefont {Abbott} \emph
  {et~al.}}]{Abbott:2016nmj}%
  \BibitemOpen
  \bibfield  {author} {\bibinfo {author} {\bibfnamefont {B.~P.}\ \bibnamefont
  {Abbott}} \emph {et~al.} (\bibinfo {collaboration} {LIGO Scientific,
  Virgo}),\ }\href {\doibase 10.1103/PhysRevLett.116.241103} {\bibfield
  {journal} {\bibinfo  {journal} {Phys. Rev. Lett.}\ }\textbf {\bibinfo
  {volume} {116}},\ \bibinfo {pages} {241103} (\bibinfo {year}
  {2016}{\natexlab{b}})},\ \Eprint {http://arxiv.org/abs/1606.04855}
  {arXiv:1606.04855 [gr-qc]} \BibitemShut {NoStop}%
\bibitem [{\citenamefont {Abbott}\ \emph
  {et~al.}(2017{\natexlab{a}})\citenamefont {Abbott} \emph
  {et~al.}}]{Abbott:2017vtc}%
  \BibitemOpen
  \bibfield  {author} {\bibinfo {author} {\bibfnamefont {B.~P.}\ \bibnamefont
  {Abbott}} \emph {et~al.} (\bibinfo {collaboration} {LIGO Scientific,
  VIRGO}),\ }\href {\doibase 10.1103/PhysRevLett.118.221101} {\bibfield
  {journal} {\bibinfo  {journal} {Phys. Rev. Lett.}\ }\textbf {\bibinfo
  {volume} {118}},\ \bibinfo {pages} {221101} (\bibinfo {year}
  {2017}{\natexlab{a}})},\ \bibinfo {note} {[Erratum: Phys.Rev.Lett. 121,
  129901 (2018)]},\ \Eprint {http://arxiv.org/abs/1706.01812} {arXiv:1706.01812
  [gr-qc]} \BibitemShut {NoStop}%
\bibitem [{\citenamefont {Abbott}\ \emph
  {et~al.}(2017{\natexlab{b}})\citenamefont {Abbott} \emph
  {et~al.}}]{TheLIGOScientific:2017qsa}%
  \BibitemOpen
  \bibfield  {author} {\bibinfo {author} {\bibfnamefont {B.~P.}\ \bibnamefont
  {Abbott}} \emph {et~al.} (\bibinfo {collaboration} {LIGO Scientific,
  Virgo}),\ }\href {\doibase 10.1103/PhysRevLett.119.161101} {\bibfield
  {journal} {\bibinfo  {journal} {Phys. Rev. Lett.}\ }\textbf {\bibinfo
  {volume} {119}},\ \bibinfo {pages} {161101} (\bibinfo {year}
  {2017}{\natexlab{b}})},\ \Eprint {http://arxiv.org/abs/1710.05832}
  {arXiv:1710.05832 [gr-qc]} \BibitemShut {NoStop}%
\bibitem [{\citenamefont {Mann}\ \emph {et~al.}(2019)\citenamefont {Mann},
  \citenamefont {Richer}, \citenamefont {Heyl}, \citenamefont {Anderson},
  \citenamefont {Kalirai}, \citenamefont {Caiazzo}, \citenamefont {M\"ohle},
  \citenamefont {Knee},\ and\ \citenamefont {Baumgardt}}]{Mann:2018xkm}%
  \BibitemOpen
  \bibfield  {author} {\bibinfo {author} {\bibfnamefont {C.~R.}\ \bibnamefont
  {Mann}}, \bibinfo {author} {\bibfnamefont {H.}~\bibnamefont {Richer}},
  \bibinfo {author} {\bibfnamefont {J.}~\bibnamefont {Heyl}}, \bibinfo {author}
  {\bibfnamefont {J.}~\bibnamefont {Anderson}}, \bibinfo {author}
  {\bibfnamefont {J.}~\bibnamefont {Kalirai}}, \bibinfo {author} {\bibfnamefont
  {I.}~\bibnamefont {Caiazzo}}, \bibinfo {author} {\bibfnamefont
  {S.}~\bibnamefont {M\"ohle}}, \bibinfo {author} {\bibfnamefont
  {A.}~\bibnamefont {Knee}}, \ and\ \bibinfo {author} {\bibfnamefont
  {H.}~\bibnamefont {Baumgardt}},\ }\href {\doibase 10.3847/1538-4357/ab0e6d}
  {\bibfield  {journal} {\bibinfo  {journal} {Astrophys. J.}\ }\textbf
  {\bibinfo {volume} {875}},\ \bibinfo {pages} {1} (\bibinfo {year} {2019})},\
  \Eprint {http://arxiv.org/abs/1807.03307} {arXiv:1807.03307 [astro-ph.GA]}
  \BibitemShut {NoStop}%
\bibitem [{\citenamefont {Poisson}\ and\ \citenamefont
  {Israel}(1989)}]{Poisson:1989zz}%
  \BibitemOpen
  \bibfield  {author} {\bibinfo {author} {\bibfnamefont {E.}~\bibnamefont
  {Poisson}}\ and\ \bibinfo {author} {\bibfnamefont {W.}~\bibnamefont
  {Israel}},\ }\href {\doibase 10.1103/PhysRevLett.63.1663} {\bibfield
  {journal} {\bibinfo  {journal} {Phys. Rev. Lett.}\ }\textbf {\bibinfo
  {volume} {63}},\ \bibinfo {pages} {1663} (\bibinfo {year}
  {1989})}\BibitemShut {NoStop}%
\bibitem [{\citenamefont {Poisson}\ and\ \citenamefont
  {Israel}(1990)}]{Poisson:1990eh}%
  \BibitemOpen
  \bibfield  {author} {\bibinfo {author} {\bibfnamefont {E.}~\bibnamefont
  {Poisson}}\ and\ \bibinfo {author} {\bibfnamefont {W.}~\bibnamefont
  {Israel}},\ }\href {\doibase 10.1103/PhysRevD.41.1796} {\bibfield  {journal}
  {\bibinfo  {journal} {Phys. Rev. D}\ }\textbf {\bibinfo {volume} {41}},\
  \bibinfo {pages} {1796} (\bibinfo {year} {1990})}\BibitemShut {NoStop}%
\bibitem [{\citenamefont {Ori}(1991)}]{Ori:1991zz}%
  \BibitemOpen
  \bibfield  {author} {\bibinfo {author} {\bibfnamefont {A.}~\bibnamefont
  {Ori}},\ }\href {\doibase 10.1103/PhysRevLett.67.789} {\bibfield  {journal}
  {\bibinfo  {journal} {Phys. Rev. Lett.}\ }\textbf {\bibinfo {volume} {67}},\
  \bibinfo {pages} {789} (\bibinfo {year} {1991})}\BibitemShut {NoStop}%
\bibitem [{\citenamefont {Simpson}\ and\ \citenamefont
  {Visser}(2019)}]{Simpson:2018tsi}%
  \BibitemOpen
  \bibfield  {author} {\bibinfo {author} {\bibfnamefont {A.}~\bibnamefont
  {Simpson}}\ and\ \bibinfo {author} {\bibfnamefont {M.}~\bibnamefont
  {Visser}},\ }\href {\doibase 10.1088/1475-7516/2019/02/042} {\bibfield
  {journal} {\bibinfo  {journal} {JCAP}\ }\textbf {\bibinfo {volume} {02}},\
  \bibinfo {pages} {042} (\bibinfo {year} {2019})},\ \Eprint
  {http://arxiv.org/abs/1812.07114} {arXiv:1812.07114 [gr-qc]} \BibitemShut
  {NoStop}%
\bibitem [{\citenamefont {Cardoso}\ and\ \citenamefont
  {Pani}(2019)}]{Cardoso:2019rvt}%
  \BibitemOpen
  \bibfield  {author} {\bibinfo {author} {\bibfnamefont {V.}~\bibnamefont
  {Cardoso}}\ and\ \bibinfo {author} {\bibfnamefont {P.}~\bibnamefont {Pani}},\
  }\href {\doibase 10.1007/s41114-019-0020-4} {\bibfield  {journal} {\bibinfo
  {journal} {Living Rev. Rel.}\ }\textbf {\bibinfo {volume} {22}},\ \bibinfo
  {pages} {4} (\bibinfo {year} {2019})},\ \Eprint
  {http://arxiv.org/abs/1904.05363} {arXiv:1904.05363 [gr-qc]} \BibitemShut
  {NoStop}%
\bibitem [{\citenamefont {Carballo-Rubio}(2018)}]{Carballo-Rubio:2017tlh}%
  \BibitemOpen
  \bibfield  {author} {\bibinfo {author} {\bibfnamefont {R.}~\bibnamefont
  {Carballo-Rubio}},\ }\href {\doibase 10.1103/PhysRevLett.120.061102}
  {\bibfield  {journal} {\bibinfo  {journal} {Phys. Rev. Lett.}\ }\textbf
  {\bibinfo {volume} {120}},\ \bibinfo {pages} {061102} (\bibinfo {year}
  {2018})},\ \Eprint {http://arxiv.org/abs/1706.05379} {arXiv:1706.05379
  [gr-qc]} \BibitemShut {NoStop}%
\bibitem [{\citenamefont {Mark}\ \emph {et~al.}(2017)\citenamefont {Mark},
  \citenamefont {Zimmerman}, \citenamefont {Du},\ and\ \citenamefont
  {Chen}}]{Mark:2017dnq}%
  \BibitemOpen
  \bibfield  {author} {\bibinfo {author} {\bibfnamefont {Z.}~\bibnamefont
  {Mark}}, \bibinfo {author} {\bibfnamefont {A.}~\bibnamefont {Zimmerman}},
  \bibinfo {author} {\bibfnamefont {S.~M.}\ \bibnamefont {Du}}, \ and\ \bibinfo
  {author} {\bibfnamefont {Y.}~\bibnamefont {Chen}},\ }\href {\doibase
  10.1103/PhysRevD.96.084002} {\bibfield  {journal} {\bibinfo  {journal} {Phys.
  Rev. D}\ }\textbf {\bibinfo {volume} {96}},\ \bibinfo {pages} {084002}
  (\bibinfo {year} {2017})},\ \Eprint {http://arxiv.org/abs/1706.06155}
  {arXiv:1706.06155 [gr-qc]} \BibitemShut {NoStop}%
\bibitem [{\citenamefont {Cardoso}\ \emph {et~al.}(2016)\citenamefont
  {Cardoso}, \citenamefont {Hopper}, \citenamefont {Macedo}, \citenamefont
  {Palenzuela},\ and\ \citenamefont {Pani}}]{Cardoso:2016oxy}%
  \BibitemOpen
  \bibfield  {author} {\bibinfo {author} {\bibfnamefont {V.}~\bibnamefont
  {Cardoso}}, \bibinfo {author} {\bibfnamefont {S.}~\bibnamefont {Hopper}},
  \bibinfo {author} {\bibfnamefont {C.~F.~B.}\ \bibnamefont {Macedo}}, \bibinfo
  {author} {\bibfnamefont {C.}~\bibnamefont {Palenzuela}}, \ and\ \bibinfo
  {author} {\bibfnamefont {P.}~\bibnamefont {Pani}},\ }\href {\doibase
  10.1103/PhysRevD.94.084031} {\bibfield  {journal} {\bibinfo  {journal} {Phys.
  Rev. D}\ }\textbf {\bibinfo {volume} {94}},\ \bibinfo {pages} {084031}
  (\bibinfo {year} {2016})},\ \Eprint {http://arxiv.org/abs/1608.08637}
  {arXiv:1608.08637 [gr-qc]} \BibitemShut {NoStop}%
\bibitem [{\citenamefont {Hod}(2017)}]{Hod:2017cga}%
  \BibitemOpen
  \bibfield  {author} {\bibinfo {author} {\bibfnamefont {S.}~\bibnamefont
  {Hod}},\ }\href {\doibase 10.1007/JHEP06(2017)132} {\bibfield  {journal}
  {\bibinfo  {journal} {JHEP}\ }\textbf {\bibinfo {volume} {06}},\ \bibinfo
  {pages} {132} (\bibinfo {year} {2017})},\ \Eprint
  {http://arxiv.org/abs/1704.05856} {arXiv:1704.05856 [hep-th]} \BibitemShut
  {NoStop}%
\bibitem [{\citenamefont {Cardoso}\ \emph {et~al.}(2014)\citenamefont
  {Cardoso}, \citenamefont {Crispino}, \citenamefont {Macedo}, \citenamefont
  {Okawa},\ and\ \citenamefont {Pani}}]{Cardoso:2014sna}%
  \BibitemOpen
  \bibfield  {author} {\bibinfo {author} {\bibfnamefont {V.}~\bibnamefont
  {Cardoso}}, \bibinfo {author} {\bibfnamefont {L.~C.~B.}\ \bibnamefont
  {Crispino}}, \bibinfo {author} {\bibfnamefont {C.~F.~B.}\ \bibnamefont
  {Macedo}}, \bibinfo {author} {\bibfnamefont {H.}~\bibnamefont {Okawa}}, \
  and\ \bibinfo {author} {\bibfnamefont {P.}~\bibnamefont {Pani}},\ }\href
  {\doibase 10.1103/PhysRevD.90.044069} {\bibfield  {journal} {\bibinfo
  {journal} {Phys. Rev. D}\ }\textbf {\bibinfo {volume} {90}},\ \bibinfo
  {pages} {044069} (\bibinfo {year} {2014})},\ \Eprint
  {http://arxiv.org/abs/1406.5510} {arXiv:1406.5510 [gr-qc]} \BibitemShut
  {NoStop}%
\bibitem [{\citenamefont {Sebastiani}\ \emph {et~al.}(2019)\citenamefont
  {Sebastiani}, \citenamefont {Vanzo},\ and\ \citenamefont
  {Zerbini}}]{Sebastiani:2018ktb}%
  \BibitemOpen
  \bibfield  {author} {\bibinfo {author} {\bibfnamefont {L.}~\bibnamefont
  {Sebastiani}}, \bibinfo {author} {\bibfnamefont {L.}~\bibnamefont {Vanzo}}, \
  and\ \bibinfo {author} {\bibfnamefont {S.}~\bibnamefont {Zerbini}},\ }\href
  {\doibase 10.1142/S0219887819501810} {\bibfield  {journal} {\bibinfo
  {journal} {Int. J. Geom. Meth. Mod. Phys.}\ }\textbf {\bibinfo {volume}
  {16}},\ \bibinfo {pages} {1950181} (\bibinfo {year} {2019})},\ \Eprint
  {http://arxiv.org/abs/1808.06939} {arXiv:1808.06939 [gr-qc]} \BibitemShut
  {NoStop}%
\bibitem [{\citenamefont {Antoniadis}\ \emph {et~al.}(1986)\citenamefont
  {Antoniadis}, \citenamefont {Iliopoulos},\ and\ \citenamefont
  {Tomaras}}]{Antoniadis:1985pj}%
  \BibitemOpen
  \bibfield  {author} {\bibinfo {author} {\bibfnamefont {I.}~\bibnamefont
  {Antoniadis}}, \bibinfo {author} {\bibfnamefont {J.}~\bibnamefont
  {Iliopoulos}}, \ and\ \bibinfo {author} {\bibfnamefont {T.~N.}\ \bibnamefont
  {Tomaras}},\ }\href {\doibase 10.1103/PhysRevLett.56.1319} {\bibfield
  {journal} {\bibinfo  {journal} {Phys. Rev. Lett.}\ }\textbf {\bibinfo
  {volume} {56}},\ \bibinfo {pages} {1319} (\bibinfo {year}
  {1986})}\BibitemShut {NoStop}%
\bibitem [{\citenamefont {Capozziello}\ and\ \citenamefont
  {De~Laurentis}(2011)}]{Capozziello:2011et}%
  \BibitemOpen
  \bibfield  {author} {\bibinfo {author} {\bibfnamefont {S.}~\bibnamefont
  {Capozziello}}\ and\ \bibinfo {author} {\bibfnamefont {M.}~\bibnamefont
  {De~Laurentis}},\ }\href {\doibase 10.1016/j.physrep.2011.09.003} {\bibfield
  {journal} {\bibinfo  {journal} {Phys. Rept.}\ }\textbf {\bibinfo {volume}
  {509}},\ \bibinfo {pages} {167} (\bibinfo {year} {2011})},\ \Eprint
  {http://arxiv.org/abs/1108.6266} {arXiv:1108.6266 [gr-qc]} \BibitemShut
  {NoStop}%
\bibitem [{\citenamefont {Martin-Moruno}\ and\ \citenamefont
  {Nunes}(2015)}]{Martin-Moruno:2015kaa}%
  \BibitemOpen
  \bibfield  {author} {\bibinfo {author} {\bibfnamefont {P.}~\bibnamefont
  {Martin-Moruno}}\ and\ \bibinfo {author} {\bibfnamefont {N.~J.}\ \bibnamefont
  {Nunes}},\ }\href {\doibase 10.1088/1475-7516/2015/09/056} {\bibfield
  {journal} {\bibinfo  {journal} {JCAP}\ }\textbf {\bibinfo {volume} {09}},\
  \bibinfo {pages} {056} (\bibinfo {year} {2015})},\ \Eprint
  {http://arxiv.org/abs/1506.02497} {arXiv:1506.02497 [gr-qc]} \BibitemShut
  {NoStop}%
\bibitem [{\citenamefont {Bhattacharya}(2016)}]{Bhattacharya:2016lup}%
  \BibitemOpen
  \bibfield  {author} {\bibinfo {author} {\bibfnamefont {S.}~\bibnamefont
  {Bhattacharya}},\ }\href {\doibase 10.1007/s10714-016-2119-1} {\bibfield
  {journal} {\bibinfo  {journal} {Gen. Rel. Grav.}\ }\textbf {\bibinfo {volume}
  {48}},\ \bibinfo {pages} {128} (\bibinfo {year} {2016})},\ \Eprint
  {http://arxiv.org/abs/1602.04306} {arXiv:1602.04306 [gr-qc]} \BibitemShut
  {NoStop}%
\bibitem [{\citenamefont {Cardoso}\ \emph {et~al.}(2018)\citenamefont
  {Cardoso}, \citenamefont {Dias}, \citenamefont {Hartnett}, \citenamefont
  {Middleton}, \citenamefont {Pani},\ and\ \citenamefont
  {Santos}}]{Cardoso:2018tly}%
  \BibitemOpen
  \bibfield  {author} {\bibinfo {author} {\bibfnamefont {V.}~\bibnamefont
  {Cardoso}}, \bibinfo {author} {\bibfnamefont {O.~J.~C.}\ \bibnamefont
  {Dias}}, \bibinfo {author} {\bibfnamefont {G.~S.}\ \bibnamefont {Hartnett}},
  \bibinfo {author} {\bibfnamefont {M.}~\bibnamefont {Middleton}}, \bibinfo
  {author} {\bibfnamefont {P.}~\bibnamefont {Pani}}, \ and\ \bibinfo {author}
  {\bibfnamefont {J.~E.}\ \bibnamefont {Santos}},\ }\href {\doibase
  10.1088/1475-7516/2018/03/043} {\bibfield  {journal} {\bibinfo  {journal}
  {JCAP}\ }\textbf {\bibinfo {volume} {03}},\ \bibinfo {pages} {043} (\bibinfo
  {year} {2018})},\ \Eprint {http://arxiv.org/abs/1801.01420} {arXiv:1801.01420
  [gr-qc]} \BibitemShut {NoStop}%
\bibitem [{\citenamefont {Nojiri}\ \emph {et~al.}(2017)\citenamefont {Nojiri},
  \citenamefont {Odintsov},\ and\ \citenamefont {Oikonomou}}]{Nojiri:2017ncd}%
  \BibitemOpen
  \bibfield  {author} {\bibinfo {author} {\bibfnamefont {S.}~\bibnamefont
  {Nojiri}}, \bibinfo {author} {\bibfnamefont {S.~D.}\ \bibnamefont
  {Odintsov}}, \ and\ \bibinfo {author} {\bibfnamefont {V.~K.}\ \bibnamefont
  {Oikonomou}},\ }\href {\doibase 10.1016/j.physrep.2017.06.001} {\bibfield
  {journal} {\bibinfo  {journal} {Phys. Rept.}\ }\textbf {\bibinfo {volume}
  {692}},\ \bibinfo {pages} {1} (\bibinfo {year} {2017})},\ \Eprint
  {http://arxiv.org/abs/1705.11098} {arXiv:1705.11098 [gr-qc]} \BibitemShut
  {NoStop}%
\bibitem [{\citenamefont {Nojiri}\ and\ \citenamefont
  {Odintsov}(2011)}]{Nojiri:2010wj}%
  \BibitemOpen
  \bibfield  {author} {\bibinfo {author} {\bibfnamefont {S.}~\bibnamefont
  {Nojiri}}\ and\ \bibinfo {author} {\bibfnamefont {S.~D.}\ \bibnamefont
  {Odintsov}},\ }\href {\doibase 10.1016/j.physrep.2011.04.001} {\bibfield
  {journal} {\bibinfo  {journal} {Phys. Rept.}\ }\textbf {\bibinfo {volume}
  {505}},\ \bibinfo {pages} {59} (\bibinfo {year} {2011})},\ \Eprint
  {http://arxiv.org/abs/1011.0544} {arXiv:1011.0544 [gr-qc]} \BibitemShut
  {NoStop}%
\bibitem [{\citenamefont {De~Felice}\ and\ \citenamefont
  {Tsujikawa}(2010)}]{DeFelice:2010aj}%
  \BibitemOpen
  \bibfield  {author} {\bibinfo {author} {\bibfnamefont {A.}~\bibnamefont
  {De~Felice}}\ and\ \bibinfo {author} {\bibfnamefont {S.}~\bibnamefont
  {Tsujikawa}},\ }\href {\doibase 10.12942/lrr-2010-3} {\bibfield  {journal}
  {\bibinfo  {journal} {Living Rev. Rel.}\ }\textbf {\bibinfo {volume} {13}},\
  \bibinfo {pages} {3} (\bibinfo {year} {2010})},\ \Eprint
  {http://arxiv.org/abs/1002.4928} {arXiv:1002.4928 [gr-qc]} \BibitemShut
  {NoStop}%
\bibitem [{\citenamefont {Capozziello}\ \emph {et~al.}(2007)\citenamefont
  {Capozziello}, \citenamefont {Stabile},\ and\ \citenamefont
  {Troisi}}]{Capozziello:2007wc}%
  \BibitemOpen
  \bibfield  {author} {\bibinfo {author} {\bibfnamefont {S.}~\bibnamefont
  {Capozziello}}, \bibinfo {author} {\bibfnamefont {A.}~\bibnamefont
  {Stabile}}, \ and\ \bibinfo {author} {\bibfnamefont {A.}~\bibnamefont
  {Troisi}},\ }\href {\doibase 10.1088/0264-9381/24/8/013} {\bibfield
  {journal} {\bibinfo  {journal} {Class. Quant. Grav.}\ }\textbf {\bibinfo
  {volume} {24}},\ \bibinfo {pages} {2153} (\bibinfo {year} {2007})},\ \Eprint
  {http://arxiv.org/abs/gr-qc/0703067} {arXiv:gr-qc/0703067 [gr-qc]}
  \BibitemShut {NoStop}%
\bibitem [{\citenamefont {Capozziello}\ \emph {et~al.}(2010)\citenamefont
  {Capozziello}, \citenamefont {De~laurentis},\ and\ \citenamefont
  {Stabile}}]{Capozziello:2009jg}%
  \BibitemOpen
  \bibfield  {author} {\bibinfo {author} {\bibfnamefont {S.}~\bibnamefont
  {Capozziello}}, \bibinfo {author} {\bibfnamefont {M.}~\bibnamefont
  {De~laurentis}}, \ and\ \bibinfo {author} {\bibfnamefont {A.}~\bibnamefont
  {Stabile}},\ }\href {\doibase 10.1088/0264-9381/27/16/165008} {\bibfield
  {journal} {\bibinfo  {journal} {Class. Quant. Grav.}\ }\textbf {\bibinfo
  {volume} {27}},\ \bibinfo {pages} {165008} (\bibinfo {year} {2010})},\
  \Eprint {http://arxiv.org/abs/0912.5286} {arXiv:0912.5286 [gr-qc]}
  \BibitemShut {NoStop}%
\bibitem [{\citenamefont {Multam\"aki}\ and\ \citenamefont
  {Vilja}(2006)}]{PhysRevD.74.064022}%
  \BibitemOpen
  \bibfield  {author} {\bibinfo {author} {\bibfnamefont {T.}~\bibnamefont
  {Multam\"aki}}\ and\ \bibinfo {author} {\bibfnamefont {I.}~\bibnamefont
  {Vilja}},\ }\href {\doibase 10.1103/PhysRevD.74.064022} {\bibfield  {journal}
  {\bibinfo  {journal} {Phys. Rev. D}\ }\textbf {\bibinfo {volume} {74}},\
  \bibinfo {pages} {064022} (\bibinfo {year} {2006})}\BibitemShut {NoStop}%
\bibitem [{\citenamefont {Multam\"aki}\ and\ \citenamefont
  {Vilja}(2007)}]{PhysRevD.76.064021}%
  \BibitemOpen
  \bibfield  {author} {\bibinfo {author} {\bibfnamefont {T.}~\bibnamefont
  {Multam\"aki}}\ and\ \bibinfo {author} {\bibfnamefont {I.}~\bibnamefont
  {Vilja}},\ }\href {\doibase 10.1103/PhysRevD.76.064021} {\bibfield  {journal}
  {\bibinfo  {journal} {Phys. Rev. D}\ }\textbf {\bibinfo {volume} {76}},\
  \bibinfo {pages} {064021} (\bibinfo {year} {2007})}\BibitemShut {NoStop}%
\bibitem [{\citenamefont {Hollenstein}\ and\ \citenamefont
  {Lobo}(2008)}]{Hollenstein:2008hp}%
  \BibitemOpen
  \bibfield  {author} {\bibinfo {author} {\bibfnamefont {L.}~\bibnamefont
  {Hollenstein}}\ and\ \bibinfo {author} {\bibfnamefont {F.~S.~N.}\
  \bibnamefont {Lobo}},\ }\href {\doibase 10.1103/PhysRevD.78.124007}
  {\bibfield  {journal} {\bibinfo  {journal} {Phys. Rev.}\ }\textbf {\bibinfo
  {volume} {D78}},\ \bibinfo {pages} {124007} (\bibinfo {year} {2008})},\
  \Eprint {http://arxiv.org/abs/0807.2325} {arXiv:0807.2325 [gr-qc]}
  \BibitemShut {NoStop}%
\bibitem [{\citenamefont {Goswami}\ \emph {et~al.}(2014)\citenamefont
  {Goswami}, \citenamefont {Nzioki}, \citenamefont {Maharaj},\ and\
  \citenamefont {Ghosh}}]{PhysRevD.90.084011}%
  \BibitemOpen
  \bibfield  {author} {\bibinfo {author} {\bibfnamefont {R.}~\bibnamefont
  {Goswami}}, \bibinfo {author} {\bibfnamefont {A.~M.}\ \bibnamefont {Nzioki}},
  \bibinfo {author} {\bibfnamefont {S.~D.}\ \bibnamefont {Maharaj}}, \ and\
  \bibinfo {author} {\bibfnamefont {S.~G.}\ \bibnamefont {Ghosh}},\ }\href
  {\doibase 10.1103/PhysRevD.90.084011} {\bibfield  {journal} {\bibinfo
  {journal} {Phys. Rev. D}\ }\textbf {\bibinfo {volume} {90}},\ \bibinfo
  {pages} {084011} (\bibinfo {year} {2014})}\BibitemShut {NoStop}%
\bibitem [{\citenamefont {Hendi}\ \emph {et~al.}(2012)\citenamefont {Hendi},
  \citenamefont {Eslam~Panah},\ and\ \citenamefont {Mousavi}}]{Hendi:2011eg}%
  \BibitemOpen
  \bibfield  {author} {\bibinfo {author} {\bibfnamefont {S.~H.}\ \bibnamefont
  {Hendi}}, \bibinfo {author} {\bibfnamefont {B.}~\bibnamefont {Eslam~Panah}},
  \ and\ \bibinfo {author} {\bibfnamefont {S.~M.}\ \bibnamefont {Mousavi}},\
  }\href {\doibase 10.1007/s10714-011-1307-2} {\bibfield  {journal} {\bibinfo
  {journal} {Gen. Rel. Grav.}\ }\textbf {\bibinfo {volume} {44}},\ \bibinfo
  {pages} {835} (\bibinfo {year} {2012})},\ \Eprint
  {http://arxiv.org/abs/1102.0089} {arXiv:1102.0089 [hep-th]} \BibitemShut
  {NoStop}%
\bibitem [{\citenamefont {Nashed}(2006)}]{Nashed:2005kn}%
  \BibitemOpen
  \bibfield  {author} {\bibinfo {author} {\bibfnamefont {G.~G.~L.}\
  \bibnamefont {Nashed}},\ }\href {\doibase 10.1142/S0217751X06031478}
  {\bibfield  {journal} {\bibinfo  {journal} {Int. J. Mod. Phys.}\ ,\ \bibinfo
  {pages} {3181}} (\bibinfo {year} {2006})},\ \Eprint
  {http://arxiv.org/abs/gr-qc/0501002} {arXiv:gr-qc/0501002 [gr-qc]}
  \BibitemShut {NoStop}%
\bibitem [{\citenamefont {Bamba}\ \emph {et~al.}(2012)\citenamefont {Bamba},
  \citenamefont {Nojiri},\ and\ \citenamefont {Odintsov}}]{PhysRevD.85.044012}%
  \BibitemOpen
  \bibfield  {author} {\bibinfo {author} {\bibfnamefont {K.}~\bibnamefont
  {Bamba}}, \bibinfo {author} {\bibfnamefont {S.}~\bibnamefont {Nojiri}}, \
  and\ \bibinfo {author} {\bibfnamefont {S.~D.}\ \bibnamefont {Odintsov}},\
  }\href {\doibase 10.1103/PhysRevD.85.044012} {\bibfield  {journal} {\bibinfo
  {journal} {Phys. Rev. D}\ }\textbf {\bibinfo {volume} {85}},\ \bibinfo
  {pages} {044012} (\bibinfo {year} {2012})}\BibitemShut {NoStop}%
\bibitem [{\citenamefont {Nojiri}\ and\ \citenamefont
  {Odintsov}(2014)}]{Nojiri:2014jqa}%
  \BibitemOpen
  \bibfield  {author} {\bibinfo {author} {\bibfnamefont {S.}~\bibnamefont
  {Nojiri}}\ and\ \bibinfo {author} {\bibfnamefont {S.~D.}\ \bibnamefont
  {Odintsov}},\ }\href {\doibase 10.1016/j.physletb.2014.06.070} {\bibfield
  {journal} {\bibinfo  {journal} {Phys. Lett.}\ }\textbf {\bibinfo {volume}
  {B735}},\ \bibinfo {pages} {376} (\bibinfo {year} {2014})},\ \Eprint
  {http://arxiv.org/abs/1405.2439} {arXiv:1405.2439 [gr-qc]} \BibitemShut
  {NoStop}%
\bibitem [{\citenamefont {Nojiri}\ and\ \citenamefont
  {Odintsov}(2017)}]{Nojiri:2017kex}%
  \BibitemOpen
  \bibfield  {author} {\bibinfo {author} {\bibfnamefont {S.}~\bibnamefont
  {Nojiri}}\ and\ \bibinfo {author} {\bibfnamefont {S.~D.}\ \bibnamefont
  {Odintsov}},\ }\href {\doibase 10.1103/PhysRevD.96.104008} {\bibfield
  {journal} {\bibinfo  {journal} {Phys. Rev.}\ }\textbf {\bibinfo {volume}
  {D96}},\ \bibinfo {pages} {104008} (\bibinfo {year} {2017})},\ \Eprint
  {http://arxiv.org/abs/1708.05226} {arXiv:1708.05226 [hep-th]} \BibitemShut
  {NoStop}%
\bibitem [{\citenamefont {Shirafuji}\ and\ \citenamefont
  {Nashed}(1997)}]{Shirafuji:1997wy}%
  \BibitemOpen
  \bibfield  {author} {\bibinfo {author} {\bibfnamefont {T.}~\bibnamefont
  {Shirafuji}}\ and\ \bibinfo {author} {\bibfnamefont {G.~G.~L.}\ \bibnamefont
  {Nashed}},\ }\href {\doibase 10.1143/PTP.98.1355} {\bibfield  {journal}
  {\bibinfo  {journal} {Prog. Theor. Phys.}\ }\textbf {\bibinfo {volume}
  {98}},\ \bibinfo {pages} {1355} (\bibinfo {year} {1997})},\ \Eprint
  {http://arxiv.org/abs/gr-qc/9711010} {arXiv:gr-qc/9711010} \BibitemShut
  {NoStop}%
\bibitem [{\citenamefont {Bergliaffa}\ and\ \citenamefont
  {Nunes}(2011)}]{PhysRevD.84.084006}%
  \BibitemOpen
  \bibfield  {author} {\bibinfo {author} {\bibfnamefont {S.~E.~P.}\
  \bibnamefont {Bergliaffa}}\ and\ \bibinfo {author} {\bibfnamefont {Y.~E. C.
  d.~O.}\ \bibnamefont {Nunes}},\ }\href {\doibase 10.1103/PhysRevD.84.084006}
  {\bibfield  {journal} {\bibinfo  {journal} {Phys. Rev. D}\ }\textbf {\bibinfo
  {volume} {84}},\ \bibinfo {pages} {084006} (\bibinfo {year}
  {2011})}\BibitemShut {NoStop}%
\bibitem [{\citenamefont {Awad}\ \emph {et~al.}(2017)\citenamefont {Awad},
  \citenamefont {Capozziello},\ and\ \citenamefont {Nashed}}]{Awad:2017tyz}%
  \BibitemOpen
  \bibfield  {author} {\bibinfo {author} {\bibfnamefont {A.~M.}\ \bibnamefont
  {Awad}}, \bibinfo {author} {\bibfnamefont {S.}~\bibnamefont {Capozziello}}, \
  and\ \bibinfo {author} {\bibfnamefont {G.~G.~L.}\ \bibnamefont {Nashed}},\
  }\href {\doibase 10.1007/JHEP07(2017)136} {\bibfield  {journal} {\bibinfo
  {journal} {JHEP}\ }\textbf {\bibinfo {volume} {07}},\ \bibinfo {pages} {136}
  (\bibinfo {year} {2017})},\ \Eprint {http://arxiv.org/abs/1706.01773}
  {arXiv:1706.01773 [gr-qc]} \BibitemShut {NoStop}%
\bibitem [{\citenamefont {Cembranos}\ \emph {et~al.}(2014)\citenamefont
  {Cembranos}, \citenamefont {de~la Cruz-Dombriz},\ and\ \citenamefont
  {Jimeno~Romero}}]{Cembranos:2011sr}%
  \BibitemOpen
  \bibfield  {author} {\bibinfo {author} {\bibfnamefont {J.~A.~R.}\
  \bibnamefont {Cembranos}}, \bibinfo {author} {\bibfnamefont {A.}~\bibnamefont
  {de~la Cruz-Dombriz}}, \ and\ \bibinfo {author} {\bibfnamefont
  {P.}~\bibnamefont {Jimeno~Romero}},\ }\href {\doibase
  10.1142/S0219887814500017} {\bibfield  {journal} {\bibinfo  {journal} {Int.
  J. Geom. Meth. Mod. Phys.}\ }\textbf {\bibinfo {volume} {11}},\ \bibinfo
  {pages} {1450001} (\bibinfo {year} {2014})},\ \Eprint
  {http://arxiv.org/abs/1109.4519} {arXiv:1109.4519 [gr-qc]} \BibitemShut
  {NoStop}%
\bibitem [{\citenamefont {Nashed}(2007{\natexlab{a}})}]{Nashed:2006yw}%
  \BibitemOpen
  \bibfield  {author} {\bibinfo {author} {\bibfnamefont {G.~G.~L.}\
  \bibnamefont {Nashed}},\ }\href {\doibase 10.1142/S021773230702141X}
  {\bibfield  {journal} {\bibinfo  {journal} {Mod. Phys. Lett.}\ }\textbf
  {\bibinfo {volume} {A22}},\ \bibinfo {pages} {1047} (\bibinfo {year}
  {2007}{\natexlab{a}})},\ \Eprint {http://arxiv.org/abs/gr-qc/0609096}
  {arXiv:gr-qc/0609096 [gr-qc]} \BibitemShut {NoStop}%
\bibitem [{\citenamefont {Lobo}\ and\ \citenamefont
  {Oliveira}(2009)}]{PhysRevD.80.104012}%
  \BibitemOpen
  \bibfield  {author} {\bibinfo {author} {\bibfnamefont {F.~S.~N.}\
  \bibnamefont {Lobo}}\ and\ \bibinfo {author} {\bibfnamefont {M.~A.}\
  \bibnamefont {Oliveira}},\ }\href {\doibase 10.1103/PhysRevD.80.104012}
  {\bibfield  {journal} {\bibinfo  {journal} {Phys. Rev. D}\ }\textbf {\bibinfo
  {volume} {80}},\ \bibinfo {pages} {104012} (\bibinfo {year}
  {2009})}\BibitemShut {NoStop}%
\bibitem [{\citenamefont {Azadi}\ \emph {et~al.}(2008)\citenamefont {Azadi},
  \citenamefont {Momeni},\ and\ \citenamefont {Nouri-Zonoz}}]{Azadi:2008qu}%
  \BibitemOpen
  \bibfield  {author} {\bibinfo {author} {\bibfnamefont {A.}~\bibnamefont
  {Azadi}}, \bibinfo {author} {\bibfnamefont {D.}~\bibnamefont {Momeni}}, \
  and\ \bibinfo {author} {\bibfnamefont {M.}~\bibnamefont {Nouri-Zonoz}},\
  }\href {\doibase 10.1016/j.physletb.2008.10.054} {\bibfield  {journal}
  {\bibinfo  {journal} {Phys. Lett.}\ }\textbf {\bibinfo {volume} {B670}},\
  \bibinfo {pages} {210} (\bibinfo {year} {2008})},\ \Eprint
  {http://arxiv.org/abs/0810.4673} {arXiv:0810.4673 [gr-qc]} \BibitemShut
  {NoStop}%
\bibitem [{\citenamefont {Nashed}(2010{\natexlab{a}})}]{Nashed:2009hn}%
  \BibitemOpen
  \bibfield  {author} {\bibinfo {author} {\bibfnamefont {G.~G.~L.}\
  \bibnamefont {Nashed}},\ }\href {\doibase 10.1088/1674-1056/19/2/020401}
  {\bibfield  {journal} {\bibinfo  {journal} {Chin. Phys.}\ ,\ \bibinfo {pages}
  {020401}} (\bibinfo {year} {2010}{\natexlab{a}})},\ \Eprint
  {http://arxiv.org/abs/0910.5124} {arXiv:0910.5124 [gr-qc]} \BibitemShut
  {NoStop}%
\bibitem [{\citenamefont {Nashed}(2008)}]{Nashed:2008ys}%
  \BibitemOpen
  \bibfield  {author} {\bibinfo {author} {\bibfnamefont {G.~G.~L.}\
  \bibnamefont {Nashed}},\ }\href {\doibase 10.1140/epjc/s10052-007-0511-4}
  {\bibfield  {journal} {\bibinfo  {journal} {Eur. Phys. J.}\ }\textbf
  {\bibinfo {volume} {C54}},\ \bibinfo {pages} {291} (\bibinfo {year}
  {2008})},\ \Eprint {http://arxiv.org/abs/0804.3285} {arXiv:0804.3285 [gr-qc]}
  \BibitemShut {NoStop}%
\bibitem [{\citenamefont {Capozziello}\ \emph {et~al.}(2008)\citenamefont
  {Capozziello}, \citenamefont {Stabile},\ and\ \citenamefont
  {Troisi}}]{Capozziello:2007id}%
  \BibitemOpen
  \bibfield  {author} {\bibinfo {author} {\bibfnamefont {S.}~\bibnamefont
  {Capozziello}}, \bibinfo {author} {\bibfnamefont {A.}~\bibnamefont
  {Stabile}}, \ and\ \bibinfo {author} {\bibfnamefont {A.}~\bibnamefont
  {Troisi}},\ }\href {\doibase 10.1088/0264-9381/25/8/085004} {\bibfield
  {journal} {\bibinfo  {journal} {Class. Quant. Grav.}\ }\textbf {\bibinfo
  {volume} {25}},\ \bibinfo {pages} {085004} (\bibinfo {year} {2008})},\
  \Eprint {http://arxiv.org/abs/0709.0891} {arXiv:0709.0891 [gr-qc]}
  \BibitemShut {NoStop}%
\bibitem [{\citenamefont {Kainulainen}\ \emph {et~al.}(2007)\citenamefont
  {Kainulainen}, \citenamefont {Piilonen}, \citenamefont {Reijonen},\ and\
  \citenamefont {Sunhede}}]{PhysRevD.76.024020}%
  \BibitemOpen
  \bibfield  {author} {\bibinfo {author} {\bibfnamefont {K.}~\bibnamefont
  {Kainulainen}}, \bibinfo {author} {\bibfnamefont {J.}~\bibnamefont
  {Piilonen}}, \bibinfo {author} {\bibfnamefont {V.}~\bibnamefont {Reijonen}},
  \ and\ \bibinfo {author} {\bibfnamefont {D.}~\bibnamefont {Sunhede}},\ }\href
  {\doibase 10.1103/PhysRevD.76.024020} {\bibfield  {journal} {\bibinfo
  {journal} {Phys. Rev. D}\ }\textbf {\bibinfo {volume} {76}},\ \bibinfo
  {pages} {024020} (\bibinfo {year} {2007})}\BibitemShut {NoStop}%
\bibitem [{\citenamefont {Nashed}(2007{\natexlab{b}})}]{Nashed:2007cu}%
  \BibitemOpen
  \bibfield  {author} {\bibinfo {author} {\bibfnamefont {G.~G.~L.}\
  \bibnamefont {Nashed}},\ }\href {\doibase 10.1140/epjc/s10052-006-0154-x}
  {\bibfield  {journal} {\bibinfo  {journal} {Eur. Phys. J.}\ ,\ \bibinfo
  {pages} {851}} (\bibinfo {year} {2007}{\natexlab{b}})},\ \Eprint
  {http://arxiv.org/abs/0706.0260} {arXiv:0706.0260 [gr-qc]} \BibitemShut
  {NoStop}%
\bibitem [{\citenamefont {Nashed}(2015)}]{Nashed:2015pda}%
  \BibitemOpen
  \bibfield  {author} {\bibinfo {author} {\bibfnamefont {G.~L.}\ \bibnamefont
  {Nashed}},\ }\href {\doibase 10.1007/s10714-015-1917-1} {\bibfield  {journal}
  {\bibinfo  {journal} {Gen. Rel. Grav.}\ }\textbf {\bibinfo {volume} {47}},\
  \bibinfo {pages} {75} (\bibinfo {year} {2015})},\ \Eprint
  {http://arxiv.org/abs/1506.08695} {arXiv:1506.08695 [gr-qc]} \BibitemShut
  {NoStop}%
\bibitem [{\citenamefont {Cognola}\ \emph {et~al.}(2015)\citenamefont
  {Cognola}, \citenamefont {Rinaldi}, \citenamefont {Vanzo},\ and\
  \citenamefont {Zerbini}}]{PhysRevD.91.104004}%
  \BibitemOpen
  \bibfield  {author} {\bibinfo {author} {\bibfnamefont {G.}~\bibnamefont
  {Cognola}}, \bibinfo {author} {\bibfnamefont {M.}~\bibnamefont {Rinaldi}},
  \bibinfo {author} {\bibfnamefont {L.}~\bibnamefont {Vanzo}}, \ and\ \bibinfo
  {author} {\bibfnamefont {S.}~\bibnamefont {Zerbini}},\ }\href {\doibase
  10.1103/PhysRevD.91.104004} {\bibfield  {journal} {\bibinfo  {journal} {Phys.
  Rev. D}\ }\textbf {\bibinfo {volume} {91}},\ \bibinfo {pages} {104004}
  (\bibinfo {year} {2015})}\BibitemShut {NoStop}%
\bibitem [{\citenamefont {{Nojiri}}\ and\ \citenamefont
  {{Odintsov}}(2013)}]{2013CQGra..30l5003N}%
  \BibitemOpen
  \bibfield  {author} {\bibinfo {author} {\bibfnamefont {S.}~\bibnamefont
  {{Nojiri}}}\ and\ \bibinfo {author} {\bibfnamefont {S.~D.}\ \bibnamefont
  {{Odintsov}}},\ }\href {\doibase 10.1088/0264-9381/30/12/125003} {\bibfield
  {journal} {\bibinfo  {journal} {Classical and Quantum Gravity}\ }\textbf
  {\bibinfo {volume} {30}},\ \bibinfo {eid} {125003} (\bibinfo {year}
  {2013})},\ \Eprint {http://arxiv.org/abs/1301.2775} {arXiv:1301.2775
  [hep-th]} \BibitemShut {NoStop}%
\bibitem [{\citenamefont {Nashed}\ and\ \citenamefont
  {El~Hanafy}(2017)}]{Nashed:2016tbj}%
  \BibitemOpen
  \bibfield  {author} {\bibinfo {author} {\bibfnamefont {G.~G.~L.}\
  \bibnamefont {Nashed}}\ and\ \bibinfo {author} {\bibfnamefont
  {W.}~\bibnamefont {El~Hanafy}},\ }\href {\doibase
  10.1140/epjc/s10052-017-4663-6} {\bibfield  {journal} {\bibinfo  {journal}
  {Eur. Phys. J.}\ ,\ \bibinfo {pages} {90}} (\bibinfo {year} {2017})},\
  \Eprint {http://arxiv.org/abs/1612.05106} {arXiv:1612.05106 [gr-qc]}
  \BibitemShut {NoStop}%
\bibitem [{\citenamefont {{Hendi}}\ \emph {et~al.}(2014)\citenamefont
  {{Hendi}}, \citenamefont {{Panah}},\ and\ \citenamefont
  {{Corda}}}]{2014CaJPh..92...76H}%
  \BibitemOpen
  \bibfield  {author} {\bibinfo {author} {\bibfnamefont {S.~H.}\ \bibnamefont
  {{Hendi}}}, \bibinfo {author} {\bibfnamefont {B.~E.}\ \bibnamefont
  {{Panah}}}, \ and\ \bibinfo {author} {\bibfnamefont {C.}~\bibnamefont
  {{Corda}}},\ }\href {\doibase 10.1139/cjp-2013-0357} {\bibfield  {journal}
  {\bibinfo  {journal} {Canadian Journal of Physics}\ }\textbf {\bibinfo
  {volume} {92}},\ \bibinfo {pages} {76} (\bibinfo {year} {2014})},\ \Eprint
  {http://arxiv.org/abs/1309.2135} {arXiv:1309.2135 [gr-qc]} \BibitemShut
  {NoStop}%
\bibitem [{\citenamefont {Awad}\ and\ \citenamefont
  {Nashed}(2017)}]{Awad:2017sau}%
  \BibitemOpen
  \bibfield  {author} {\bibinfo {author} {\bibfnamefont {A.}~\bibnamefont
  {Awad}}\ and\ \bibinfo {author} {\bibfnamefont {G.}~\bibnamefont {Nashed}},\
  }\href {\doibase 10.1088/1475-7516/2017/02/046} {\bibfield  {journal}
  {\bibinfo  {journal} {JCAP}\ }\textbf {\bibinfo {volume} {02}},\ \bibinfo
  {pages} {046} (\bibinfo {year} {2017})},\ \Eprint
  {http://arxiv.org/abs/1701.06899} {arXiv:1701.06899 [gr-qc]} \BibitemShut
  {NoStop}%
\bibitem [{\citenamefont {Hendi}\ \emph {et~al.}(2014)\citenamefont {Hendi},
  \citenamefont {Eslam~Panah},\ and\ \citenamefont {Saffari}}]{Hendi:2014mba}%
  \BibitemOpen
  \bibfield  {author} {\bibinfo {author} {\bibfnamefont {S.~H.}\ \bibnamefont
  {Hendi}}, \bibinfo {author} {\bibfnamefont {B.}~\bibnamefont {Eslam~Panah}},
  \ and\ \bibinfo {author} {\bibfnamefont {R.}~\bibnamefont {Saffari}},\ }\href
  {\doibase 10.1142/S0218271814500886} {\bibfield  {journal} {\bibinfo
  {journal} {Int. J. Mod. Phys.}\ }\textbf {\bibinfo {volume} {D23}},\ \bibinfo
  {pages} {1450088} (\bibinfo {year} {2014})},\ \Eprint
  {http://arxiv.org/abs/1408.5570} {arXiv:1408.5570 [hep-th]} \BibitemShut
  {NoStop}%
\bibitem [{\citenamefont {Nashed}(2013)}]{Nashed:uja}%
  \BibitemOpen
  \bibfield  {author} {\bibinfo {author} {\bibfnamefont {G.~G.~L.}\
  \bibnamefont {Nashed}},\ }\href {\doibase 10.1007/s10714-013-1566-1}
  {\bibfield  {journal} {\bibinfo  {journal} {Gen. Rel. Grav.}\ }\textbf
  {\bibinfo {volume} {45}},\ \bibinfo {pages} {1887} (\bibinfo {year}
  {2013})},\ \Eprint {http://arxiv.org/abs/1502.05219} {arXiv:1502.05219
  [gr-qc]} \BibitemShut {NoStop}%
\bibitem [{\citenamefont {Nashed}(2010{\natexlab{b}})}]{Nashed:2015qza}%
  \BibitemOpen
  \bibfield  {author} {\bibinfo {author} {\bibfnamefont {G.~G.~L.}\
  \bibnamefont {Nashed}},\ }\href {\doibase 10.1007/s10509-010-0375-1}
  {\bibfield  {journal} {\bibinfo  {journal} {Astrophys. Space Sci.}\ }\textbf
  {\bibinfo {volume} {330}},\ \bibinfo {pages} {173} (\bibinfo {year}
  {2010}{\natexlab{b}})},\ \Eprint {http://arxiv.org/abs/1503.01379}
  {arXiv:1503.01379 [gr-qc]} \BibitemShut {NoStop}%
\bibitem [{\citenamefont {Babichev}\ and\ \citenamefont
  {Langlois}(2010)}]{PhysRevD.81.124051}%
  \BibitemOpen
  \bibfield  {author} {\bibinfo {author} {\bibfnamefont {E.}~\bibnamefont
  {Babichev}}\ and\ \bibinfo {author} {\bibfnamefont {D.}~\bibnamefont
  {Langlois}},\ }\href {\doibase 10.1103/PhysRevD.81.124051} {\bibfield
  {journal} {\bibinfo  {journal} {Phys. Rev. D}\ }\textbf {\bibinfo {volume}
  {81}},\ \bibinfo {pages} {124051} (\bibinfo {year} {2010})}\BibitemShut
  {NoStop}%
\bibitem [{\citenamefont {Nashed}(2018)}]{Nashed:2018piz}%
  \BibitemOpen
  \bibfield  {author} {\bibinfo {author} {\bibfnamefont {G.~G.~L.}\
  \bibnamefont {Nashed}},\ }\href {\doibase 10.1155/2018/7323574} {\bibfield
  {journal} {\bibinfo  {journal} {Adv. High Energy Phys.}\ }\textbf {\bibinfo
  {volume} {2018}},\ \bibinfo {pages} {7323574} (\bibinfo {year}
  {2018})}\BibitemShut {NoStop}%
\bibitem [{\citenamefont {{Nashed}}(2018{\natexlab{a}})}]{2018EPJP..133...18N}%
  \BibitemOpen
  \bibfield  {author} {\bibinfo {author} {\bibfnamefont {G.~G.~L.}\
  \bibnamefont {{Nashed}}},\ }\href {\doibase 10.1140/epjp/i2018-11849-7}
  {\bibfield  {journal} {\bibinfo  {journal} {European Physical Journal Plus}\
  }\textbf {\bibinfo {volume} {133}},\ \bibinfo {eid} {18} (\bibinfo {year}
  {2018}{\natexlab{a}})}\BibitemShut {NoStop}%
\bibitem [{\citenamefont {{Nashed}}(2018{\natexlab{b}})}]{2018IJMPD..2750074N}%
  \BibitemOpen
  \bibfield  {author} {\bibinfo {author} {\bibfnamefont {G.~G.~L.}\
  \bibnamefont {{Nashed}}},\ }\href {\doibase 10.1142/S0218271818500748}
  {\bibfield  {journal} {\bibinfo  {journal} {International Journal of Modern
  Physics D}\ }\textbf {\bibinfo {volume} {27}},\ \bibinfo {eid} {1850074}
  (\bibinfo {year} {2018}{\natexlab{b}})}\BibitemShut {NoStop}%
\bibitem [{\citenamefont {Babichev}\ and\ \citenamefont
  {Langlois}(2009)}]{PhysRevD.80.121501}%
  \BibitemOpen
  \bibfield  {author} {\bibinfo {author} {\bibfnamefont {E.}~\bibnamefont
  {Babichev}}\ and\ \bibinfo {author} {\bibfnamefont {D.}~\bibnamefont
  {Langlois}},\ }\href {\doibase 10.1103/PhysRevD.80.121501} {\bibfield
  {journal} {\bibinfo  {journal} {Phys. Rev. D}\ }\textbf {\bibinfo {volume}
  {80}},\ \bibinfo {pages} {121501} (\bibinfo {year} {2009})}\BibitemShut
  {NoStop}%
\bibitem [{\citenamefont {Hendi}\ and\ \citenamefont
  {Momeni}(2011)}]{Hendi:2012nj}%
  \BibitemOpen
  \bibfield  {author} {\bibinfo {author} {\bibfnamefont {S.~H.}\ \bibnamefont
  {Hendi}}\ and\ \bibinfo {author} {\bibfnamefont {D.}~\bibnamefont {Momeni}},\
  }\href {\doibase 10.1140/epjc/s10052-011-1823-y} {\bibfield  {journal}
  {\bibinfo  {journal} {Eur. Phys. J.}\ }\textbf {\bibinfo {volume} {C71}},\
  \bibinfo {pages} {1823} (\bibinfo {year} {2011})},\ \Eprint
  {http://arxiv.org/abs/1201.0061} {arXiv:1201.0061 [gr-qc]} \BibitemShut
  {NoStop}%
\bibitem [{\citenamefont {Myrzakulov}\ \emph {et~al.}(2016)\citenamefont
  {Myrzakulov}, \citenamefont {Sebastiani}, \citenamefont {Vagnozzi},\ and\
  \citenamefont {Zerbini}}]{Myrzakulov:2015kda}%
  \BibitemOpen
  \bibfield  {author} {\bibinfo {author} {\bibfnamefont {R.}~\bibnamefont
  {Myrzakulov}}, \bibinfo {author} {\bibfnamefont {L.}~\bibnamefont
  {Sebastiani}}, \bibinfo {author} {\bibfnamefont {S.}~\bibnamefont
  {Vagnozzi}}, \ and\ \bibinfo {author} {\bibfnamefont {S.}~\bibnamefont
  {Zerbini}},\ }\href {\doibase 10.1088/0264-9381/33/12/125005} {\bibfield
  {journal} {\bibinfo  {journal} {Class. Quant. Grav.}\ }\textbf {\bibinfo
  {volume} {33}},\ \bibinfo {pages} {125005} (\bibinfo {year} {2016})},\
  \Eprint {http://arxiv.org/abs/1510.02284} {arXiv:1510.02284 [gr-qc]}
  \BibitemShut {NoStop}%
\bibitem [{\citenamefont {L\"u}\ \emph
  {et~al.}(2015{\natexlab{a}})\citenamefont {L\"u}, \citenamefont {Perkins},
  \citenamefont {Pope},\ and\ \citenamefont {Stelle}}]{PhysRevD.92.124019}%
  \BibitemOpen
  \bibfield  {author} {\bibinfo {author} {\bibfnamefont {H.}~\bibnamefont
  {L\"u}}, \bibinfo {author} {\bibfnamefont {A.}~\bibnamefont {Perkins}},
  \bibinfo {author} {\bibfnamefont {C.~N.}\ \bibnamefont {Pope}}, \ and\
  \bibinfo {author} {\bibfnamefont {K.~S.}\ \bibnamefont {Stelle}},\ }\href
  {\doibase 10.1103/PhysRevD.92.124019} {\bibfield  {journal} {\bibinfo
  {journal} {Phys. Rev. D}\ }\textbf {\bibinfo {volume} {92}},\ \bibinfo
  {pages} {124019} (\bibinfo {year} {2015}{\natexlab{a}})}\BibitemShut
  {NoStop}%
\bibitem [{\citenamefont {El~Hanafy}\ and\ \citenamefont
  {Nashed}(2016)}]{Hanafy:2015yya}%
  \BibitemOpen
  \bibfield  {author} {\bibinfo {author} {\bibfnamefont {W.}~\bibnamefont
  {El~Hanafy}}\ and\ \bibinfo {author} {\bibfnamefont {G.~G.~L.}\ \bibnamefont
  {Nashed}},\ }\href {\doibase 10.1007/s10509-016-2662-y} {\bibfield  {journal}
  {\bibinfo  {journal} {Astrophys. Space Sci.}\ }\textbf {\bibinfo {volume}
  {361}},\ \bibinfo {pages} {68} (\bibinfo {year} {2016})},\ \Eprint
  {http://arxiv.org/abs/1507.07377} {arXiv:1507.07377 [gr-qc]} \BibitemShut
  {NoStop}%
\bibitem [{\citenamefont {L\"u}\ \emph
  {et~al.}(2015{\natexlab{b}})\citenamefont {L\"u}, \citenamefont {Perkins},
  \citenamefont {Pope},\ and\ \citenamefont {Stelle}}]{PhysRevLett.114.171601}%
  \BibitemOpen
  \bibfield  {author} {\bibinfo {author} {\bibfnamefont {H.}~\bibnamefont
  {L\"u}}, \bibinfo {author} {\bibfnamefont {A.}~\bibnamefont {Perkins}},
  \bibinfo {author} {\bibfnamefont {C.~N.}\ \bibnamefont {Pope}}, \ and\
  \bibinfo {author} {\bibfnamefont {K.~S.}\ \bibnamefont {Stelle}},\ }\href
  {\doibase 10.1103/PhysRevLett.114.171601} {\bibfield  {journal} {\bibinfo
  {journal} {Phys. Rev. Lett.}\ }\textbf {\bibinfo {volume} {114}},\ \bibinfo
  {pages} {171601} (\bibinfo {year} {2015}{\natexlab{b}})}\BibitemShut
  {NoStop}%
\bibitem [{\citenamefont {Hassa\"{\i}ne}\ and\ \citenamefont
  {Mart\'{\i}nez}(2007)}]{PhysRevD.75.027502}%
  \BibitemOpen
  \bibfield  {author} {\bibinfo {author} {\bibfnamefont {M.}~\bibnamefont
  {Hassa\"{\i}ne}}\ and\ \bibinfo {author} {\bibfnamefont {C.}~\bibnamefont
  {Mart\'{\i}nez}},\ }\href {\doibase 10.1103/PhysRevD.75.027502} {\bibfield
  {journal} {\bibinfo  {journal} {Phys. Rev. D}\ }\textbf {\bibinfo {volume}
  {75}},\ \bibinfo {pages} {027502} (\bibinfo {year} {2007})}\BibitemShut
  {NoStop}%
\bibitem [{\citenamefont {Sebastiani}\ and\ \citenamefont
  {Zerbini}(2011)}]{Sebastiani:2010kv}%
  \BibitemOpen
  \bibfield  {author} {\bibinfo {author} {\bibfnamefont {L.}~\bibnamefont
  {Sebastiani}}\ and\ \bibinfo {author} {\bibfnamefont {S.}~\bibnamefont
  {Zerbini}},\ }\href {\doibase 10.1140/epjc/s10052-011-1591-8} {\bibfield
  {journal} {\bibinfo  {journal} {Eur. Phys. J.}\ }\textbf {\bibinfo {volume}
  {C71}},\ \bibinfo {pages} {1591} (\bibinfo {year} {2011})},\ \Eprint
  {http://arxiv.org/abs/1012.5230} {arXiv:1012.5230 [gr-qc]} \BibitemShut
  {NoStop}%
\bibitem [{\citenamefont {Brevik}\ \emph {et~al.}(2004)\citenamefont {Brevik},
  \citenamefont {Nojiri}, \citenamefont {Odintsov},\ and\ \citenamefont
  {Vanzo}}]{Brevik:2004sd}%
  \BibitemOpen
  \bibfield  {author} {\bibinfo {author} {\bibfnamefont {I.~H.}\ \bibnamefont
  {Brevik}}, \bibinfo {author} {\bibfnamefont {S.}~\bibnamefont {Nojiri}},
  \bibinfo {author} {\bibfnamefont {S.~D.}\ \bibnamefont {Odintsov}}, \ and\
  \bibinfo {author} {\bibfnamefont {L.}~\bibnamefont {Vanzo}},\ }\href
  {\doibase 10.1103/PhysRevD.70.043520} {\bibfield  {journal} {\bibinfo
  {journal} {Phys. Rev. D}\ }\textbf {\bibinfo {volume} {70}},\ \bibinfo
  {pages} {043520} (\bibinfo {year} {2004})},\ \Eprint
  {http://arxiv.org/abs/hep-th/0401073} {arXiv:hep-th/0401073} \BibitemShut
  {NoStop}%
\bibitem [{\citenamefont {Cognola}\ \emph {et~al.}(2005)\citenamefont
  {Cognola}, \citenamefont {Elizalde}, \citenamefont {Nojiri}, \citenamefont
  {Odintsov},\ and\ \citenamefont {Zerbini}}]{Cognola:2005de}%
  \BibitemOpen
  \bibfield  {author} {\bibinfo {author} {\bibfnamefont {G.}~\bibnamefont
  {Cognola}}, \bibinfo {author} {\bibfnamefont {E.}~\bibnamefont {Elizalde}},
  \bibinfo {author} {\bibfnamefont {S.}~\bibnamefont {Nojiri}}, \bibinfo
  {author} {\bibfnamefont {S.~D.}\ \bibnamefont {Odintsov}}, \ and\ \bibinfo
  {author} {\bibfnamefont {S.}~\bibnamefont {Zerbini}},\ }\href {\doibase
  10.1088/1475-7516/2005/02/010} {\bibfield  {journal} {\bibinfo  {journal}
  {JCAP}\ }\textbf {\bibinfo {volume} {02}},\ \bibinfo {pages} {010} (\bibinfo
  {year} {2005})},\ \Eprint {http://arxiv.org/abs/hep-th/0501096}
  {arXiv:hep-th/0501096} \BibitemShut {NoStop}%
\bibitem [{\citenamefont {Saffari}\ and\ \citenamefont
  {Rahvar}(2008)}]{Saffari:2007zt}%
  \BibitemOpen
  \bibfield  {author} {\bibinfo {author} {\bibfnamefont {R.}~\bibnamefont
  {Saffari}}\ and\ \bibinfo {author} {\bibfnamefont {S.}~\bibnamefont
  {Rahvar}},\ }\href {\doibase 10.1103/PhysRevD.77.104028} {\bibfield
  {journal} {\bibinfo  {journal} {Phys. Rev. D}\ }\textbf {\bibinfo {volume}
  {77}},\ \bibinfo {pages} {104028} (\bibinfo {year} {2008})},\ \Eprint
  {http://arxiv.org/abs/0708.1482} {arXiv:0708.1482 [astro-ph]} \BibitemShut
  {NoStop}%
\bibitem [{\citenamefont {de~la Cruz-Dombriz}\ \emph
  {et~al.}(2009)\citenamefont {de~la Cruz-Dombriz}, \citenamefont {Dobado},\
  and\ \citenamefont {Maroto}}]{delaCruz-Dombriz:2009pzc}%
  \BibitemOpen
  \bibfield  {author} {\bibinfo {author} {\bibfnamefont {A.}~\bibnamefont
  {de~la Cruz-Dombriz}}, \bibinfo {author} {\bibfnamefont {A.}~\bibnamefont
  {Dobado}}, \ and\ \bibinfo {author} {\bibfnamefont {A.~L.}\ \bibnamefont
  {Maroto}},\ }\href {\doibase 10.1103/PhysRevD.80.124011} {\bibfield
  {journal} {\bibinfo  {journal} {Phys. Rev. D}\ }\textbf {\bibinfo {volume}
  {80}},\ \bibinfo {pages} {124011} (\bibinfo {year} {2009})},\ \bibinfo {note}
  {[Erratum: Phys.Rev.D 83, 029903 (2011)]},\ \Eprint
  {http://arxiv.org/abs/0907.3872} {arXiv:0907.3872 [gr-qc]} \BibitemShut
  {NoStop}%
\bibitem [{\citenamefont {Nashed}(2014)}]{Nashed:2014sea}%
  \BibitemOpen
  \bibfield  {author} {\bibinfo {author} {\bibfnamefont {G.~G.~L.}\
  \bibnamefont {Nashed}},\ }\href {\doibase 10.1209/0295-5075/105/10001}
  {\bibfield  {journal} {\bibinfo  {journal} {EPL}\ }\textbf {\bibinfo {volume}
  {105}},\ \bibinfo {pages} {10001} (\bibinfo {year} {2014})},\ \Eprint
  {http://arxiv.org/abs/1501.00974} {arXiv:1501.00974 [gr-qc]} \BibitemShut
  {NoStop}%
\bibitem [{\citenamefont {Larranaga}(2012)}]{Larranaga:2011fv}%
  \BibitemOpen
  \bibfield  {author} {\bibinfo {author} {\bibfnamefont {A.}~\bibnamefont
  {Larranaga}},\ }\href {\doibase 10.1007/s12043-012-0278-5} {\bibfield
  {journal} {\bibinfo  {journal} {Pramana}\ }\textbf {\bibinfo {volume} {78}},\
  \bibinfo {pages} {697} (\bibinfo {year} {2012})},\ \Eprint
  {http://arxiv.org/abs/1108.6325} {arXiv:1108.6325 [gr-qc]} \BibitemShut
  {NoStop}%
\bibitem [{\citenamefont {de~la Cruz-Dombriz}\ and\ \citenamefont
  {Saez-Gomez}(2012)}]{delaCruz-Dombriz:2012bni}%
  \BibitemOpen
  \bibfield  {author} {\bibinfo {author} {\bibfnamefont {A.}~\bibnamefont
  {de~la Cruz-Dombriz}}\ and\ \bibinfo {author} {\bibfnamefont
  {D.}~\bibnamefont {Saez-Gomez}},\ }\href {\doibase 10.3390/e14091717}
  {\bibfield  {journal} {\bibinfo  {journal} {Entropy}\ }\textbf {\bibinfo
  {volume} {14}},\ \bibinfo {pages} {1717} (\bibinfo {year} {2012})},\ \Eprint
  {http://arxiv.org/abs/1207.2663} {arXiv:1207.2663 [gr-qc]} \BibitemShut
  {NoStop}%
\bibitem [{\citenamefont {Moon}\ \emph {et~al.}(2011)\citenamefont {Moon},
  \citenamefont {Myung},\ and\ \citenamefont {Son}}]{Moon:2011hq}%
  \BibitemOpen
  \bibfield  {author} {\bibinfo {author} {\bibfnamefont {T.}~\bibnamefont
  {Moon}}, \bibinfo {author} {\bibfnamefont {Y.~S.}\ \bibnamefont {Myung}}, \
  and\ \bibinfo {author} {\bibfnamefont {E.~J.}\ \bibnamefont {Son}},\ }\href
  {\doibase 10.1007/s10714-011-1225-3} {\bibfield  {journal} {\bibinfo
  {journal} {Gen. Rel. Grav.}\ }\textbf {\bibinfo {volume} {43}},\ \bibinfo
  {pages} {3079} (\bibinfo {year} {2011})},\ \Eprint
  {http://arxiv.org/abs/1101.1153} {arXiv:1101.1153 [gr-qc]} \BibitemShut
  {NoStop}%
\bibitem [{\citenamefont {{Buchdahl}}(1970)}]{1970MNRAS.150....1B}%
  \BibitemOpen
  \bibfield  {author} {\bibinfo {author} {\bibfnamefont {H.~A.}\ \bibnamefont
  {{Buchdahl}}},\ }\href {\doibase 10.1093/mnras/150.1.1} {\bibfield  {journal}
  {\bibinfo  {journal} {mnras}\ }\textbf {\bibinfo {volume} {150}},\ \bibinfo
  {pages} {1} (\bibinfo {year} {1970})}\BibitemShut {NoStop}%
\bibitem [{\citenamefont {{Amendola}}\ and\ \citenamefont
  {{Tsujikawa}}(2010)}]{2010deto.book.....A}%
  \BibitemOpen
  \bibfield  {author} {\bibinfo {author} {\bibfnamefont {L.}~\bibnamefont
  {{Amendola}}}\ and\ \bibinfo {author} {\bibfnamefont {S.}~\bibnamefont
  {{Tsujikawa}}},\ }\href@noop {} {\emph {\bibinfo {title} {Dark Energy :
  Theory and Observations by Luca Amendola and Shinji Tsujikawa.~Cambridge
  University Press, 2010.~ISBN: 9780521516006}}}\ (\bibinfo {year}
  {2010})\BibitemShut {NoStop}%
\bibitem [{\citenamefont {Capozziello}\ \emph {et~al.}(2003)\citenamefont
  {Capozziello}, \citenamefont {Cardone}, \citenamefont {Carloni},\ and\
  \citenamefont {Troisi}}]{Capozziello:2003gx}%
  \BibitemOpen
  \bibfield  {author} {\bibinfo {author} {\bibfnamefont {S.}~\bibnamefont
  {Capozziello}}, \bibinfo {author} {\bibfnamefont {V.~F.}\ \bibnamefont
  {Cardone}}, \bibinfo {author} {\bibfnamefont {S.}~\bibnamefont {Carloni}}, \
  and\ \bibinfo {author} {\bibfnamefont {A.}~\bibnamefont {Troisi}},\ }\href
  {\doibase 10.1142/S0218271803004407} {\bibfield  {journal} {\bibinfo
  {journal} {Int. J. Mod. Phys.}\ }\textbf {\bibinfo {volume} {D12}},\ \bibinfo
  {pages} {1969} (\bibinfo {year} {2003})},\ \Eprint
  {http://arxiv.org/abs/astro-ph/0307018} {arXiv:astro-ph/0307018 [astro-ph]}
  \BibitemShut {NoStop}%
\bibitem [{\citenamefont {Capozziello}(2002)}]{Capozziello:2002rd}%
  \BibitemOpen
  \bibfield  {author} {\bibinfo {author} {\bibfnamefont {S.}~\bibnamefont
  {Capozziello}},\ }\href {\doibase 10.1142/S0218271802002025} {\bibfield
  {journal} {\bibinfo  {journal} {Int. J. Mod. Phys.}\ ,\ \bibinfo {pages}
  {483}} (\bibinfo {year} {2002})},\ \Eprint
  {http://arxiv.org/abs/gr-qc/0201033} {arXiv:gr-qc/0201033 [gr-qc]}
  \BibitemShut {NoStop}%
\bibitem [{\citenamefont {Nojiri}\ and\ \citenamefont
  {Odintsov}(2003)}]{Nojiri:2003ft}%
  \BibitemOpen
  \bibfield  {author} {\bibinfo {author} {\bibfnamefont {S.}~\bibnamefont
  {Nojiri}}\ and\ \bibinfo {author} {\bibfnamefont {S.~D.}\ \bibnamefont
  {Odintsov}},\ }\href {\doibase 10.1103/PhysRevD.68.123512} {\bibfield
  {journal} {\bibinfo  {journal} {Phys. Rev. D}\ }\textbf {\bibinfo {volume}
  {68}},\ \bibinfo {pages} {123512} (\bibinfo {year} {2003})},\ \Eprint
  {http://arxiv.org/abs/hep-th/0307288} {arXiv:hep-th/0307288} \BibitemShut
  {NoStop}%
\bibitem [{\citenamefont {Carroll}\ \emph {et~al.}(2004)\citenamefont
  {Carroll}, \citenamefont {Duvvuri}, \citenamefont {Trodden},\ and\
  \citenamefont {Turner}}]{Carroll:2003wy}%
  \BibitemOpen
  \bibfield  {author} {\bibinfo {author} {\bibfnamefont {S.~M.}\ \bibnamefont
  {Carroll}}, \bibinfo {author} {\bibfnamefont {V.}~\bibnamefont {Duvvuri}},
  \bibinfo {author} {\bibfnamefont {M.}~\bibnamefont {Trodden}}, \ and\
  \bibinfo {author} {\bibfnamefont {M.~S.}\ \bibnamefont {Turner}},\ }\href
  {\doibase 10.1103/PhysRevD.70.043528} {\bibfield  {journal} {\bibinfo
  {journal} {Phys. Rev.}\ ,\ \bibinfo {pages} {043528}} (\bibinfo {year}
  {2004})},\ \Eprint {http://arxiv.org/abs/astro-ph/0306438}
  {arXiv:astro-ph/0306438 [astro-ph]} \BibitemShut {NoStop}%
\bibitem [{\citenamefont {{Cognola}}\ \emph {et~al.}(2005)\citenamefont
  {{Cognola}}, \citenamefont {{Elizalde}}, \citenamefont {{Nojiri}},
  \citenamefont {{Odintsov}},\ and\ \citenamefont
  {{Zerbini}}}]{2005JCAP...02..010C}%
  \BibitemOpen
  \bibfield  {author} {\bibinfo {author} {\bibfnamefont {G.}~\bibnamefont
  {{Cognola}}}, \bibinfo {author} {\bibfnamefont {E.}~\bibnamefont
  {{Elizalde}}}, \bibinfo {author} {\bibfnamefont {S.}~\bibnamefont
  {{Nojiri}}}, \bibinfo {author} {\bibfnamefont {S.~D.}\ \bibnamefont
  {{Odintsov}}}, \ and\ \bibinfo {author} {\bibfnamefont {S.}~\bibnamefont
  {{Zerbini}}},\ }\href {\doibase 10.1088/1475-7516/2005/02/010} {\bibfield
  {journal} {\bibinfo  {journal} {jcap}\ }\textbf {\bibinfo {volume} {2}},\
  \bibinfo {eid} {010} (\bibinfo {year} {2005})},\ \Eprint
  {http://arxiv.org/abs/hep-th/0501096} {hep-th/0501096} \BibitemShut {NoStop}%
\bibitem [{\citenamefont {Koivisto}\ and\ \citenamefont
  {Kurki-Suonio}(2006)}]{Koivisto:2005yc}%
  \BibitemOpen
  \bibfield  {author} {\bibinfo {author} {\bibfnamefont {T.}~\bibnamefont
  {Koivisto}}\ and\ \bibinfo {author} {\bibfnamefont {H.}~\bibnamefont
  {Kurki-Suonio}},\ }\href {\doibase 10.1088/0264-9381/23/7/009} {\bibfield
  {journal} {\bibinfo  {journal} {Class. Quant. Grav.}\ }\textbf {\bibinfo
  {volume} {23}},\ \bibinfo {pages} {2355} (\bibinfo {year} {2006})},\ \Eprint
  {http://arxiv.org/abs/astro-ph/0509422} {arXiv:astro-ph/0509422 [astro-ph]}
  \BibitemShut {NoStop}%
\bibitem [{\citenamefont {Astorino}\ and\ \citenamefont
  {Vigano}(2021)}]{Astorino:2021dju}%
  \BibitemOpen
  \bibfield  {author} {\bibinfo {author} {\bibfnamefont {M.}~\bibnamefont
  {Astorino}}\ and\ \bibinfo {author} {\bibfnamefont {A.}~\bibnamefont
  {Vigano}},\ }\href {\doibase 10.1016/j.physletb.2021.136506} {\bibfield
  {journal} {\bibinfo  {journal} {Phys. Lett. B}\ }\textbf {\bibinfo {volume}
  {820}},\ \bibinfo {pages} {136506} (\bibinfo {year} {2021})},\ \Eprint
  {http://arxiv.org/abs/2104.07686} {arXiv:2104.07686 [gr-qc]} \BibitemShut
  {NoStop}%
\bibitem [{\citenamefont {Ronveaux}(2003)}]{RONVEAUX2003177}%
  \BibitemOpen
  \bibfield  {author} {\bibinfo {author} {\bibfnamefont {A.}~\bibnamefont
  {Ronveaux}},\ }\href {\doibase https://doi.org/10.1016/S0096-3003(02)00331-4}
  {\bibfield  {journal} {\bibinfo  {journal} {Applied Mathematics and
  Computation}\ }\textbf {\bibinfo {volume} {141}},\ \bibinfo {pages} {177 }
  (\bibinfo {year} {2003})},\ \bibinfo {note} {advanced Special Functions and
  Related Topics in Differential Equations, Third Melfi Workshop, Proceedings
  of the Melfi School on Advanced Topics in Mathematics and
  Physics}\BibitemShut {NoStop}%
\bibitem [{\citenamefont {Maier}(2005)}]{MAIER2005171}%
  \BibitemOpen
  \bibfield  {author} {\bibinfo {author} {\bibfnamefont {R.~S.}\ \bibnamefont
  {Maier}},\ }\href {\doibase https://doi.org/10.1016/j.jde.2004.07.020}
  {\bibfield  {journal} {\bibinfo  {journal} {Journal of Differential
  Equations}\ }\textbf {\bibinfo {volume} {213}},\ \bibinfo {pages} {171 }
  (\bibinfo {year} {2005})}\BibitemShut {NoStop}%
\bibitem [{\citenamefont {Sheykhi}(2012)}]{PhysRevD.86.024013}%
  \BibitemOpen
  \bibfield  {author} {\bibinfo {author} {\bibfnamefont {A.}~\bibnamefont
  {Sheykhi}},\ }\href {\doibase 10.1103/PhysRevD.86.024013} {\bibfield
  {journal} {\bibinfo  {journal} {Phys. Rev. D}\ }\textbf {\bibinfo {volume}
  {86}},\ \bibinfo {pages} {024013} (\bibinfo {year} {2012})}\BibitemShut
  {NoStop}%
\bibitem [{\citenamefont {Sheykhi}(2010)}]{Sheykhi:2010zz}%
  \BibitemOpen
  \bibfield  {author} {\bibinfo {author} {\bibfnamefont {A.}~\bibnamefont
  {Sheykhi}},\ }\href {\doibase 10.1140/epjc/s10052-010-1372-9} {\bibfield
  {journal} {\bibinfo  {journal} {Eur. Phys. J.}\ }\textbf {\bibinfo {volume}
  {C69}},\ \bibinfo {pages} {265} (\bibinfo {year} {2010})},\ \Eprint
  {http://arxiv.org/abs/1012.0383} {arXiv:1012.0383 [hep-th]} \BibitemShut
  {NoStop}%
\bibitem [{\citenamefont {Hendi}\ \emph {et~al.}(2010)\citenamefont {Hendi},
  \citenamefont {Sheykhi},\ and\ \citenamefont {Dehghani}}]{Hendi:2010gq}%
  \BibitemOpen
  \bibfield  {author} {\bibinfo {author} {\bibfnamefont {S.~H.}\ \bibnamefont
  {Hendi}}, \bibinfo {author} {\bibfnamefont {A.}~\bibnamefont {Sheykhi}}, \
  and\ \bibinfo {author} {\bibfnamefont {M.~H.}\ \bibnamefont {Dehghani}},\
  }\href {\doibase 10.1140/epjc/s10052-010-1483-3} {\bibfield  {journal}
  {\bibinfo  {journal} {Eur. Phys. J.}\ }\textbf {\bibinfo {volume} {C70}},\
  \bibinfo {pages} {703} (\bibinfo {year} {2010})},\ \Eprint
  {http://arxiv.org/abs/1002.0202} {arXiv:1002.0202 [hep-th]} \BibitemShut
  {NoStop}%
\bibitem [{\citenamefont {Sheykhi}\ \emph {et~al.}(2010)\citenamefont
  {Sheykhi}, \citenamefont {Dehghani},\ and\ \citenamefont
  {Hendi}}]{PhysRevD.81.084040}%
  \BibitemOpen
  \bibfield  {author} {\bibinfo {author} {\bibfnamefont {A.}~\bibnamefont
  {Sheykhi}}, \bibinfo {author} {\bibfnamefont {M.~H.}\ \bibnamefont
  {Dehghani}}, \ and\ \bibinfo {author} {\bibfnamefont {S.~H.}\ \bibnamefont
  {Hendi}},\ }\href {\doibase 10.1103/PhysRevD.81.084040} {\bibfield  {journal}
  {\bibinfo  {journal} {Phys. Rev. D}\ }\textbf {\bibinfo {volume} {81}},\
  \bibinfo {pages} {084040} (\bibinfo {year} {2010})}\BibitemShut {NoStop}%
\bibitem [{\citenamefont {Cognola}\ \emph {et~al.}(2011)\citenamefont
  {Cognola}, \citenamefont {Gorbunova}, \citenamefont {Sebastiani},\ and\
  \citenamefont {Zerbini}}]{PhysRevD.84.023515}%
  \BibitemOpen
  \bibfield  {author} {\bibinfo {author} {\bibfnamefont {G.}~\bibnamefont
  {Cognola}}, \bibinfo {author} {\bibfnamefont {O.}~\bibnamefont {Gorbunova}},
  \bibinfo {author} {\bibfnamefont {L.}~\bibnamefont {Sebastiani}}, \ and\
  \bibinfo {author} {\bibfnamefont {S.}~\bibnamefont {Zerbini}},\ }\href
  {\doibase 10.1103/PhysRevD.84.023515} {\bibfield  {journal} {\bibinfo
  {journal} {Phys. Rev. D}\ }\textbf {\bibinfo {volume} {84}},\ \bibinfo
  {pages} {023515} (\bibinfo {year} {2011})}\BibitemShut {NoStop}%
\bibitem [{\citenamefont {Zheng}\ and\ \citenamefont
  {Yang}(2018)}]{Zheng:2018fyn}%
  \BibitemOpen
  \bibfield  {author} {\bibinfo {author} {\bibfnamefont {Y.}~\bibnamefont
  {Zheng}}\ and\ \bibinfo {author} {\bibfnamefont {R.-J.}\ \bibnamefont
  {Yang}},\ }\href {\doibase 10.1140/epjc/s10052-018-6167-4} {\bibfield
  {journal} {\bibinfo  {journal} {Eur. Phys. J.}\ }\textbf {\bibinfo {volume}
  {C78}},\ \bibinfo {pages} {682} (\bibinfo {year} {2018})},\ \Eprint
  {http://arxiv.org/abs/1806.09858} {arXiv:1806.09858 [gr-qc]} \BibitemShut
  {NoStop}%
\bibitem [{\citenamefont {Kim}\ and\ \citenamefont {Kim}(2012)}]{Kim:2012cma}%
  \BibitemOpen
  \bibfield  {author} {\bibinfo {author} {\bibfnamefont {W.}~\bibnamefont
  {Kim}}\ and\ \bibinfo {author} {\bibfnamefont {Y.}~\bibnamefont {Kim}},\
  }\href {\doibase 10.1016/j.physletb.2012.11.017} {\bibfield  {journal}
  {\bibinfo  {journal} {Phys. Lett.}\ }\textbf {\bibinfo {volume} {B718}},\
  \bibinfo {pages} {687} (\bibinfo {year} {2012})},\ \Eprint
  {http://arxiv.org/abs/1207.5318} {arXiv:1207.5318 [gr-qc]} \BibitemShut
  {NoStop}%
\bibitem [{\citenamefont {{D'Inverno}}(1992)}]{1992ier..book.....D}%
  \BibitemOpen
  \bibfield  {author} {\bibinfo {author} {\bibfnamefont {R.~A.}\ \bibnamefont
  {{D'Inverno}}},\ }\href@noop {} {\emph {\bibinfo {title} {Internationale
  Elektronische Rundschau}}}\ (\bibinfo {year} {1992})\BibitemShut {NoStop}%
\bibitem [{\citenamefont {Misner}\ \emph {et~al.}(1973)\citenamefont {Misner},
  \citenamefont {Thorne},\ and\ \citenamefont {Wheeler}}]{Misner:1974qy}%
  \BibitemOpen
  \bibfield  {author} {\bibinfo {author} {\bibfnamefont {C.~W.}\ \bibnamefont
  {Misner}}, \bibinfo {author} {\bibfnamefont {K.~S.}\ \bibnamefont {Thorne}},
  \ and\ \bibinfo {author} {\bibfnamefont {J.~A.}\ \bibnamefont {Wheeler}},\
  }\href@noop {} {\emph {\bibinfo {title} {{Gravitation}}}}\ (\bibinfo
  {publisher} {W. H. Freeman},\ \bibinfo {address} {San Francisco},\ \bibinfo
  {year} {1973})\BibitemShut {NoStop}%
\bibitem [{\citenamefont {Heusler}(1998)}]{Heusler:1997ui}%
  \BibitemOpen
  \bibfield  {author} {\bibinfo {author} {\bibfnamefont {M.}~\bibnamefont
  {Heusler}},\ }\href {\doibase 10.1007/978-3-540-49535-2_7} {\bibfield
  {journal} {\bibinfo  {journal} {Lect. Notes Phys.}\ }\textbf {\bibinfo
  {volume} {514}},\ \bibinfo {pages} {157} (\bibinfo {year}
  {1998})}\BibitemShut {NoStop}%
\bibitem [{\citenamefont {Robinson}(1975)}]{Robinson:1975bv}%
  \BibitemOpen
  \bibfield  {author} {\bibinfo {author} {\bibfnamefont {D.~C.}\ \bibnamefont
  {Robinson}},\ }\href {\doibase 10.1103/PhysRevLett.34.905} {\bibfield
  {journal} {\bibinfo  {journal} {Phys. Rev. Lett.}\ }\textbf {\bibinfo
  {volume} {34}},\ \bibinfo {pages} {905} (\bibinfo {year} {1975})}\BibitemShut
  {NoStop}%
\bibitem [{\citenamefont {'t~Hooft}\ and\ \citenamefont
  {Veltman}(1974)}]{tHooft:1974toh}%
  \BibitemOpen
  \bibfield  {author} {\bibinfo {author} {\bibfnamefont {G.}~\bibnamefont
  {'t~Hooft}}\ and\ \bibinfo {author} {\bibfnamefont {M.~J.~G.}\ \bibnamefont
  {Veltman}},\ }\href@noop {} {\bibfield  {journal} {\bibinfo  {journal} {Ann.
  Inst. H. Poincare Phys. Theor. A}\ }\textbf {\bibinfo {volume} {20}},\
  \bibinfo {pages} {69} (\bibinfo {year} {1974})}\BibitemShut {NoStop}%
\bibitem [{\citenamefont {Glampedakis}\ and\ \citenamefont
  {Babak}(2006)}]{Glampedakis:2005cf}%
  \BibitemOpen
  \bibfield  {author} {\bibinfo {author} {\bibfnamefont {K.}~\bibnamefont
  {Glampedakis}}\ and\ \bibinfo {author} {\bibfnamefont {S.}~\bibnamefont
  {Babak}},\ }\href {\doibase 10.1088/0264-9381/23/12/013} {\bibfield
  {journal} {\bibinfo  {journal} {Class. Quant. Grav.}\ }\textbf {\bibinfo
  {volume} {23}},\ \bibinfo {pages} {4167} (\bibinfo {year} {2006})},\ \Eprint
  {http://arxiv.org/abs/gr-qc/0510057} {arXiv:gr-qc/0510057} \BibitemShut
  {NoStop}%
\bibitem [{\citenamefont {Papadopoulos}\ and\ \citenamefont
  {Kokkotas}(2018)}]{Papadopoulos:2018nvd}%
  \BibitemOpen
  \bibfield  {author} {\bibinfo {author} {\bibfnamefont {G.~O.}\ \bibnamefont
  {Papadopoulos}}\ and\ \bibinfo {author} {\bibfnamefont {K.~D.}\ \bibnamefont
  {Kokkotas}},\ }\href {\doibase 10.1088/1361-6382/aad7f4} {\bibfield
  {journal} {\bibinfo  {journal} {Class. Quant. Grav.}\ }\textbf {\bibinfo
  {volume} {35}},\ \bibinfo {pages} {185014} (\bibinfo {year} {2018})},\
  \Eprint {http://arxiv.org/abs/1807.08594} {arXiv:1807.08594 [gr-qc]}
  \BibitemShut {NoStop}%
\bibitem [{\citenamefont {Carson}\ and\ \citenamefont
  {Yagi}(2020)}]{Carson:2020dez}%
  \BibitemOpen
  \bibfield  {author} {\bibinfo {author} {\bibfnamefont {Z.}~\bibnamefont
  {Carson}}\ and\ \bibinfo {author} {\bibfnamefont {K.}~\bibnamefont {Yagi}},\
  }\href {\doibase 10.1103/PhysRevD.101.084030} {\bibfield  {journal} {\bibinfo
   {journal} {Phys. Rev. D}\ }\textbf {\bibinfo {volume} {101}},\ \bibinfo
  {pages} {084030} (\bibinfo {year} {2020})},\ \Eprint
  {http://arxiv.org/abs/2002.01028} {arXiv:2002.01028 [gr-qc]} \BibitemShut
  {NoStop}%
\bibitem [{\citenamefont {Astashenok}\ \emph
  {et~al.}(2021{\natexlab{a}})\citenamefont {Astashenok}, \citenamefont
  {Capozziello}, \citenamefont {Odintsov},\ and\ \citenamefont
  {Oikonomou}}]{Astashenok:2021xpm}%
  \BibitemOpen
  \bibfield  {author} {\bibinfo {author} {\bibfnamefont {A.~V.}\ \bibnamefont
  {Astashenok}}, \bibinfo {author} {\bibfnamefont {S.}~\bibnamefont
  {Capozziello}}, \bibinfo {author} {\bibfnamefont {S.~D.}\ \bibnamefont
  {Odintsov}}, \ and\ \bibinfo {author} {\bibfnamefont {V.~K.}\ \bibnamefont
  {Oikonomou}},\ }\href {\doibase 10.1209/0295-5075/134/59001} {\bibfield
  {journal} {\bibinfo  {journal} {EPL}\ }\textbf {\bibinfo {volume} {134}},\
  \bibinfo {pages} {59001} (\bibinfo {year} {2021}{\natexlab{a}})},\ \Eprint
  {http://arxiv.org/abs/2106.01234} {arXiv:2106.01234 [gr-qc]} \BibitemShut
  {NoStop}%
\bibitem [{\citenamefont {Astashenok}\ \emph
  {et~al.}(2021{\natexlab{b}})\citenamefont {Astashenok}, \citenamefont
  {Capozziello}, \citenamefont {Odintsov},\ and\ \citenamefont
  {Oikonomou}}]{Astashenok:2021peo}%
  \BibitemOpen
  \bibfield  {author} {\bibinfo {author} {\bibfnamefont {A.~V.}\ \bibnamefont
  {Astashenok}}, \bibinfo {author} {\bibfnamefont {S.}~\bibnamefont
  {Capozziello}}, \bibinfo {author} {\bibfnamefont {S.~D.}\ \bibnamefont
  {Odintsov}}, \ and\ \bibinfo {author} {\bibfnamefont {V.~K.}\ \bibnamefont
  {Oikonomou}},\ }\href {\doibase 10.1016/j.physletb.2021.136222} {\bibfield
  {journal} {\bibinfo  {journal} {Phys. Lett. B}\ }\textbf {\bibinfo {volume}
  {816}},\ \bibinfo {pages} {136222} (\bibinfo {year} {2021}{\natexlab{b}})},\
  \Eprint {http://arxiv.org/abs/2103.04144} {arXiv:2103.04144 [gr-qc]}
  \BibitemShut {NoStop}%
\bibitem [{\citenamefont {Astashenok}\ \emph {et~al.}(2020)\citenamefont
  {Astashenok}, \citenamefont {Capozziello}, \citenamefont {Odintsov},\ and\
  \citenamefont {Oikonomou}}]{Astashenok:2020qds}%
  \BibitemOpen
  \bibfield  {author} {\bibinfo {author} {\bibfnamefont {A.~V.}\ \bibnamefont
  {Astashenok}}, \bibinfo {author} {\bibfnamefont {S.}~\bibnamefont
  {Capozziello}}, \bibinfo {author} {\bibfnamefont {S.~D.}\ \bibnamefont
  {Odintsov}}, \ and\ \bibinfo {author} {\bibfnamefont {V.~K.}\ \bibnamefont
  {Oikonomou}},\ }\href {\doibase 10.1016/j.physletb.2020.135910} {\bibfield
  {journal} {\bibinfo  {journal} {Phys. Lett. B}\ }\textbf {\bibinfo {volume}
  {811}},\ \bibinfo {pages} {135910} (\bibinfo {year} {2020})},\ \Eprint
  {http://arxiv.org/abs/2008.10884} {arXiv:2008.10884 [gr-qc]} \BibitemShut
  {NoStop}%
\end{thebibliography}
%

\end{document}